\documentclass{article}
\usepackage{graphicx} 

\usepackage{mathtools}

\usepackage{color,xcolor,ucs}
\usepackage{mathtools}   \usepackage{tikz} 
\usepackage{ amssymb }
\usepackage{extarrows} 
\usepackage{pgf,tikz}
\usepackage{float}
\usetikzlibrary{positioning}
\usetikzlibrary{shapes.geometric}
\usetikzlibrary{shapes.misc}
\usetikzlibrary{arrows}
\usepackage{caption}
\usepackage{mathrsfs}
\usetikzlibrary{arrows,shapes,automata,backgrounds,petri,positioning}
\usetikzlibrary{decorations.pathmorphing}
\usetikzlibrary{decorations.shapes}
\usetikzlibrary{decorations.text}
\usetikzlibrary{decorations.fractals}
\usetikzlibrary{decorations.footprints}
\usetikzlibrary{shadows}
\usetikzlibrary{calc}
\usetikzlibrary{spy}
\usepackage{amsmath}
\usepackage{array}
\usepackage{ amssymb }
\usepackage{braket}
\usepackage{qcircuit}
\usepackage{soul}
\usepackage{braket} 
\usepackage{relsize}

\usepackage{amsmath}
\usepackage{ amssymb }
\usepackage{braket}
\usepackage{qcircuit}
\usepackage{soul}
\usepackage{braket}

\begin{document}


\title{{\LARGE Optimal, and approximately optimal, quantum strategies for $\mathrm{XOR}^{*}$ and $\mathrm{FFL}$ games}}
\author{Pete Rigas \footnote{Newport Beach, CA 92625, pbr43@cornell.edu, ORCiD: 0009-0003-1053-9720}}
\date{}

 \maketitle 


\abstract{We analyze optimal, and approximately optimal, quantum strategies in two non-local settings based upon previous investigations of the XOR and CHSH games. By building upon previous arguments due to Ostrev in  2016, {\color{blue}[37]}, which characterized approximately optimal, and optimal, strategies that players Alice and Bob can adopt for increasing their respective winning probabilities, we identify additional applications of the framework for analyzing prospective quantum advantage in other game-theoretic settings. In particular, while it is possible for Alice and Bob to realize quantum advantage if they adopt strategies that are dependent upon entanglement, two-dimensional resource systems, and reversible transformations, the magnitude of the duality gap between the performance of classical and quantum strategies has not yet been classified from the perspective of error bounds. Such bounds are not only dependent upon the winning probability if Alice and Bob leverage entanglement in their strategies but also upon a suitably defined representation-theoretic intertwining operation. For the Fortnow-Feige-Lovasz (FFL) game, the Ostrev framework, {\color{blue}[37]}, is applicable, which consists of the following steps: (1) constructing a suitable, nonzero, linear transformation corresponding to the representation-theoretic intertwining operation; (2) demonstrating that the intertwining operation has unit Frobenius norm; (3)  constructing error bounds, through computations of the states, $\big\{  A_k \otimes \textbf{I} \big\}  \ket{\psi}$, and $\big\{ \textbf{I} \otimes   \big( \frac{\pm B_{kl} + B_{lk}}{\sqrt{2}} \big)   \big\}  \ket{\psi}$ for the optimal state $\ket{\psi}$ of a quantum strategy and the collection of responses $A$ and $B$ from Alice and Bob, respectively; (4) constructing additional error bound to account for the inclusion of additional $A$, or $B$ terms in tensor products of operators; (5) obtaining Frobenius norm upper bounds. As alluded to in several comparisons with multiplayer game-theoretic settings, it continues to remain of interest to further adapt this framework to other games with less regular structures. \textit{{Keywords}: quantum games, non-locality, quantum computation, entangled states, verification} \footnote{\textbf{MSC Class}: 81P02; 81Q02}}

\section{Introduction}

\subsection{Overview}

While quantum computation has attracted wide attention within the Mathematics and Physics communities as efforts have been dedicated towards: classifying the computational complexity of training variational quantum algorithms, {\color{blue}[3]}; variational inference and classification, {\color{blue}[2,7,16]}; solving systems of differential equations with kernel methods, {\color{blue}[38]}; in addition to energy applications, {\color{blue}[39]}, other topics of theoretical interest which are significant for the notions of optimal strategies for quantum games introduced in {\color{blue}[37]}, continue to remain of interest to explore. Along these lines, several related questions of interest not only have connections with the recoverability of quantum information, {\color{blue}[50]}, but also with the protocols used by autonomous agents for reaching decisions through measurement and entanglement, {\color{blue}[51]}. Historically, while there have been decades-long debates between Einstein and Bohr regarding the fundamental nature of the universe is probabilistic or deterministic, {\color{blue}[49]}, probabilistic frameworks, leveraged for computational and information-theoretic purposes alike (particularly through the quantum channel, {\color{blue}[48]}), whether quantum advantage extends to a wider variety of algorithms than those that have been examined in the NISQ era is of great value to further characterize. In the best case scenario, one could expect for there to be valuable Mathematical structures of algorithms, or information processing tasks such as the one considered in this work, which could shed light upon aspects underlying the geometric and topological behavior of spacetime, {\color{blue}[1]}, aspects of quantum memory over a channel, {\color{blue}[4]}, correlation through noise, {\color{blue}[36]}, amongst several related questions, {\color{blue}[8},{\color{blue}25},{\color{blue}44]}. 

Albeit the fact that quantum computation in the NISQ era has provided precise theoretical predictions underlying, for example: the transition dynamics and probability theory {\color{blue}[41,43]}; group theory, and related areas, {\color{blue}[6,34,35,43,47]} simulation of waveguide modes, {\color{blue}[13]}; metrology and cyber-physical systems, {\color{blue}[17,26]}; stochastic optimization, {\color{blue}[18]}; and even various speedups in runtime appearing in problems from the aerospace and engineering domains {\color{blue}[19]}, whether similar computational speedups in runtime will continue to be discovered is unknown. Generally speaking, in several game-theoretic settings advantages of quantum information have previously been identified through optimal strategies, also known as strategic advantage, {\color{blue}[52]}, in which Alice and Bob can distribute information across a quantum channel in the presence of noise or also through the action of an adversary. In several previous works from the literature, {\color{blue}[10},{\color{blue}11},{\color{blue}14},{\color{blue}16},{\color{blue}27},{\color{blue}40]}, agents have increased their winning probabilities of several games of interest which is quantified through a performance gap between the respective winning probabilities using Classical and quantum strategies. However, although quantum advantage has been identified through the construction of objects specifically related to: lower bounds on the entanglement, {\color{blue}[43]}; experimental realizations on near-term quantum hardware, {\color{blue}[10]}; decoherence, {\color{blue}[15]}; suitably defined embeddings, {\color{blue}[16]}; the physicality of quantum information, {\color{blue}[9},{\color{blue}12]}, formulating error bounds and associated commutation relations with respect to a suitably defined linear operator, is an alternative perspective which has thus far only been obtained for XOR games, {\color{blue}[37]}.

Recent efforts devoted towards characterizing nonlinear transformations, {\color{blue}[13]}, appear in lower bounds for quantum learning {\color{blue}[11]}, expressions for gate decomposition {\color{blue}[19]}, as well as for data classification with quantum convolutional neural networks {\color{blue}[12]}. Notwithstanding of such developments, suitably defined transformations introduced under different assumptions are of significance as has been already demonstrated for rank-one approximations of a game, {\color{blue}[8]}, in which various properties of the entangled optimal value are obtained. From a similar perspective of leveraging fundamental properties of entanglement for quantum advantage, in this work we formulate how likely the optimal value is to change by reversing (alternatively, anticommuting) the direction of the action of $T$ operators discussed further in the next section. As a result we establish several points of comparisons between the XOR, XOR* and FFL games. Clearly despite the fact that each game has an identical number of participants, the underlying probability distribution which determines the answer that Alice and Bob  provide to the referree, particularly from the optimality perspective, has not yet been examined beyond the XOR setting. Along these lines we are interested in focusing on obtaining inequalities which are upper bounded with Frobenius norms of various transformations of interest.

To further contribute to the fields of quantum computation and quantum information theory, we build upon previous work which has demonstrated: applications of quantum algorithms focused on solving stochastic differential equations, {\color{blue}[28,38]}; other computationally intensive fields, {\color{blue}[30,31,33,34,39,45]}, ranging from simulated annealing, comparisons between classical and quantum correlations and density-functional theory, to rigorously formulate how  connections between quantum physics and game theory are related to the several expected advantages of entangled quantum information. Opposed to classical information, quantum information not only remains a topic of great theoretical interest, but also has a long history through Einstein's characterization of 'spooky action a distance', {\color{blue}[49]}. In the particular game-theoretic setting considered in this work the entangled structure, albeit initially guaranteeing a winning probability that is $11$ percent higher, can also be related to other games through appropriate senses of duality. Such notions are widely used throughout the Mathematics and Physics literature, especially in the study of phase transitions which have been characterized in various information-theoretic senses through entropy, {\color{blue}[11},{\color{blue}15},{\color{blue}36},{\color{blue}46]}.

Error bounds that have previously been obtained in {\color{blue}[37]} develop arguments for analyzing how two players in the XOR game, Alice and Bob, can interact to maximize a linear functional, in which it is possible that either Alice and Bob win, or that Alice and Bob lose depending upon the third party of the Referee (or Verifier) which examines their responses after each player makes a measurement. Crucially such objects are greatly dependent upon the Frobenius norm operator and its relation to the maximum winning probability using entanglement. Even if entanglement of quantum information is expected to outperform standard properties of classical information, quantifying the expected performance gap from the perspective of computations with respect to the Frobenius norm has not yet been explored beyond the two-player XOR setting. As such, we demonstrate how the quantitative observations first provided in {\color{blue}[37]} can be adapted to a broader variety of game-theoretic settings that have not yet been examined in the literature. 

Within the framework of XOR games for which optimal, and approximately optimal, superior performance of quantum strategies as opposed to classical strategies is ultimately reliant upon entanglement. Qualitatively, albeit the fact that quantum strategies are expected to outperform classical strategies for other games rather than the XOR and CHSH games, the way in which such advantage can be formally described continues to remain of interest. To address potential research directions expressed at the end of {\color{blue}[37]} related to characterizing optimal, and approximately optimal. quantum strategies for other games we state the inequalities, expressed in terms of the Frobenius norm, for strategies that Alice and Bob can adopt as counterparts to strategies that have already been described for the $\mathrm{CHSH}\big( n \big)$ and XOR games for $n\geq 2$ where $n \in \textbf{N}$. Through the discussion of related games, included those under parallel repetition, {\color{blue}[5]}, through the formulation of coloring vertices of a cycle, we characterize error bounds for the $\mathrm{XOR}^{*}$ and FFL games. Following this discussion of the quantum objects involved in optimal, and nearly optimal, strategies, we identify other XOR games of interest, first beginning with the infinite $\mathrm{CHSH^{*}\big( n \big)}$ family, from which arguments are adapted for the FFL game, which has quantum and classical success biases equal to $2 / 3$.

In comparison to the XOR game, the FFL game exhibits a difference choice of the optimal value for the quantum and classical biases of the two players Alice and Bob, which raises differences in several inequalities that are formed based upon the optimal value. When playing an FFL game together, the fact that the optimal value equals $2 / 3$ implies that the interactions between Alice and Bob before the referee determines the outcome of the game are determined by different prefactors preceding $n^2 \sqrt{\epsilon}$ terms appearing in \textbf{Lemma} \textit{1}, \textbf{Lemma} \textit{7} and \textbf{Lemma} \textit{9}. To characterize $\epsilon$-optimality of $\ket{\psi_{\mathrm{FFL}}}$ states from a suitably defined notion of duality between $\ket{\psi_{\mathrm{XOR}}}$ and $\ket{\psi_{\mathrm{XOR}^{*}}}$ states, arguments for obtaining error bounds are provided in \textbf{Lemma} \textit{2}, \textbf{Lemma} \textit{3}, \textbf{Lemma} \textit{5} and \textbf{Lemma} \textit{6}. To exhibit how the arguments in the two-player setting are related to the multiplayer XOR setting we describe how the Bell states are related to a system of Frobenius norms, as opposed to two Frobenius norms. In such settings, having players in groups larger than two implies that additional sets of inequalities, and new arguments altogether, must be developed to characterize optimality.

\subsection{This paper's contributions}

\noindent This paper characterizes optimal strategies for XOR* and FFL games. In comparison to a previously obtained optimality framework provided in {\color{blue}[37]} for the XOR game from structures of the CHSH game we not only discuss how other notions (particularly, through duality from a single entangled bit) can relate optimal values of an original game to that of its dual but also to error bounds which quantitatively describe how the optimal value changes under different strategies. Central to our approach are the construction of novel error bounds which relate the Frobenius norm of a suitably defined operator to tensor products of responses which Alice and Bob construct for responding to the referee. Although error bounds constructed in the XOR setting, {\color{blue}[37]}, have captured the manner in which the optimal value (denoted $\omega_{\mathrm{XOR}}$) depends upon the left and right actions with respect to the tensor product, other classes of operators for other games could exhibit different properties with respect to their Frobenius norm. To contextualize how such quantitative motivations are related to aspects of quantum information theory, in the next section we provide an overview focusing on the following areas: (1) optimal strategies for the CHSH game; (2) well-posedness of semidefinite programs, along with corresponding primal feasible solutions; (3) estimates for the decay of the optimal value (also known as the winning probability for a game) under entanglement; (4) an overview of previous results in {\color{blue}[37]} for optimal quantum strategies; (5) formulations of error bounds in game-theoretic settings similar to those of the XOR* and FFL games; (6) paper organization.

In the third section we explicitly state the duality notion provided in {\color{blue}[5]} which allows one to relate the optimal value of the XOR game to the dual optimal value of the XOR* game. Although the optimal values for the XOR and XOR* game are equal the fact that the classical-quantum gap is zero provides information regarding other games. The FFL game is of particular interest because the probability distribution,

{\small

\begin{align*}
 \pi_{\mathrm{FFL}} \big( 0 , 0 \big) =  \pi_{\mathrm{FFL}} \big( 0 , 1 \big) = \pi_{\mathrm{FFL}} \big( 1 , 0 \big)    =  1 / 3  , \end{align*}

    \begin{align*} 
     \pi \equiv    \pi_{\mathrm{FFL}} \big( 1 , 1 \big)    =   0       , \\ 
\end{align*}

}

\noindent corresponding to an iid sample $\big( i , j \big) \sim \pi_{\mathrm{FFL}} \times \pi_{\mathrm{FFL}}$ is similar to that of the XOR game,

{\small

\begin{align*}
     \pi_{\mathrm{XOR}} \big(  0 , 0   \big) =  \pi_{\mathrm{XOR}} \big( 0 , 1   \big) =    \pi_{\mathrm{XOR}} \big( 1 , 0   \big) =       \pi_{\mathrm{XOR}} \big( 1 , 1  \big)    =      1 / 4              , \\ 
\end{align*}

}

\noindent corresponding to an iid sample $\big( i , j \big) \sim \pi_{\mathrm{XOR}} \times \pi_{\mathrm{XOR}}$. To this end we demonstrate how the quantities,

{\small

\begin{align*}
    \bigg\{  \frac{A_i + A_j}{2} \bigg\}_{ \{ ( i , j ) : 1 \leq i \leq d_A , 1 \leq j \leq d_B  \}       }   ,  \\ \end{align*}

    \begin{align*} \bigg\{  \frac{\pm B_{kl} + B_{lk}}{\sqrt{2}} \bigg\}_{ \{ ( k , l ) , k \neq l  : 1 \leq k \leq d_A , 1 \leq l \leq d_B  \}       }      ,  \\ 
\end{align*}

}

\noindent are related to the construction of error bounds which will be further described in the next section.

\section{Game-theoretic objects}

\subsection{quantum objects for optimal CHSH strategies}

We define several quantities for an overview of strategies for the infinite $\mathrm{CHSH(n)}$ family of XOR games. First, from the Frobenius norm,

\begin{align*}
  \big|\big| A \big|\big|_F \equiv \sqrt{\overset{m}{\underset{i=1}{\sum}} \overset{n}{\underset{j=1}{\sum}} \big| a_{ij} \big|^2 } = \sqrt{\mathrm{Tr} \big[ A^{\dagger} A \big] }  \text{, } 
\end{align*}

\noindent of an $m \times n$ matrix $A$ with entries $a_{ij}$. The Frobenius norm is widely used throughout quantum Information theory, with connections to the following inequalities,

{\small

\begin{align*}
    \big| \big| \rho - \sigma \big| \big|_1 = \mathrm{Tr} \big[ \big( \rho - \sigma \big)^{\dagger} \big( \rho - \sigma \big)   \big] \equiv \mathrm{Tr} \big| \rho - \sigma \big| , \\ \\  \tag{\textit{l-1 distance between two quantum states}} \end{align*}
    
    \begin{align*} F \big( \rho , \sigma \big) = \big| \big| \sqrt{\rho} \sqrt{\sigma} \big| \big|_1       ,  \end{align*}

    \begin{align*} \tag{\textit{Representation of the fidelity with respect to the l-1 norm}} \end{align*}
    
    \begin{align*}   2 \big( 1 - F \big( \rho , \sigma \big) \big) \leq \big| \big| \rho - \sigma \big| \big|_1 \leq 2 \sqrt{1 - F \big( \rho , \sigma \big)}           , \\ \\ \tag{\textit{Fucsh-van de Graaf inequality}} \end{align*}
    
    \begin{align*}        \big| \big| \mathcal{E} \big( \rho \big) - \mathcal{E} \big( \sigma \big) \big| \big| \leq \big| \big| \rho - \sigma \big| \big|_1     , \\ \\ \tag{\textit{Data-Processing inequality}} \end{align*}
    
    \begin{align*}    \underset{1 \leq i \leq n}{\bigwedge} \big| \big| \mathcal{E} \big( \rho_{i} \big) - \mathcal{E} \big( \sigma_{i } \big) \big| \big|  \equiv    \big| \big| \mathcal{E} \big( \rho_{1\cdots n} \big) - \mathcal{E} \big( \sigma_{1\cdots n } \big) \big| \big| \leq \big| \big| \rho_{1 \cdots n } - \sigma_{1 \cdots n} \big| \big|_1 = \underset{1 \leq i \leq n}{\bigwedge} \big| \big| \rho_{i} - \sigma_{i} \big| \big|     , \\ \\ \tag{\textit{Data-Processing inequality under the repetition operator $\bigwedge$}} \\ 
\end{align*}

}

\noindent where $\mathcal{E}$ denotes a quantum channel and quantum states $\rho$ and $\sigma$. 

\bigskip

\noindent There exists a \textit{linear bijection} $\mathcal{L}$ between the tensor product space, $\textbf{C}^{d_A} \otimes \textbf{C}^{d_B}$, and the space of $d_A \times d_B$ matrices with complex entries, $\mathrm{Mat}_{d_A , d_B} \big( \textbf{C} \big)$, satisfying (\textbf{Lemma} \textit{1}, {\color{blue}[37]}),

\begin{itemize}
\item[$\bullet$] {\textit{Image of the tensor product of two quantum states under} $\mathcal{L}$}: $\forall \ket{u} \in \textbf{C}^{d_A}, \ket{w} \in \textbf{C}^{d_B}, \exists \ket{u^{*}} \in \textbf{C}^{d_B} : \mathcal{L} \big( \ket{u} \otimes \ket{w} \big) = \ket{u} \bra{u^{*}}  \text{, }$ 

\bigskip

\item[$\bullet$] {\textit{Product of a matrix with the image of a quantum state under} $\mathcal{L}$}: $\forall \ket{u} \in \textbf{C}^{d_A}, \exists A \in \mathrm{Mat}_{d_A} \big( \textbf{C} \big) : A \mathcal{L} \big( \ket{u} \big) = \mathcal{L} \big( A \otimes I \ket{u} \big)\text{, }$

\bigskip

\item[$\bullet$] {\textit{Product of the image of a quantum state under $\mathcal{L}$ with the transpose of a matrix}}:  $\forall \ket{w} \in \textbf{C}^{d_B}, \exists B \in \mathrm{Mat}_{d_B} \big( \textbf{C} \big) : \mathcal{L} \big( \ket{w} \big) B^T = \mathcal{L} \big( I \otimes B \ket{w} \big)  \text{, }$

\bigskip

\item[$\bullet$] {\textit{Frobenius norm equality}}: $\forall \ket{w} \in \textbf{C}^{d_B} : \big|\big| \mathcal{L} \big(   \ket{w}     \big) 
 \big|\big|_F = \ket{w}  \text{. } $
\end{itemize}

\noindent where the basis of $\textbf{C}^{d_A} \otimes \textbf{C}^{d_B}$ is of the form $\ket{i} \otimes \ket{j}$,  and the basis for $\mathrm{Mat}_{d_A, d_B} \big( \textbf{C} \big)$ is of the form $\ket{i}\bra{j}$, for $1 \leq i \leq d_A$ and $1 \leq j \leq d_B$.

To quantitatively determine how optimal a quantum strategy for a game is over a Classical strategy, one primary focus of this is the computation of quantities of the form,

{\small

\begin{align*}
 \big| \big|    \bigotimes T^{\mathrm{XOR}}      \big| \big|_F    , \\ \\  \big| \big|   \bigwedge T^{\mathrm{XOR}}      \big| \big|_F  , \\ \\  \big| \big|  T^{\mathrm{XOR}\wedge \cdots \wedge \mathrm{XOR}}      \big| \big|_F  , \\ \\  \big| \big|    T^{\mathrm{FFL}\wedge  \mathrm{FFL}}      \big| \big|_F  , \\ 
\end{align*}

}

\noindent from the collection of suitable operators,

{\small

\begin{align*}
 \bigotimes T^{\mathrm{XOR}} : \underset{N \text{ copies}}{\bigotimes} \big\{   \textbf{C}^{2 \lceil \frac{n}{2} \rceil } \big\}   \longrightarrow        \underset{1 \leq i \leq N}{\bigotimes} \big\{  \textbf{C}^{d_i} \big\}  \text{, }  \end{align*}

 \begin{align*} \bigwedge T^{\mathrm{XOR}} :  \big\{ \textbf{C}^{2 \lceil \frac{n}{2} \rceil } \big\} \bigwedge     \big\{  \textbf{C}^{2 \lceil \frac{n}{2} \rceil } \big\}  \bigwedge \cdots \bigwedge  \big\{ \textbf{C}^{2 \lceil \frac{n}{2} \rceil }  \big\} \longrightarrow        \big\{ \textbf{C}^{d_A} \big\}  \bigwedge \big\{ \textbf{C}^{d_B} \big\}  \bigwedge \big\{ \textbf{C}^{d^{(1)}_B} \big\} \bigwedge\cdots \\ \bigwedge \big\{ \textbf{C}^{d^{(n-2)}_B}  \big\} \text{, } \end{align*}

 \begin{align*}    T^{ \{ \mathrm{XOR}\wedge \cdots \wedge \mathrm{XOR} \} } :     \textbf{C}^{ \{ 2 \lceil \frac{n}{2} \rceil \wedge 2 \lceil \frac{n}{2} \rceil \wedge \cdots \wedge 2 \lceil \frac{n}{2} \rceil \} }  \longrightarrow        \textbf{C}^{ \{ d_A \wedge d_B \wedge d^{(1)}_B \wedge \cdots \wedge {d^{(n-2)}_B } \} }      \text{, } \end{align*}

 \begin{align*}      T^{\mathrm{FFL}\wedge  \mathrm{FFL}} :     \textbf{C}^{ \{ 2 \lceil \frac{n}{2} \rceil \wedge 2 \lceil \frac{n}{2} \rceil       \}    }  \longrightarrow        \textbf{C}^{ \{ d_A \wedge d_B    \}   }     \text{. }
\end{align*}

}

\noindent In an identical way that the dimensions $d_A$ and $d_B$ were introduced for the respective dimensions of the Hilbert spaces belonging to Alice and Bob, by denoting an additional parameter $d_C$ corresponding to the Hilbert space of a third player in the XOR setting, the operator, along with ones in related settings, is a mapping of the form,

{\small

\begin{align*}
      T^{3\mathrm{XOR}} :  \big\{ \textbf{C}^{3 \lceil \frac{n}{3} \rceil } \big\}  \otimes      \big\{ \textbf{C}^{3 \lceil \frac{n}{3} \rceil }   \big\} \otimes  \big\{ \textbf{C}^{3 \lceil \frac{n}{3} \rceil }    \big\}   \longrightarrow        \big\{ \textbf{C}^{d_A} \big\} \otimes \big\{ \textbf{C}^{d_B} \big\} \otimes \big\{ \textbf{C}^{d_C} \big\}     \text{.  }
\end{align*}

}

From the four properties above of $\mathcal{L}$, for two finite sets $S$ and $T$, also define the map $V : S \times T \longrightarrow \big\{ - 1 , 1 \big\}$. From a product probability distribution $\pi$ over $S \times T$, the game proceeds with the Referee examining the responses of Alice and Bob depending upon the entangled state that they share, in which, after sampling a pair $\big( S , T \big) \sim \pi$, and sending one question $s$ to Alice and another question $t$ to Bob,

\begin{align*}
  \big\{   V \big(  s , t \big)   ab  \equiv 1 \big\} \Longleftrightarrow \big\{  \text{Alice and Bob win}   \big\} , \\   \big\{  V \big(  s , t \big)   ab \equiv -1 \big\} \Longleftrightarrow  \big\{ \text{Alice and Bob lose}  \big\}     ,
\end{align*}

\noindent in which, depending upon whether $V \big( s ,t \big) \equiv 1$, or $V \big( s , t \big) \equiv -1$, Alice and Bob must either give the same answers, and opposing answers, to win, respectively. The quantities $a$ denote the answer which Alice provides to the Referee after receiving question $s$, while $b$ denotes the answer which Bob provides to the Referee after receiving question $t$. Before describing the structure of the two-player bias, we provide an overview of scoring functions $V$ used by the referee for other games which are related to the FFL game. In particular, introduce, from the set of all possible questions, and answers, 

\begin{align*}
  Q_1 \times \cdots \times Q_i \times \cdots \times Q_N  \text{, } \\ \\ A_1 \times \cdots \times  A_i \times  \cdots \times A_N \text{, }
\end{align*}

\noindent respectively. The referee's scoring function (or alternatively, the predicate), in the case of an arbitrary number of questions which can be distributed to each participant,

\begin{align*}
  p \big( q_1, \cdots, q_i, \cdots, q_n \big)   \text{, }
\end{align*}

\noindent with scoring function,

\begin{align*}
  V \big( a_1, \cdots, a_i, \cdots, a_n | q_1, \cdots, q_i, \cdots, q_n \big)   \text{. }
\end{align*}

\noindent While quantum-classical gaps in the winning probability have been previously identified for the Torpedo game in two and three dimensions, in which, related to analytical expressions for the optimal value of the Odd-Cycle game, {\color{blue}[5]}, the optimal values of the Torpedo game satisfy, for a dimension $d$ taken to be $2$ or $3$,

{\small \begin{align*}
  {\omega^{\mathrm{Torpedo}}_{Q,\text{ } \{ d\equiv 2\} }} \big/ {\omega^{\mathrm{Torpedo}}_{C,\text{ } \{ d \equiv 2 \} }}  \approx 1.053      \text{, }  \\ \end{align*}

  \begin{align*} {\omega^{\mathrm{Torpedo}}_{Q,\text{ } \{ d\equiv 3\} }} \big/ {\omega^{\mathrm{Torpedo}}_{C,\text{ } \{ d \equiv 3 \} }}   \approx 1.091       , 
\end{align*} }

\noindent where,

{\small \begin{align*}
\omega^{\mathrm{Torpedo}}_{C,\text{ } \{ d\equiv 2 \} } \equiv 3 \big/ {4} \text{, } \end{align*}

\begin{align*} \omega^{\mathrm{Torpedo}}_{C, \text{ } \{ d\equiv 3\} } \equiv {11} \big/ {12}    . 
\end{align*} }

\noindent corresponds to the classical values, and,

{\small \begin{align*}
 \omega^{\mathrm{Torpedo}}_{Q,\text{ } \{ d\equiv 2 \} } \approx 0.79   \text{, } \end{align*}

 \begin{align*} \omega^{\mathrm{Torpedo}}_{Q,\text{ } \{ d\equiv 3 \} } \equiv 1 \text{. }
\end{align*}  }

\bigskip 

\noindent corresponds to the quantum values. In the case of the FFL game, as a possible game could stand to benefit from the optimality perspective for the XOR/CHSH game, {\color{blue}[37]}, specifically through the adaptation of a duality notion between games that is provided in {\color{blue}[5]}. Under parallel repetition, the operation of repeating multiple instances of a game of interest in parallel, the XOR optimal value, as opposed to the FFL optimal value, satisfies,

\begin{align*}
  \omega_{\mathrm{XOR} \wedge  \mathrm{XOR}} \big( G_{\mathrm{XOR} \wedge \mathrm{XOR}} \big)  \equiv \omega \big(  \mathrm{XOR} \wedge \mathrm{XOR} \big)  \equiv \underset{\# \text{ of strong parallel repetitions } j }{\prod}   \omega \big( \mathrm{XOR}\big)^{j} \\    \\ \equiv \underset{1 \leq j \leq 2}{\prod}   \omega \big( \mathrm{XOR}\big)^{j}   \equiv \big\{ \omega \big( \mathrm{XOR} \big) \big\}^2 \text{,} \end{align*}

\noindent while,

\begin{align*}
  \omega_{\mathrm{FFL} \wedge  \mathrm{FFL}} \big( G_{\mathrm{FFL}} \big) \equiv \omega_{\mathrm{FFL}} \big( G_{\mathrm{FFL}} \wedge G_{\mathrm{FFL}} \big)   \equiv \omega \big(  \mathrm{FFL}  \wedge \mathrm{FFL} \big) \equiv  \omega \big( \mathrm{FFL}\big) \equiv 2 / 3  \text{.} \end{align*}

\noindent

 \bigskip

\noindent Equipped with $V$ and $\pi$, there exists a \textit{game matrix} $G$, so that $G_{st} = V \big( s, t \big) \pi \big( s, t \big)$. Subject to the normalization that the sum over all rows and columns equal $1$, ie $\sum_{st} G_{st} \equiv 1$, a \textit{quantum strategy} for the XOR game is denoted with $\mathcal{S}$, with corresponding state $\ket{\psi} \in \textbf{C}^{d_A} \otimes \textbf{C}^{d_B}$. For an XOR game $G$ and strategy $\mathcal{S}$, define,

\begin{align*}
      \beta \big( G , \mathcal{S}  \big)     \equiv \underset{s \in S}{\sum} \underset{t \in T}{\sum} G_{st} \bra{\psi} A_s \otimes B_t \ket{\psi}  \text{, } 
\end{align*}

\noindent as the \textit{success bias}, where the summation runs over all rows and columns $s$ and $t$ of $G$, with the observables in the tensor product taking the form,

{\small \begin{align*}
 A_S \equiv \underset{s \in S}{\bigcup} A_s \equiv \big\{   s  \in S :  A_s \in \big\{ - 1 , + 1 \big\}    \big\}  \subsetneq \textbf{C}^{d_A}      \text{, } \\  \\   B_T \equiv \underset{t \in T}{\bigcup} B_t \equiv \big\{ t \in T  :  B_t \in \big\{ -1 , + 1 \big\}  \big\}  \subsetneq \textbf{C}^{d_B}  \text{, } \\ \\  \widetilde{A}_S \equiv \underset{s \in S}{\bigcup} \widetilde{A}_s \equiv \big\{   s  \in S :  \widetilde{A}_s \in \big\{ - 1 , + 1 \big\}    \big\}  \subsetneq \textbf{C}^{2^{\lceil n / 2 \rceil}} , \\ \\  \widetilde{B}_S \equiv \underset{s \in S}{\bigcup} \widetilde{B}_s \equiv \big\{   t  \in T :  \widetilde{B}_t \in \big\{ - 1 , + 1 \big\}    \big\}  \subsetneq \textbf{C}^{2^{\lceil n / 2 \rceil}}    .   
\end{align*} } 

\noindent The quantity above is related to the probability of winning the XOR game given $\mathcal{S}$, denoted as $\omega \big( G , \mathcal{S} \big)$, as,

\begin{align*}
       \beta \big( G , \mathcal{S}  \big)  = 2   \omega \big( G , \mathcal{S} \big) - 1      \text{. } 
\end{align*}

\noindent As a supremum over all possible $\mathcal{S}$ for $G$, define,

\begin{align*}
 \beta \big( G \big) \equiv  \underset{\mathcal{S}}{\mathrm{sup}} \big\{ \beta \big( G , \mathcal{S} \big) \big\} \text{, } 
\end{align*}

\noindent corresponding to the optimal quantum strategy. From the optimal strategy $\beta \big( G \big)$, the notion of approximately optimal strategies can be introduced, in which for some strictly positive $\epsilon$,

\begin{align*}
\big( 1 - \epsilon \big) \beta \big( G \big)   \leq   \beta \big( G , \mathcal{S} \big)        \leq  \beta \big( G \big) \text{. }
\end{align*}

\noindent In light of the approximately optimal strategies for $G$, when the Referee chooses a question to ask Bob after one of $\big\{ 1 , \cdots , n\big\}$ possible questions which can be raised to Alice, there are $\big\{\forall ij , \exists i \neq j: i , j \in \big\{ 1 , \cdots , n \big\}\big\}$ possible questions that can be asked to Bob. From each possible combination of questions that can be raised to Alice and then Bob, one can form orthonormal bases $\ket{i}$ and $\ket{ij}$, for the \textit{game matrix}, which are of the form,

\begin{align*}
  G =  \frac{1}{4 {n \choose 2}}   \underset{1\leq i \leq j \leq n}{\sum}   \bigg[    \ket{i}\bra{ij}  +   \ket{j} \bra{ij} + \ket{i} \bra{ji} - \ket{j} \bra{ji}             \bigg]    \text{. } 
\end{align*}

\noindent from which the \textit{optimal success bias} for $G$ takes the form, under the correspondence from the superposition of bra-ket states above,

\begin{align*}
    \ket{i}\bra{ij} \longleftrightarrow A_i B_{ij}       \text{, } \\    \ket{j} \bra{ij}  \longleftrightarrow A_j B_{ij}   \text{, } \\  \ket{i}\bra{ji}  \longleftrightarrow  A_i B_{ji}  \text{, } \\                  - \ket{j} \bra{ji}  
 \longleftrightarrow  - A_j B_{ji} \text{. }
\end{align*}

\noindent Taking the summation over quantum states for all $i,j$, provides,

\begin{align*}
  \underset{A_i , B_{jk} , \psi}{\mathrm{sup}} \text{ }  \frac{1}{4 {n \choose 2}}   \underset{1\leq i \leq j \leq n}{\sum}  \bra{\psi} \bigg[       A_i B_{ij} +  A_j B_{ij}  +   A_i B_{ji}  -   A_j B_{ji}  \bigg]            \ket{\psi} 
               \text{, } 
\end{align*}

\noindent which is equal to $1  / {\sqrt{2}}$, {\color{blue}[27]}. This numerical value for the optimal value of the XOR game is related to well-posedness of a semidefinite program that will be further described, through constraints that are taken with respect to the partial ordering '$\succcurlyeq$' over the positive semidefinie cone. Equipped with the generating matrix $G$ for a game of interest, the ordering over the positive semidefinite cone allows one to approximate the optimal value through constraints of the form,

{\small

\begin{align*}
   \underset{1 \leq i \leq n^3}{\sum}      y_{3\mathrm{XOR},i} E_{3\mathrm{XOR},ii}     \succcurlyeq           G_{3\mathrm{XOR},\mathrm{Sym}}        \text{, } \tag{3XOR, Sym} \\  \\ \underset{1 \leq i \leq n^4}{\sum}   y_{4\mathrm{XOR},i} E_{4\mathrm{XOR},ii}    \succcurlyeq G_{4\mathrm{XOR},\mathrm{Sym}} \text{, }  \tag{4XOR, Sym} \\  \\  \underset{1 \leq i \leq n^5}{\sum}  y_{5\mathrm{XOR},i} E_{5\mathrm{XOR},ii}    \succcurlyeq  G_{5\mathrm{XOR},\mathrm{Sym}} \text{, } \tag{5XOR, Sym} \\ \\  \underset{1 \leq i \leq n^N}{\sum}    y_{N\mathrm{XOR},i} E_{N\mathrm{XOR},ii}       \succcurlyeq G_{N\mathrm{XOR},\mathrm{Sym}} \tag{NXOR, Sym}  ,         \\ 
\end{align*}

}

\noindent where the above objects $E$ and $G$ are subject to the conditions,

{\small 

\begin{align*}
     \underset{1 \leq i \leq n^N}{\sum}   y_{N\mathrm{XOR},i} E_{N\mathrm{XOR},ii} - G_{N\mathrm{XOR},\mathrm{Sym}}     \text{, }  \\ \\ \tag{\textit{Positive semidefiniteness over the positive semidefinite cone}}  \end{align*}
     
     \begin{align*} G_{N\mathrm{XOR},\mathrm{Sym}}  \equiv G_{\mathrm{Sym}} \equiv \frac{1}{2} \begin{bmatrix}    0 &   G^{\textbf{T}} \\ G & 0\end{bmatrix} \equiv \frac{1}{2} \begin{bmatrix}    0 &   G_{N\mathrm{XOR}}^{\textbf{T}} \\ G_{N\mathrm{XOR}} & 0\end{bmatrix}   , \\ \\  \tag{\textit{Symmetrized representation of the game matrix}}     \\ 
\end{align*}}

\noindent respectively. The above representation for $G_{N\mathrm{XOR},\mathrm{Sym}}$ is important for several reasons, including: (1) relating the entries of the game matrix $G$ to those of the dual objective function that are used to formally define the semidefinite program,

{\small

\begin{align*}
      \underset{\overset{m}{\underset{i=1}{\sum}} y_i F_i \succcurlyeq G}{\mathrm{inf}}   \big\{ \vec{c} \cdot \vec{y}    \big\}   \text{, }           
\end{align*}

}

\noindent given the choice of $\vec{c}$ such that the dual objective function multiplied by a representation $F$ is larger than $G$ with respect to the ordering of the positive semidefinite cone (ie, $y_i F_i \succcurlyeq G$ for all $i$); (3) connections between the Schmidt decompositions,

{\small

\begin{align*}
\overset{s 2^{\lfloor \frac{n}{2} \rfloor}}{\underset{i=1}{\sum} } \sqrt{\lambda^{\mathrm{XOR}}_i}  \big\{   \ket{u_i} \otimes \ket{v_i}  \big\}     ,  \\ \\   \overset{s 2^{\lfloor \frac{n}{2} \rfloor}}{\underset{i=1}{\sum} } \sqrt{\lambda^{\mathrm{XOR}^{*}}_i}  \big\{   \ket{u_i} \otimes \ket{v_i}  \big\}  , \\ \\    \overset{s 2^{\lfloor \frac{n}{2} \rfloor}}{\underset{i=1}{\sum} } \sqrt{\lambda^{\mathrm{FFL}}_i}  \big\{   \ket{u_i} \otimes \ket{v_i}  \big\}  , 
\end{align*}

}

\noindent for the XOR, $\mathrm{XOR}^{*}$ and FFL games, respectively, given the choice of orthonormal bases through the linear combinations,

{\small

\begin{align*}
  \underset{1 \leq i \leq d_A}{\underset{\textbf{C}^{d_A}}{\mathrm{span}}} \big\{     \ket{u^{\mathrm{XOR}^*}_i }  \big\}   , \\ \\   \underset{1 \leq j \leq d_B}{\underset{\textbf{C}^{d_B}}{\mathrm{span}}} \big\{   \ket{v^{\mathrm{XOR}^*}_j }  \big\}   , \\ \\  \underset{1 \leq j \leq d_B}{\underset{\textbf{C}^{d_B}}{\mathrm{span}}} \big\{   \ket{u^{\mathrm{FFL}}_i }  \big\}  , \\ \\ \underset{1 \leq j \leq d_B}{\underset{\textbf{C}^{d_B}}{\mathrm{span}}} \big\{   \ket{v^{\mathrm{FFL}}_j }  \big\}  ; 
\end{align*}

}

\noindent (4) generalizations to related game-theoretic settings. 

Albeit the fact that the first property above can be related to the positive semidefinite property for the dual feasible objective $y_{3\mathrm{XOR}}$ appearing in,

{\small

\begin{align*}
          \underset{1 \leq i \leq n^3}{\sum}   y_{3\mathrm{XOR},i} E_{3\mathrm{XOR},ii} - G_{3\mathrm{XOR},\mathrm{Sym}}               , \\ 
\end{align*}

}

\noindent one makes use of a computation for the superposition,

{\small 

\begin{align*}
  \bigg\{   \big( C_{3\mathrm{XOR}} \omega_{3\mathrm{XOR}} \big)  n \bigg[ \underset{1 \leq j \leq 2}{\prod}  \big( n - j \big) \bigg]  \bigg\}^{-1}            \underset{i_3 \in \mathcal{Q}_3}{\underset{i_2 \in \mathcal{Q}_2 }{\underset{i_1 \in \mathcal{Q}_1}{\sum}}}   \bigg\{   \big(    u^{\prime}_{i_1i_2 i_3}   -    v^{\prime}_{i_1i_2 i_3}  \big) \big(   u^{\prime}_{i_1 i_2 i_3}   -    v^{\prime}_{i_1 i_2 i_3}    \big)^{\textbf{T}}     +  \big(    u^{\prime}_{i_2 i_1 i_3 }  \\  -    v^{\prime}_{i_2 i_1 i_3}  \big)   \big( u^{\prime}_{i_2 i_1 i_3 }   -    v^{\prime}_{i_2 i_1 i_3  }  \big)^{\textbf{T}}  +     \big(    u^{\prime}_{i_1 i_3 i_2}   -    v^{\prime}_{i_1 i_3 i_2}   \big)   \big(   u^{\prime}_{i_1 i_3 i_2}     -    v^{\prime}_{i_1 i_3 i_2}   \big)^{\textbf{T}}     \bigg\}       \text{,} \\ 
\end{align*}

}

\noindent where,

{\small

\begin{align*}
   u^{\prime}_{i_1 i_2 i_3 } \equiv \frac{1}{\sqrt{3}} \bigg\{  \underset{1 \leq j \leq 3}{\sum} \ket{\text{Player } j \text{ state}} \bigg\}  \text{, }   \\ \\      v^{\prime}_{i_1 i_2 i_3} \equiv \ket{i_1 i_2 i_3 } = \ket{i_1} \otimes \ket{i_2} \otimes \ket{i_3} ,  \\    \end{align*}

   \begin{align*} u^{\prime}_{i_2 i_1 i_3 } \equiv \frac{1}{\sqrt{3}} \bigg\{  \ket{\text{Player 1 state}} - \ket{\text{Player 2 state}} + \ket{\text{Player } 3 \text{ state}} \bigg\}  \text{, }  \end{align*}

   \begin{align*} v^{\prime}_{i_2 i_1 i_3} \equiv \ket{i_2 i_1 i_3} = \ket{i_2} \otimes \ket{i_1} \otimes \ket{i_3} ,  \\ \end{align*}

   \begin{align*}    u^{\prime}_{i_1 i_3 i_2 } \equiv \frac{1}{\sqrt{3}} \bigg\{   \underset{1 \leq j \leq 2}{\sum} \ket{\text{Player } j \text{ state}}  - \ket{\text{Player } 3 \text{ state}} \bigg\}        \text{, }   \end{align*}

   \begin{align*} v^{\prime}_{i_1 i_3 i_2} \equiv \ket{i_1 i_3 i_2}  = \ket{i_1} \otimes \ket{i_3} \otimes \ket{i_2}     . \\ 
\end{align*}

}

\noindent Underlying the structure of the two-player XOR game and the three-player XOR game above is the fact that the optimal value satisfies the following approximality condition. More specifically from the numerical value of $1 / {\sqrt{2}}$ attained by the supremum above, approximately optimal strategies satisfy,

\begin{align*}
    \frac{1}{\sqrt{2}} \big( 1 - \epsilon \big) \leq    \underset{A_i , B_{jk} , \psi}{\mathrm{sup}}   \frac{1}{4 {n \choose 2}}   \underset{1\leq i \leq j \leq n}{\sum}  \bra{\psi} \bigg[       A_i B_{ij} +  A_j B_{ij}  +   A_i B_{ji}  -   A_j B_{ji}  \bigg]            \ket{\psi}  \leq \frac{1}{\sqrt{2}}          \text{, } \\ \\  \tag{$0$}
\end{align*}

\noindent for some strictly positive $\epsilon$. 

\subsection{Semidefinite programs and their well-posedness through primal feasible solutions $Z$}

Games matrices, amongst other objects, for multiplayer games and their corresponding optimal values are closely related to the following:

{\small 
\[   \left\{\!\begin{array}{ll@{}>{{}}l} 
{\text{Question superset}} \equiv  \mathcal{Q} \equiv \underset{\# \text{ players}}{\bigcup} \mathcal{Q} \sim \bigg\{ \underset{\# \text{ players}}{\bigotimes}  \pi \equiv \pi^{\underset{\# \text{ players}}{\bigotimes} } \bigg\} \text{, } \\       \\ {\text{Probability measures over the first player's responses}} \equiv         \textbf{P}_{\mathcal{Q}_1} \big[ \cdot \big]      \text{, } \\ \\   {\text{Conditional Probability measures over the last player's responses}} \equiv         \textbf{P}_{\mathcal{Q}_1} \big[ \cdot \big| \cdot \big]              \text{, } \\ \vdots \\         {\text{Probability measures over the last player's responses}} \equiv         \textbf{P}_{\mathcal{Q}_N} \big[ \cdot \big]      \text{, } \\ \\ {\text{Conditional Probability measures over the first player's responses}} \equiv         \textbf{P}_{\mathcal{Q}_N} \big[ \cdot \big| \cdot \big]              \text{, }     
\\ \\   {\text{Joint probability measure over all possible responses from the first and second}} \\ {\text{players} }    \equiv \textbf{P}_{\mathcal{Q}_{12} } \big[ \cdot \big]  \equiv         \textbf{P}_{\mathcal{Q}_{1} \mathcal{Q}_2 } \big[ \cdot \big]     \equiv \textbf{P}_{\mathcal{Q}\big|_{12} } \big[ \cdot \big]  \text{, } 
    \\ \\ {\text{Conditional joint probability measure over all possible responses from the first}} \\   {\text{and second players} } \equiv \textbf{P}_{\mathcal{Q}_{12} } \big[ \cdot \big| \cdot \big]  \equiv         \textbf{P}_{\mathcal{Q}_{1} \mathcal{Q}_2 } \big[ \cdot \big| \cdot  \big]     \equiv \textbf{P}_{\mathcal{Q}\big|_{12} } \big[ \cdot \big| \cdot  \big]   \text{, } \\     \vdots \\  {\text{Joint probability measure over all possible responses from the first and second}} \\ {\text{players} }    \equiv \textbf{P}_{\mathcal{Q}_{12\cdots N} } \big[ \cdot \big]  \equiv         \textbf{P}_{\mathcal{Q}_{1} \mathcal{Q}_2 \cdots \mathcal{Q}_N} \big[ \cdot \big]     \equiv \textbf{P}_{\mathcal{Q}\big|_{12\cdots N} } \big[ \cdot \big]  \text{, } 
    \\ \\ {\text{Conditional joint probability measure over all possible responses from the first}} \\ {\text{and second players} } \equiv \textbf{P}_{\mathcal{Q}_{12\cdots N} } \big[ \cdot \big| \cdot \big]  \equiv         \textbf{P}_{\mathcal{Q}_{1} \mathcal{Q}_2 \cdots \mathcal{Q}_N} \big[ \cdot \big| \cdot  \big]     \equiv \textbf{P}_{\mathcal{Q}\big|_{12\cdots N} } \big[ \cdot \big| \cdot  \big]   . 
\end{array}\right. 
\]  }

\bigskip



\noindent As a generalization of the $\mathrm{CHSH}$ game, the $\mathrm{CHSH(n)}$ game admits important connections with semidefinite programs, which are manipulated in {\color{blue}[37]} for characterizing optimal strategies, in which we fix two real symmetric matrices $A$ and $B$, with inner product,

{\small \begin{align*}
  A \cdot B \equiv \mathrm{Tr} \big( A B \big) \equiv \underset{ \text{ rows, columns} \text{ } ij}{\sum} A_{ij} B_{ij}   \text{, } 
\end{align*} }

\noindent from which the semidefinite program is an optimization problem of the form, for a \textit{primal feasible} $Z$,

{\small \begin{align*}
 \underset{\forall i = 1 , \cdots , m \text{ } :\text{ }  \{ F_i \cdot Z = c_i \} }{\underset{Z \succcurlyeq 0}{\mathrm{sup}}} \big\{   G \cdot Z   \big\}    \text{, } \tag{1}
\end{align*} } 

\noindent where the conditions under which the supremum are taken are such that $Z$ is positive semidefinite, and that $F_i = F^T_i$, and $c_i \in \textbf{R} \text{ }  \forall i$. Straightforwardly, for the multiplayer XOR game primal feasible solutions for semidefinite programs before, and after, parallel repetition take the form,

{\small \begin{align*}
  \underset{\forall Z_{\mathrm{XOR}\wedge \cdots \wedge \mathrm{XOR}} \succcurlyeq 0 , 1 \leq i \leq m, \{ F^{(j)}_{\wedge i} \cdot Z_{\mathrm{XOR} \wedge \cdots \wedge {\mathrm{XOR}}} \equiv C^{(j)}_i \} }{\mathrm{sup}}    \big\{ G_{\mathrm{XOR}} Z_{\mathrm{XOR}} \big\}      ,   \\ \\  \underset{1 \leq j \leq n}{\bigwedge} \bigg\{   \underset{\forall Z_{\mathrm{XOR}\wedge \cdots \wedge \mathrm{XOR}} \succcurlyeq 0 , 1 \leq i \leq m, \{ F^{(j)}_{\wedge i} \cdot Z_{\mathrm{XOR} \wedge \cdots \wedge {\mathrm{XOR}}} \equiv C^{(j)}_i \} }{\mathrm{sup}}    \big\{ G_{\mathrm{XOR}} Z_{\mathrm{XOR}}    \big\}   \bigg\}   ,    \\ 
\end{align*}}

\noindent respectively.  The semidefinite program above admits a dual semidefinite program, which can be readily obtained by interchanging the conditions for which the supremum of $G \cdot Z$ is obtained with $Z \succcurlyeq 0$, in which,

{\small \begin{align*}
     \underset{\overset{m}{\underset{i=1}{\sum}} y_i F_i \succcurlyeq G}{\mathrm{inf}}   \big\{ \vec{c} \cdot \vec{y}    \big\}   \text{, }      \tag{2} 
\end{align*} }

\noindent for the \textit{dual feasible objective function} $\vec{y}$. The solution to $\mathrm{(1)}$ is denoted with $v_{\mathrm{primal}}$, while the solution to $\mathrm{(2)}$ is denoted with $v_{\mathrm{dual}}$.

Dual semidefinite programs of the above form are informative not only for the two-player setting but also for the multiplayer setting. Straightforwardly given an analagous formulation of the semidefinite program with primal feasible solution $Z_{\mathrm{3XOR}}$,

{\small

\begin{align*}
   \underset{\exists c_i \in \textbf{R}:  \{ F_{3 \mathrm{XOR},i} \cdot G_{3\mathrm{XOR}} \equiv c_{3\mathrm{XOR},i\} }   }{\underset{\forall Z^{\prime} \succcurlyeq 0  \text{, } 1 \leq i \leq 3 , }{\mathrm{sup} }}  \big\{  G_{3\mathrm{XOR}}  \cdot Z_{3\mathrm{XOR}}  \big\}        \text{, }
\end{align*}

}

\bigskip

\noindent for the 3XOR game matrix $G_{\mathrm{3XOR}}$ one can explicitly write out the entries of the dual objective $y$ expressed above for the 3XOR game, $y_{\mathrm{3XOR}}$,

{\small

\[
\vec{y_{3\mathrm{XOR}}}  \equiv  \left\{\!\begin{array}{ll@{}>{{}}l}
      y_{3\mathrm{XOR},1} \equiv \cdots \equiv y_{3\mathrm{XOR},n} \equiv \omega_{3\mathrm{XOR}} \big\{  {3! n} \big\}^{-1}  \text{, } \\  \\   y_{3\mathrm{XOR},n+1} \equiv \cdots \equiv    y_{3\mathrm{XOR},n^2}  \equiv \omega_{3\mathrm{XOR}} \big\{  {3! n ( n-1 ) }  \big\}^{-1} \text{, } \\ \\   y_{3\mathrm{XOR},n^2+1} \equiv \cdots \equiv    y_{3\mathrm{XOR},n^3}  \equiv \omega_{3\mathrm{XOR}}  \big\{ {3! n ( n-1 ) ( n-2 ) } \big\}^{-1}   \text{, } 
      \end{array}\right.
\]

}

\bigskip

\noindent from the dual semidefinite program,

{\small \begin{align*}
     \underset{\underset{1 \leq i \leq n^3}{\sum} y_{3\mathrm{XOR},i} F_{3\mathrm{XOR},i} \succcurlyeq G_{3\mathrm{XOR}}}{\mathrm{inf}} \big\{  \vec{c_{3 \mathrm{XOR}}}\cdot \vec{y_{3\mathrm{XOR}}} \big\}       \text{. }
\end{align*} }

\noindent As one can anticipate, the combinatorial factors appearing in the closed form expression for the above dual objective of the 3XOR game mirror those for other XOR games, particularly the NXOR game with the dual objective,

{\small \[
 \vec{y_{N\mathrm{XOR}}} \equiv \left\{\!\begin{array}{ll@{}>{{}}l} y_{N\mathrm{XOR},1} \equiv   \cdots \equiv    y_{N\mathrm{XOR},n} \equiv {\omega_{N\mathrm{XOR}}} \big\{       {N! n}       \big\}^{-1}          \text{, } \\   \\ y_{N\mathrm{XOR},n+1} \equiv   \cdots \equiv    y_{N\mathrm{XOR},n^2} \equiv {\omega_{N\mathrm{XOR}}} \big\{       N! n ( n -1 )        \big\}^{-1}          \text{, }  \\ \vdots \\    y_{N\mathrm{XOR},1} \equiv   \cdots \equiv    y_{N\mathrm{XOR},n} \equiv {\omega_{N\mathrm{XOR}}} \big\{       N! n ( n-1 ) \times \cdots \times (  n - (N-1) )        \big\}^{-1}   \text{. }
\end{array}\right.
\]

}

\bigskip

\noindent In addition, given $\mathrm{(1)}$ and $\mathrm{(2)}$, one can formulate the duality gap, (\textbf{Theorem} \textit{1}, {\color{blue}[37]}),

{\small \begin{align*}
  \bigg[  \overset{m}{\underset{i=1}{\sum}} \text{ }   y_i F_i - G   \bigg]  \cdot Z \geq 0  \text{, }
\end{align*}
 }

\noindent which vanishes iff $v_{\mathrm{primal}} \equiv v_{\mathrm{dual}}$. The final component that we introduce relates to representation theory, specifically through the intertwining operation that is achieved from operators of type $T$ below, which are defined as,

{\small

\begin{align*}
 \bigotimes T^{\mathrm{XOR}} \text{, }  \end{align*}

 \begin{align*} \bigwedge T^{\mathrm{XOR}}  \text{, } \end{align*}

 \begin{align*}     T^{ \{ \mathrm{XOR}\wedge \cdots \wedge \mathrm{XOR} \} }   \text{, } \end{align*}

 \begin{align*}      T^{\mathrm{FFL}\wedge  \mathrm{FFL}}    \text{, }
\end{align*}

}

\noindent respectively. Fundamentally underlying the action of such operators is a correspondence between,

{\small \begin{align*}
\bigg\{ \underset{s \in S}{\bigcup} A_s \equiv \big\{   s  \in S :  A_s \in \big\{ - 1 , + 1 \big\}    \big\}  \bigg\} \bigotimes \bigg\{  \underset{t \in T}{\bigcup} B_t \equiv \big\{ t \in T  :  B_t \in \big\{ -1 , + 1 \big\}  \big\}  \bigg\}  \text{, } \end{align*} }

\noindent supported over $\textbf{C}^{d_A} \otimes \textbf{C}^{d_B}$, and,

{\small \begin{align*} \bigg\{  \underset{s \in S}{\bigcup} \widetilde{A}_s \equiv \big\{   s  \in S :  \widetilde{A}_s \in \big\{ - 1 , + 1 \big\}    \big\}  \bigg\} \bigotimes  \bigg\{ \underset{t \in T}{\bigcup} \widetilde{B}_t \equiv \big\{   t  \in T :  \widetilde{B}_t \in \big\{ - 1 , + 1 \big\}    \big\} \bigg\}   ,  
\end{align*} } 

\noindent supported over $\textbf{C}^{2^{\lceil n / 2 \rceil}} \otimes \textbf{C}^{2^{\lceil n / 2 \rceil}}$, respectively.

To define a collection of linear operators, given some index set $\mathcal{I}$, for $V \subsetneq \textbf{C}^d$, denote the collection of linear operators with $\big\{ A_i : i \in \mathcal{I} \big\}$, while for $W \subsetneq \textbf{C}^d$, denote the collection of linear operators with $\big\{ B_i : i \in \mathcal{I} \big\}$. From each $A_i$ and $B_i$, equipped with a linear operator $T : V \longrightarrow W$ by \textbf{Lemma} \textit{3}, {\color{blue}[37]}: $\mathrm{Im} \big( T \big)$ is invariant under $\big\{ B_i : i \in \mathcal{I} \big\}$; $\mathrm{Ker} \big( T \big)$ is invariant under $\big\{ A_i : i \in \mathcal{I} \big\}$; each eigenspace of $T$ is invariant under $\big\{ A_i : i \in \mathcal{I} \big\}$. Each $i \mapsto A_i$, and $i \mapsto B_i$ either has the structure of a group homomorphism, or the structure of an algebra homomorphism which is related to the representation-theoretic intertwiners,

{\small \[
  \left\{\!\begin{array}{ll@{}>{{}}l} T^{\mathrm{XOR}^{*}} , \\ \\ T^{\mathrm{FFL}}  , 
\end{array}\right.
\]

}

\noindent considered in this work.

\bigskip

\noindent Such operators associated with original and dual semidefinite programs of the forms described above are related to families of operators of the form,

{\small

\begin{align*}
 \bigotimes T^{\mathrm{XOR}} : \underset{N \text{ copies}}{\bigotimes} \big\{   \textbf{C}^{2 \lceil \frac{n}{2} \rceil } \big\}   \longrightarrow        \underset{1 \leq i \leq N}{\bigotimes} \big\{  \textbf{C}^{d_i} \big\}  \text{, }  \end{align*}

 \begin{align*} \bigwedge T^{\mathrm{XOR}} :  \big\{ \textbf{C}^{2 \lceil \frac{n}{2} \rceil } \big\} \bigwedge     \big\{  \textbf{C}^{2 \lceil \frac{n}{2} \rceil } \big\}  \bigwedge \cdots \bigwedge  \big\{ \textbf{C}^{2 \lceil \frac{n}{2} \rceil }  \big\} \longrightarrow        \big\{ \textbf{C}^{d_A} \big\}  \bigwedge \big\{ \textbf{C}^{d_B} \big\}  \bigwedge \big\{ \textbf{C}^{d^{(1)}_B} \big\} \bigwedge\cdots \\ \bigwedge \big\{ \textbf{C}^{d^{(n-2)}_B}  \big\}   \text{, } \end{align*}

 \begin{align*}     T^{ \{ \mathrm{XOR}\wedge \cdots \wedge \mathrm{XOR} \} } :     \textbf{C}^{ \{ 2 \lceil \frac{n}{2} \rceil \wedge 2 \lceil \frac{n}{2} \rceil \wedge \cdots \wedge 2 \lceil \frac{n}{2} \rceil \} }  \longrightarrow        \textbf{C}^{ \{ d_A \wedge d_B \wedge d^{(1)}_B \wedge \cdots \wedge {d^{(n-2)}_B } \} }    \text{, } \end{align*}

 \begin{align*}      T^{\mathrm{FFL}\wedge  \mathrm{FFL}} :     \textbf{C}^{ \{ 2 \lceil \frac{n}{2} \rceil \wedge 2 \lceil \frac{n}{2} \rceil \} }  \longrightarrow        \textbf{C}^{ \{ d_A \wedge d_B \} }     \text{, }
\end{align*}

}

\noindent corresponding to linear operators for tensor products of the XOR game, parallel repetition of the XOR game, parallel repetition of the XOR and FFL games, respectively. As it will be demonstrated, operators of the above form are intimately linked to the anticommutation of the terms,

{\small
\begin{align*}
     \big\{  A_k \otimes \textbf{I} \big\}  \ket{\psi}          , \end{align*}

     \begin{align*} \bigg\{    \textbf{I} \otimes   \bigg( \frac{\pm B_{kl} + B_{lk}}{\sqrt{2}} \bigg)   \bigg\}  \ket{\psi}        , \\ 
\end{align*} 
}

\noindent appearing in the superposition,

{\small
\begin{align*}
  \bigg[  \big\{  A_k \otimes \textbf{I} \big\}    -  \bigg\{    \textbf{I} \otimes   \bigg( \frac{\pm B_{kl} + B_{lk}}{\sqrt{2}} \bigg)   \bigg\}    \bigg]  \ket{\psi}         , \\ 
\end{align*} 
}

\noindent with respect to the standard tensor product operation, $\otimes$, where tensors $A$ and $B$ respectively denote the possible responses prepared by Alice and Bob. Notationwise, subscripts on either of the tensors $A$ or $B$ denote the question that Alice or Bob responded to (specifically, $A_k$ denotes the space of possible responses that Alice and prepare for question $k$ whereas $B_{kl}$ and $B_{lk}$, respectively, denote the space of possible responses that Bob can prepare for questions $k$, given the responses from Alice for question $l$ and vice versa). Under the action of a suitably chosen linear operator $T$, one would expect that a superposition of the form,

{\small

\begin{align*}
  \big\{  A_i \otimes \textbf{I} \big\}  T     -          T \big\{  \widetilde{A}_i  \otimes \textbf{I} \big\}      , \\ 
\end{align*}

}

\noindent could be strictly upper bounded dependent, up to constants, with the Frobenius norm,

{\small

\begin{align*}
\big| \big| T \big| \big|_F    . \\ 
\end{align*}

}

\noindent While previous work, {\color{blue}[37]}, has demonstrated that the up to constants factor of the above Frobenius norm are $12$, and $17$, corresponding to the action of $T$ on $A$ and $B$, respectively, the up to constants factors on $\big| \big| T \big| \big|_F$ are dependent upon the probability that Alice and Bob respond to an iid sample drawn according to $\pi$. In particular, if $\pi \equiv \pi_{\mathrm{XOR}} \big( i , j \big) = \pi_{\mathrm{XOR}}$, 

{\small

\begin{align*}
        \pi \big(  0 , 0   \big) =  \pi \big( 0 , 1   \big) =    \pi \big( 1 , 0   \big) =       \pi \big( 1 , 1  \big)    =      1 / 4              , \\ 
\end{align*}

}

\noindent while if $\pi \equiv \pi_{\mathrm{FFL}} \big( i , j \big) = \pi_{\mathrm{FFL}}$,

{\small

\begin{align*}
    \pi \big( 0 , 0 \big) =  \pi \big( 0 , 1 \big) = \pi \big( 1 , 0 \big)    =  1 / 3  , \end{align*}

    \begin{align*} 
        \pi \big( 1 , 1 \big)    =   0       , \\ 
\end{align*}

}

\noindent To further demonstrate the dependence of multiplicative constants to the Frobenius norm, by making use of the above expressions for $T$, as well as the computation of,

{\small \begin{align*}
   \big\{   A_i \otimes \textbf{I} \big\}  \ket{\psi} =  \bigg\{  \textbf{I} \otimes \frac{B_{ij} + B_{ji}}{\sqrt{2}} \bigg\}  \ket{\psi} \text{, }     \end{align*}

   \begin{align*} \big\{  A_j \otimes \textbf{I} \big\}  \ket{\psi} = \bigg\{   \textbf{I} \otimes \frac{B_{ij} - B_{ji}}{\sqrt{2}} \bigg\}  \ket{\psi}    \text{, } \\ 
\end{align*} }

\noindent for $1 \leq i \leq n = d_A $ and $1 \leq j \leq n-1 = d_B$ with $j \neq i$ we discuss how previous approaches have characterized the optimal value. Albeit the fact that the above probability distributions for the XOR and FFL games could be regarded as uniform, in the sense that the probability that Alice and Bob answer the majority of questions is either $1 / 4$ or $1 / 3$ the probability distribution over iid samples of $\pi$, where $X = Y = \big\{ 1 , \cdots , \big\}$ and $A = B = \big\{ 1 , \cdots , k \big\}$,

{\small

\[ \pi \big( x , y \big) =   \left\{\!\begin{array}{ll@{}>{{}}l} 
       1 / \big(2n \big) \Longleftrightarrow  \big\{ x = y \big\} \textit{, namely $x,y \in V \big( H \big) = V$}       , \\ \\  1 / \big(4 m \big) \Longleftrightarrow \big\{  \big\{ x,y  \big\} \textit{ is an edge of the graph $H$} \big\} \textit{, namely $\big\{ x , y \big\} \in E \big( H \big) = E$}       , \\ \\    0 \Longleftrightarrow \big\{ \textit{otherwise} \big\}     , 
\end{array}\right. 
\] 

}

\bigskip 

\noindent if $\pi \equiv \pi_{\mathrm{CG}} \big( x , y \big) = \pi_{\mathrm{CG}}$, where $\mathrm{CG}$ is the coloring game, $x,y$ denote the vertices of an undirected graph $H = \big( V , E \big)$, $n = \big| V \big|$ and $m \geq 1$.

\subsection{The optimal value under entanglement}

\begin{itemize}
\item[$\bullet$] \textit{Optimal values for expanded games under anchoring}. Fix $0 < \alpha < 1$. Then the optimal value, $\omega^*$, of an expanded game $G$ under parallel repetition,  in comparison to the original optimal value, $\omega$, of $G$ before performing parallel repetition, and subsequently, anchoring, satisfies,

\begin{align*}
 \big\{ \textit{Optimal value of the anchored game $G_{\bot}$} \big\}  \equiv \omega^{*} \big( G_{\bot} \big) = 1 - \big( 1 - \alpha \big)^2 \big[ 1 - \omega^* \big( G \big) \big]     \text{, }
\end{align*}

\noindent for the $\alpha$-anchored game, $G_{\bot}$,

{\small

\begin{align*}
   \textbf{P} \big[ \textit{Alice's response to an iid sample of the referee's probability distribution is $\bot$} \big] = \alpha   , \\ \\   \textbf{P} \big[ \textit{Bob's response to an iid sample of the referee's probability distribution is $\bot$} \big] = \alpha  . 
\end{align*}

}

\bigskip 

\item[$\bullet$] \textit{Parallel repetition of} $\alpha$-\textit{anchoring}. Fix the same choice of $\alpha$ in the previous result, as well as $\epsilon$ sufficiently small so that the entangled optimal value, $\omega^{*} \big( G_{\bot} \big)$, satisfies $\omega^{*} \big( G_{\bot} \big)< 1 - \epsilon$. In particular, parallel repetition of $\alpha$-anchored expanded games defined above yields the following exponentially decaying estimate,

\begin{align*}
\big\{ \textit{Parallel repetition of $\omega^{*} \big( G_{\bot} \big)$} \big\} \equiv  \omega^{*} \big( G^n_{\bot} \big) \leq \big\{ {4} \big/ {\epsilon}   \big\}    \cdot    \mathrm{exp} \big\{   - c \alpha^{48} \epsilon^{17} n \big/ {s}   \big\}     \text{, }
\end{align*}

\noindent for $s \equiv \mathrm{sup} \big\{ \mathrm{log} \big( \big| \mathcal{A} \mathcal{B} \big| , 1 \big\}$, ie the supremum of the natural logarithm of the product of the alphabets from Alice and Bob and $1$, as well as strictly positive $c$.

\bigskip

\item[$\bullet$] \textit{Generalizations of the previous item for expressions of optimal values under anchoring}. Following the authors of [2,3,4], the authors of [30] argued that a generalization of the anchored parallel repetition on the optimal value satisfies,

{\small \begin{align*}
 \big\{    \textit{Generalization of the optimal value of $\omega^{*} \big( G^n_{\bot} \big)$}        \big\} \equiv  \omega^* \big( {\widetilde{G}}^k \big)  , \end{align*} }
 
 \noindent equals, 
 
 {\small \begin{align*}  \big\{  1 - \big[ 1  - \omega^* \big( \widetilde{G} \big) \big]^5 \big\}^{\Omega [ k ( \mathrm{log} ( | \mathcal{A} | | \mathcal{B} \big| ) )^{-1} ]}  \text{, }
\end{align*} }

\noindent for the sets, and strictly positive parameter,

{\small \begin{align*}
 \mathcal{A} \equiv \textit{Set of answers from first player}   \text{, } \end{align*}

 \begin{align*} \mathcal{B} \equiv \textit{Set of answers from second player}  \text{, } \end{align*}

 \begin{align*} k \equiv \textit{Number of parallel repetition operations} > 0  \text{, }
\end{align*}
 }

\noindent as well as the function,

{\small \begin{align*}
  \Omega \equiv \underset{a \in \mathcal{A}, b \in \mathcal{B} }{\bigcup} \big\{ \text{functions} \big( a , b \big) \big\}  \equiv \text{functions} \big( \mathcal{A} , \mathcal{B} \big)         \text{, }
\end{align*} }

\item[$\bullet$] \textit{Anchoring and parallel repetition of the optimal value}. Simultaneously, one can combine each of the operations described in the previous two items above as follows. To generalize upper bounds for the optimal value $\omega^{*} \big( G^n_{\bot} \big)$ from an upper bound for $\omega^{*} \big( \widetilde{G}^k \big)$, one analyzes:

\begin{itemize}
    \item[$\bullet$]   \textit{The decay of the winning probability under parallel repetition with one-sided anchoring}. Introduce,
    
{\small

\begin{align*}
       \Omega^{\prime} \equiv \big\{ \text{Player variables} \big\}      , \end{align*}

       \begin{align*} W_C \equiv \big\{ \text{Players win the expanded game for all coordinates } C \big\}   \text{,} \\ 
\end{align*}

}

    \noindent corresponding to the set of dependency-breaking variables and winning condition across all coordinates $C$, from which one-sided anchoring, along the lines of a previously mentioned condition,

{\small \begin{align*}
  \textbf{P} \big[ \Omega^{\prime} \big| \textbf{X}_i \equiv x, W_C \big] \approx \textbf{P} \big[ \Omega^{\prime}  \big| \textbf{X}_i \equiv x, \textbf{Y}_i \equiv y,  W_C  \big] \approx \textbf{P} \big[ \Omega^{\prime}    \big|  \textbf{Y}_i \equiv y,  W_C \big]   \text{, } \\ 
\end{align*} }

    \noindent stipulates that there exists $y, y^{*}$, for which, $\textbf{P} \big[ y^{*} \big]$ is constant, as well as the conditional probability distribution,

{\small \begin{align*}
    \textbf{P} \big[ y \big| x \big] \text{, }
\end{align*} }

    \noindent equaling the probability distribution,

{\small \begin{align*}
    \textbf{P} \big[ y  \big] \text{, }
\end{align*} }

    \noindent or alternatively, that there exists $x, x^{*}$, for which, $\textbf{P} \big[ x^{*} \big]$ is constant, as well as the conditional probability distribution,

{\small \begin{align*}
    \textbf{P} \big[ x \big| y \big] \text{, }
\end{align*} }

    \noindent equaling the probability distribution,

{\small \begin{align*}
    \textbf{P} \big[ x  \big] \text{. }
\end{align*}  }

 \end{itemize}
\end{itemize}

\subsection{Theorems for optimality of quantum strategies}

After having defined the \textit{sucess bias}, the probability of winning an XOR game, and the optimal \textit{sucess bias}, and an $n \times m$ \textit{game matrix} $G$, from a $\pm 1$ observable that Alice and Bob pick as their answers to the question that the Referee draws from the product distribution, the optimality condition,

{\small \begin{align*}
   \big( 1 - \epsilon \big) \beta \big( G \big) \leq \overset{n}{\underset{i=1}{\sum} } \overset{m}{\underset{j=1}{\sum}}  G_{ij} \bra{\psi} A_i \otimes B_j \ket{\psi} \leq \beta \big( G \big)     \text{, } 
\end{align*} } 

\noindent holds iff,

{\small \begin{align*}
    \overset{r}{\underset{k=1}{\sum} } \big|\big|    \big( u_k " \cdot "  \vec{A} \otimes \textbf{I} \big) \ket{\psi}       -              \big(  \textbf{I} \otimes v_k " \cdot "  \vec{B} \big)  \ket{\psi}    \big|\big|^2 \leq \beta \big( G \big) \epsilon  \text{, }
\end{align*} }

\noindent for $\{ u_i \}_{i \in \mathcal{I}} \in \textbf{R}^n$, and $\{ v_i \}_{i \in \mathcal{I}} \in \textbf{R}^m$ (\textbf{Theorem} \textit{3}, {\color{blue}[37]}) where "$\cdot$" denotes the standard inner product between two vectors. By formulating the optimiality condition as shown above in terms of a summation over $k$, \textit{approximately optimal} quantum strategies are known to behave as follows, which is of interest to classify further for other non-local games related to the XOR game. Before stating the main results provided in {\color{blue}[37]}, it is important to recall the following objects:

\begin{itemize}
    \item[$\bullet$] {\textit{Product norm of player responses}}: Under the identifications,

    {\small \begin{align*}
 \\   \ket{\underset{\text{Odd number of players}}{\prod} i_j } \longleftrightarrow  \underset{\text{Odd number of players}}{\prod}  \ket{ i_j }   \text{, } \\ \\  \bra{\underset{\text{Even number of players}}{\prod} i_j } \longleftrightarrow \underset{\text{Even number of players}}{\prod} \bra{ i_j }   \text{, } \\ 
     \end{align*} }

    \noindent The inner product of responses from a group of $N$ players can be expressed as,

{\small \begin{align*}
\\  \bigg[ \ket{\underset{\text{Odd number of players}}{\prod} i_j } \bigg]  \bigg[ \bra{\underset{\text{Even number of players}}{\prod} i_j } \bigg]    \equiv   \bigg[ \ket{\underset{\text{Odd } j, 1 \leq j \leq N}{\prod} i_j } \bigg]  \\ \times \bigg[ \bra{\underset{\text{Even } j, 1 \leq j \leq N}{\prod} i_j } \bigg] \\   \end{align*}

 \begin{align*}  \equiv  \bigg[  \ket{i_N i_{N-2} \times \cdots \times i_1 } \bigg]  \bigg[  \bra{i_2 i_4 \times \cdots \times i_{N-1}} \bigg]    \equiv  \bigg[   \ket{i_N} \ket{i_{N-2}} \times \cdots \times \ket{i_1} \bigg]  \\ \times  \bigg[  \bra{i_2} \bra{i_4} \times \cdots  \times \bra{i_{N-1}} \bigg] \\  \end{align*}

 \begin{align*} \equiv \ket{i_N}  \bigg[ \cdots \times  \bigg[ \cdots \times \bigg[  \ket{i_3}  \bigg[ \ket{i_1} \bra{i_2} \bigg]   \bra{i_4} \bigg]   \times \cdots  \bigg]  \times  \cdots \bigg]   \bra{i_{N-1}}  \end{align*}

 \begin{align*}  \equiv      \ket{\text{Player } N \text{ responds to question } i_N \text{ given } (N-1) \text{ previous responses}}  \bigg[ \cdots  \times \bigg[ \cdots \\  \times \bigg[ \ket{\text{Player } 1 \text{ responds to the first question}}  \bra{\text{Player } 2 \text{ responds to the second question}} \bigg]   \\ \cdots \times \bigg]  \cdots \bigg]   \\ \times    \bra{\text{Player } (N-1) \text{ responds to question } i_N \text{ given } (N-2) \text{ previous responses}}  \text{. } \\ 
\end{align*} }

\noindent For the $3\mathrm{XOR}$, and $4\mathrm{XOR}$, games, the outer product for $N$ players take the form,

\begin{align*}
\ket{i_3}  \big[ \ket{i_1} \bra{i_2} \big]    \text{, } \\  \\ \ket{i_3}  \big[ \ket{i_1} \bra{i_2} \big]  \bra{i_4} \text{, }
\end{align*}

\noindent respectively.

    \bigskip

     \item[$\bullet$] {\textit{Tensor observables for players of the game}}. To define the multiplayer bias, which will be used to characterize exact, and approximate, optimality up to some parameter $\epsilon$ taken to be sufficiently small, define,

     \begin{align*}
         \bigotimes \big\{ \text{Player tensor observables} \big\}  \equiv \big\{  \text{Alice's observables} \big\}   \bigotimes \big\{  \text{Bob's observables} \big\}   \\ \bigotimes  \big\{  \text{Eve's observables} \big\}     \text{, }
     \end{align*}

     \noindent corresponding to the Hilbert space spanned by the possible set of responses for three players Alice, Bob and Eve.

\bigskip

      \item[$\bullet$] {\textit{Intertwining operation}}. For tensor products of player observables, error bounds for the two-player $\mathrm{XOR}$ game consist of \textit{interchanging} the order in which the observables that each player forms appear in tensor products, such as the one provided over all player observables above. In error bounds that will follow, the representation-theoretic intertwining operations $T$ are not only valuable for obtaining upper bounds with respect to $\big| \big| \cdot \big| \big|_F$ but also for upper bounding series of inequalities with respect to the Frobenius norm such as,

{\small

\begin{align*}
        \bigg| \bigg|      \bigg[  \bigg\{  \bigg(   \underset{1 \leq i \leq n}{\prod} A^{j_i}_i \bigg)    + \bigg(   \underset{i +1 \equiv i \oplus 1}{\underset{1 \leq i \leq n}{\prod}}  A^{j_i}_i \bigg)   \bigg\} \bigotimes \textbf{I} \bigotimes \textbf{I}  \bigg] \ket{\psi }           \bigg| \bigg|_F             , \\ \end{align*}
        
        \begin{align*}  \bigg| \bigg|    \bigg[   \textbf{I} \bigotimes  \bigg\{       \bigg(  \underset{\mathcal{Q}_3 , \mathcal{Q}_3 +1 \equiv \mathcal{Q}_3 \oplus 1}{\underset{\mathcal{Q}_2 , \mathcal{Q}_2 +1 \equiv \mathcal{Q}_2 \oplus 1}{\underset{\mathcal{Q}_1 , \mathcal{Q}_1 + 1 \equiv \mathcal{Q}_1 \oplus 1}{\prod}}}    C^{l_{ijk}}_{ijk}  \bigg) + \bigg(   \underset{\mathcal{Q}_1 , \mathcal{Q}_2 , \mathcal{Q}_3}{\prod}   C^{l_{ijk}}_{ijk}       \bigg)  \bigg\} \bigotimes \textbf{I }   \bigg] \ket{\psi }                  \bigg| \bigg|_F     , \\ \end{align*}
        
        \begin{align*} \bigg| \bigg|    \bigg[   \textbf{I} \bigotimes \textbf{I} \bigotimes \bigg\{    \bigg(     \underset{\mathcal{Q}_1 , \mathcal{Q}_2 , \mathcal{Q}_3}{\prod}   C^{(N-2),l_{ijk}}_{ijk}      \bigg)     + \bigg(  \underset{\mathcal{Q}_3 , \mathcal{Q}_3 +1 \equiv \mathcal{Q}_3 \oplus 1}{\underset{\mathcal{Q}_2 , \mathcal{Q}_2 +1 \equiv \mathcal{Q}_2 \oplus 1}{\underset{\mathcal{Q}_1 , \mathcal{Q}_1 + 1 \equiv \mathcal{Q}_1 \oplus 1}{\prod}}}    C^{(N-2), l_{ijk}}_{ijk}  \bigg)         \bigg\}      \bigg] \ket{\psi }             \bigg| \bigg|_F    ,     \\ 
\end{align*}

}

\noindent where,

{\small

\[
 \left\{\!\begin{array}{ll@{}>{{}}l}
 \\  \underset{1 \leq i \leq n}{\prod} A^{j_i}_i = A^{j_1}_1 \times \cdots \times A^{j_n}_n  , \\ \\   \underset{i +1 \equiv i \oplus 1}{\underset{1 \leq i \leq n}{\prod}}  A^{j_i}_i  = A^{j_1 \oplus 1}_1 \cdot A^{j_2} \times \cdots \times A^{j_n}_n              , \\ \\ \underset{\mathcal{Q}_3 , \mathcal{Q}_3 +1 \equiv \mathcal{Q}_3 \oplus 1}{\underset{\mathcal{Q}_2 , \mathcal{Q}_2 +1 \equiv \mathcal{Q}_2 \oplus 1}{\underset{\mathcal{Q}_1 , \mathcal{Q}_1 + 1 \equiv \mathcal{Q}_1 \oplus 1}{\prod}}}    C^{l_{ijk}}_{ijk}  =     \underset{i \neq j \neq k}{\underset{\mathcal{Q}_3 , \mathcal{Q}_3 +1 \equiv \mathcal{Q}_3 \oplus 1}{\underset{\mathcal{Q}_2 , \mathcal{Q}_2 +1 \equiv \mathcal{Q}_2 \oplus 1}{\underset{\mathcal{Q}_1 , \mathcal{Q}_1 + 1 \equiv \mathcal{Q}_1 \oplus 1}{\prod}}}}   C^{l_{\oplus i, \oplus j, \oplus k}}_{ i,  j, k}                                                    ,  \\ \\ \underset{\mathcal{Q}_1 , \mathcal{Q}_2 , \mathcal{Q}_3}{\prod}   C^{l_{ijk}}_{ijk}  =     C^{l_{\mathcal{Q}_1, \mathcal{Q}_2, \mathcal{Q}_3}}_{\mathcal{Q}_1 \mathcal{Q}_2 \mathcal{Q}_3}            , 
\\ \\ \underset{\mathcal{Q}_1 , \mathcal{Q}_2 , \mathcal{Q}_3}{\prod}   C^{(N-2),l_{ijk}}_{ijk}  =      C^{(N-2),l_{\mathcal{Q}_1 \mathcal{Q}_2 \mathcal{Q}_3}}_{\mathcal{Q}_1 \mathcal{Q}_2 \mathcal{Q}_3}   , \\ \\ \\      \underset{\mathcal{Q}_3 , \mathcal{Q}_3 +1 \equiv \mathcal{Q}_3 \oplus 1}{\underset{\mathcal{Q}_2 , \mathcal{Q}_2 +1 \equiv \mathcal{Q}_2 \oplus 1}{\underset{\mathcal{Q}_1 , \mathcal{Q}_1 + 1 \equiv \mathcal{Q}_1 \oplus 1}{\prod}}}    C^{(N-2), l_{ijk}}_{ijk}  =  \underset{i \neq j \neq k}{\underset{\mathcal{Q}_3 , \mathcal{Q}_3 +1 \equiv \mathcal{Q}_3 \oplus 1}{\underset{\mathcal{Q}_2 , \mathcal{Q}_2 +1 \equiv \mathcal{Q}_2 \oplus 1}{\underset{\mathcal{Q}_1 , \mathcal{Q}_1 + 1 \equiv \mathcal{Q}_1 \oplus 1}{\prod}}}}   C^{(N-2), l_{\oplus i, \oplus j, \oplus k}}_{i,j,k}   \\   \\ \end{array}\right.
\]

}




\bigskip 

\noindent The above operation is applied under many circumstances, not only for games with more participants but also for games which can be formulated under parallel repetition.

\bigskip

   \item[$\bullet$] {\textit{$\epsilon$-deviations from optimality}}. Given the existence of a sufficiently small parameter, besides differences in the formulation of error bounds, the bias, and optimal value, satisfying $\epsilon$-approximate optimality, reads,

{\small \begin{align*}
  \big( 1 - \epsilon \big) \beta \big( G \big)  \leq  \underset{\text{Questions}}{\sum} \bra{\text{Optimal Strategy}} \bigg[    \underset{\# \text{ Players }}{\bigotimes}   \text{Tensors of player observables}   \bigg]  \\ \times    \ket{\text{Optimal Strategy}}  \leq      \beta \big( G \big)   \text{, } \\ 
\end{align*} }

\noindent for the game $G$.
         
\end{itemize}

\bigskip

\noindent \textbf{Theorem} \textit{1} (\textit{approximately optimal quantum strategies for the nonlocal XOR game}, \textbf{Theorem} \textit{4}, {\color{blue}[37]}). For $\pm$ observables $A_i$ and $B_{jk}$, given a bipartite state $\psi$, TFAE:

\begin{itemize}
 \item[$\bullet$] {\textit{First characterization of approximate optimality}:} An $\epsilon$-approximate $\mathrm{CHSH}\big( n \big)$ satisfies $\mathrm{(0)}$.

 \item[$\bullet$] {\textit{Second characterization of approximate optimality}:} For an $\epsilon$-approximate quantum strategy, 

  {\small \begin{align*}
     \underset{1 \leq i < j \leq n}{\sum} \bigg[ \text{ }  \bigg| \bigg|   \bigg[         \bigg\{  \frac{A_i + A_j}{\sqrt{2}} \bigg\}  \otimes \textbf{I} \bigg] \ket{\psi}    -   \big\{   \textbf{I} \otimes B_{ij} \big\}  \ket{\psi}   \bigg|\bigg|^2 + \bigg| \bigg|   \bigg[ \big\{  \frac{A_i - A_j}{\sqrt{2}} \big\}   \otimes \textbf{I}  \bigg] \ket{\psi}    -   \big\{   \textbf{I} \otimes B_{ji} \big\}   \\ \times \ket{\psi}     \bigg|\bigg|^2 \text{ }  \bigg]  \leq 2n \big( n - 1 \big) \epsilon \text{. }
 \end{align*} } 
 \item[$\bullet$] {\textit{Reversing the order of the tensor product for observables}:} Related to the inequality for $\epsilon$-approximate strategies above, another inequality,

{\small \begin{align*}
    \underset{1 \leq i < j \leq n}{\sum} \bigg[  \text{ } \bigg| \bigg| \big\{  A_i \otimes \textbf{I} \big\}  \ket{\psi}    - \bigg[ \textbf{I} \otimes \bigg\{  \frac{B_{ij} + B_{ji}}{\sqrt{2}} \bigg\}   \bigg] \ket{\psi}   \bigg| \bigg|^2  + \bigg| \bigg|    \big\{  A_j  \otimes \textbf{I} \big\}  \ket{\psi}  -   \bigg[   \textbf{I} \otimes \bigg\{  \frac{B_{ij} - B_{ji}}{\sqrt{2}} \bigg\}  \bigg] \\ \times \ket{\psi}    \big| \big|^2     \text{ }   \bigg] \leq 2n \big( n - 1 \big) \epsilon \text{, } 
\end{align*} } 

 \noindent also holds. 
 
 \item[$\bullet$] {\textit{Characterization of exact optimality}}: For $\epsilon \equiv 0$,

{\small \begin{align*}
    \underset{1 \leq i < j \leq n}{\sum} \bigg[ \text{ }  \bigg| \bigg|  \bigg[  \bigg\{  \frac{A_i + A_j}{\sqrt{2}} \bigg\}  \otimes \textbf{I} \bigg] \ket{\psi}    -  \big\{  \textbf{I}\otimes B_{ij} \big\}  \ket{\psi}   \bigg| \bigg|^2  \bigg]  =  -  \underset{1 \leq i < j \leq n}{\sum} \bigg[  \text{ }   \bigg| \bigg|     \bigg[  \bigg\{  \frac{A_i - A_j}{\sqrt{2}} \bigg\}   \otimes \textbf{I} \bigg]  \ket{\psi} \\  -   \big\{  \textbf{I} \otimes B_{ji} \big\} \ket{\psi}    \bigg| \bigg|^2  \bigg]      \text{, } 
\end{align*} } 

\noindent corresponding to the first inequality, and,

{\small \begin{align*}
       \underset{1 \leq i < j \leq n}{\sum} \bigg[ \text{ }  \bigg| \bigg|       \big\{   A_i \otimes \textbf{I} \big\}   \ket{\psi}    - \bigg[  \textbf{I} \otimes \bigg\{  \frac{B_{ij} + B_{ji}}{\sqrt{2}} \bigg\}   \bigg]  \ket{\psi}   \bigg|\bigg|^2 \bigg] = -  \underset{1 \leq i < j \leq n}{\sum}   \bigg[  \bigg| \bigg|  \big\{    A_j  \otimes \textbf{I} \big\}  \ket{\psi} \\  -  \bigg[   \textbf{I}  \otimes \bigg\{  \frac{B_{ij} - B_{ji}}{\sqrt{2}} \bigg\}  \bigg]  \ket{\psi}     \bigg|\bigg|^2  \bigg]             \text{, } 
\end{align*} }

\noindent corresponding to the second inequality.
 
\end{itemize}

\noindent

\bigskip

\noindent For optimal $\mathrm{CHSH}\big(n\big)$ strategies, given a \textit{Schmidt decomposition} of the bipartite state, the decomposition,

\begin{align*}
    \overset{s 2^{\lfloor \frac{n}{2} \rfloor}}{\underset{i=1}{\sum} } \sqrt{\lambda_i}  \big\{  \ket{u_i} \otimes \ket{v_i}  \big\}   \text{, } 
\end{align*}

\noindent into a tensor product over the quantum states which respectively corresponding to each $u_i$ and $v_i$ are equal in blocks $\lambda$ between each term of the summation over $i$ above, as (\textbf{Theorem} \textit{5}, {\color{blue}[37]}),

\begin{align*}
        \lambda_i = \cdots = \lambda_{i+1} \text{, } \forall i    \text{. } 
\end{align*}

\noindent Moreover, with respect to some basis $\{ \ket{u_i} \}_{i \in \mathcal{I}}$ of $\textbf{C}^{d_A}$, the observable $A_i$ decomposes as,

\begin{align*}
   A_i   =  \mathrm{diag} \big(   A^{(1)}_i , \cdots , A^{(s)}_i , C_i       \big)    \text{, } 
\end{align*}

\noindent where each block diagonal component $A^{(s)}_i$, of size $2^{\lfloor \frac{n}{2} \rfloor} \times 2^{\lfloor \frac{n}{2} \rfloor}$, is such that each $A_i$ acts on $\mathrm{span} \big( \ket{u_{(j-1)2^{\lfloor \frac{n}{2} \rfloor}+1}} , \cdots$ $, \ket{u_{j 2^{\lfloor \frac{n}{2} \rfloor}}} \big)$ for $1 \leq i \leq n$, with,

\begin{align*}
  C_i \equiv \big\{ \pm 1 \big\}   \text{, } 
\end{align*}

\noindent $\forall \big( s - 1 \big)  2^{\lfloor \frac{n}{2} \rfloor} \leq i \leq s 2^{\lfloor \frac{n}{2} \rfloor}$. For a basis $\{ \ket{v_i} \}_{i \in \mathcal{I}}$ of $\textbf{C}^{d_B}$, the observable $B_{jk}$ decomposes as,

\begin{align*}
    B_{jk}  =  \mathrm{diag} \big(  B^{(1)}_{jk} , \cdots , B^{(s)}_{jk} , D_{jk}       \big)       \text{, } 
\end{align*}

\noindent where each block diagonal component $B^{(s)}_{jk}$, also of size $2^{\lfloor \frac{n}{2} \rfloor} \times 2^{\lfloor \frac{n}{2} \rfloor}$, is such that each $B_{jk}$ acts on $\mathrm{span} \big(   \ket{v_{ (j - 1 )  2^{\lfloor \frac{n}{2} \rfloor+1}}} , \cdots , \ket{v_{j 2^{\lfloor \frac{n}{2} \rfloor}}} \big)$ for $1 \leq j \neq k \leq n$, with,

\begin{align*}
  D_{jk} \equiv \big\{ \pm 1 \big\}  \text{, } 
\end{align*}

\noindent $\forall \big( s - 1 \big)  2^{\lfloor \frac{n}{2} \rfloor}  \leq jk \leq s  2^{\lfloor \frac{n}{2} \rfloor}$. Each $B_{jk}$ along the diagonal admits a decomposition in terms of $A_j$ and $A_k$, in which, for $1 \leq l \leq j \neq k \in  \big\{ 1 , \cdots , n \big\}$,

{\small \begin{align*}
  B^{(l)}_{jk}    \equiv  \bigg( \frac{1}{\sqrt{2}} A^{(l)}_j  +      \frac{1}{\sqrt{2}} A^{(l)}_k   \bigg)^{\mathrm{T}} = \frac{1}{\sqrt{2}} \bigg( A^{(l)}_j + A^{(l)}_k  
 \bigg)^{\mathrm{T}}  \equiv \frac{1}{\sqrt{2}}  \bigg( \big\{  A^{(l)}_j\big\}^{\mathrm{T}} +  \big\{ A^{(l)}_k \big\}^{\mathrm{T}} \bigg)  \text{, } \\    B^{(l)}_{kj}   \equiv  \bigg(   \frac{1}{\sqrt{2}}   A^{(l)}_j   - \frac{1}{\sqrt{2}}    A^{(l)}_k          \bigg)^{\mathrm{T}} = \frac{1}{\sqrt{2}} \bigg(    A^{(l)}_j   -   A^{(l)}_k     \bigg)^{\mathrm{T}}  \equiv \frac{1}{\sqrt{2}}  \bigg( \big\{    A^{(l)}_j 
 \big\}^{\mathrm{T}} -   \big\{ A^{(l)}_k  \big\}^{\mathrm{T}}   \bigg)  \text{. } 
\end{align*}      }

\subsection{Error bounds in closely related game-theoretic settings to those of the XOR* and FFL games}

\noindent The above representations for each $B^{(l)}$ operator above are of interest for error bounds, such as the ones obtained in this work, as stated in the following results:

\bigskip

\noindent \textbf{Lemma} \textit{2.5.1} (\textit{parallel repetition of }$\sqrt{\epsilon^{\wedge}}$- \textit{FFL approximality, akin to the error bound provided in} \textbf{Lemma} \textit{8}, {\color{blue}[37]}). Denote,

{\small
 \[ \left\{\!\begin{array}{ll@{}>{{}}l}  
  A_k \wedge A_{k^{\prime}}  , \\ \\    \ket{\psi_{\mathrm{FFL} \wedge \mathrm{FFL}}}      , \\ \\     B_{kl} \wedge B_{k^{\prime} l^{\prime}} , \\ \\     B_{lk} \wedge B_{k^{\prime} l^{\prime}}       , \\ \\  B_{kl} \wedge B_{k^{\prime} l^{\prime}}   ,  \\ \\  B_{lk} \wedge B_{k^{\prime} l^{\prime}}    , 
\end{array}\right. 
\]  }

\bigskip

\noindent as the parallel repetition operation $\wedge$ between $A_k$ and $A_{k^{\prime}}$, the optimal FFL strategy corresponding to two rounds of parallel repetition operation, the parallel repetition operation $\wedge$ applied to $B_{kl}$ and $B_{k^{\prime} l^{\prime}}$, the parallel repetition operation $\wedge$ applied to $B_{lk}$ and $B_{k^{\prime} l^{\prime}}$, the parallel repetition operation $\wedge$ applied to $B_{kl}$ and $B_{k^{\prime} l^{\prime}}$ and the parallel repetition operation $\wedge$ applied to $B_{lk}$ and $B_{k^{\prime} l^{\prime}}$, respectively. One has,

{\small \begin{align*}
   \bigg| \bigg|               \bigg\{  \big(  A_k  \wedge A_{k^{\prime}} \big) \otimes \textbf{I} \bigg\}  \ket{\psi_{\mathrm{FFL} \wedge \mathrm{FFL}}}    -  \bigg\{  \textbf{I} \otimes \bigg(     \frac{\pm \big(  B_{kl} \wedge B_{k^{\prime} l^{\prime}} \big)  + \big( B_{lk} \wedge B_{k^{\prime} l^{\prime}} \big) }{\big| \pm \big(  B_{kl} \wedge B_{k^{\prime} l^{\prime}} \big)  + \big( B_{lk} \wedge B_{k^{\prime} l^{\prime}} \big) \big| }           \bigg) \bigg\}  \\ \times \ket{\psi_{\mathrm{FFL} \wedge \mathrm{FFL}}}              \bigg| \bigg|    < 20 \sqrt{N \epsilon^{\wedge}}     \text{. } 
\end{align*} }

\bigskip

\noindent \textbf{Lemma} \textit{2.5.2} (\textit{parallel repetition operation in the 2XOR game, akin to the error bound provided in} \textbf{Lemma} \textit{8}, {\color{blue}[37]}). Fix $\epsilon^{\wedge}_{2\mathrm{XOR}}$ sufficiently small. Denote,

{\small
 \[ \left\{\!\begin{array}{ll@{}>{{}}l}  
      A_k  \wedge A_{k^{\prime}} \wedge \cdots \wedge A_{k^{\prime\cdots \prime}}     , \\ \\  \ket{\psi_{2\mathrm{XOR} \wedge \cdots \wedge 2\mathrm{XOR}}}         , \\ \\  B_{kl} \wedge B_{k^{\prime} l^{\prime}} \wedge \cdots \wedge B_{k^{\prime\cdots\prime}l^{\prime\cdots \prime}}   , \\ \\          B_{lk} \wedge B_{ l^{\prime} k^{\prime}} \wedge \cdots \wedge B_{l^{\prime\cdots \prime} k^{\prime\cdots\prime} }  , \\ \\ B_{kl} \wedge B_{k^{\prime} l^{\prime}} \wedge \cdots \wedge B_{k^{\prime\cdots\prime}l^{\prime\cdots \prime}}    , \\ \\   B_{lk} \wedge B_{ l^{\prime} k^{\prime}} \wedge \cdots \wedge B_{l^{\prime\cdots \prime} k^{\prime\cdots\prime} }        , 
\end{array}\right. 
\]  }

 \bigskip 

\noindent as the parallel repetition operation $\wedge$ between $A_k, A_{k^{\prime}}, \cdots , A_{k^{\prime\cdots \prime}}$, the optimal 2XOR strategy corresponding to an arbitrary number of parallel repetition operations, the parallel repetition operation $\wedge$ applied to $B_{kl}, B_{k^{\prime} l^{\prime}}, \cdots , B_{k^{\prime\cdots\prime}l^{\prime\cdots \prime}}$, the parallel repetition operation $\wedge$ applied to $B_{lk}, B_{ l^{\prime} k^{\prime}} ,\cdots , B_{l^{\prime\cdots \prime} k^{\prime\cdots\prime} }$, the parallel repetition operation $\wedge$ applied to $B_{kl},B_{k^{\prime} l^{\prime}}, \cdots, B_{k^{\prime\cdots\prime}l^{\prime\cdots \prime}}  $ and the parallel repetition operation $\wedge$ applied to $B_{lk}, B_{k^{\prime} l^{\prime}}, \cdots, B_{l^{\prime\cdots \prime} k^{\prime\cdots\prime} }$, respectively. One has,

{\small \begin{align*}
   \bigg| \bigg|               \bigg\{  \big(  A_k  \wedge A_{k^{\prime}} \wedge \cdots \wedge A_{k^{\prime\cdots \prime}} \big) \otimes \textbf{I} \bigg\}  \ket{\psi_{2\mathrm{XOR} \wedge \cdots \wedge 2\mathrm{XOR}}}   \end{align*}

   \begin{align*} -  \bigg\{  \textbf{I} \otimes \bigg(     \frac{\pm \big(  B_{kl} \wedge B_{k^{\prime} l^{\prime}} \wedge \cdots \wedge B_{k^{\prime\cdots\prime}l^{\prime\cdots \prime}}  \big)  + \big( B_{lk} \wedge B_{ l^{\prime} k^{\prime}} \wedge \cdots \wedge B_{l^{\prime\cdots \prime} k^{\prime\cdots\prime} }  \big) }{\big| \pm \big(  B_{kl} \wedge B_{k^{\prime} l^{\prime}} \wedge \cdots \wedge B_{k^{\prime\cdots\prime}l^{\prime\cdots \prime}}  \big)  + \big( B_{lk} \wedge B_{ l^{\prime} k^{\prime}} \wedge \cdots \wedge B_{l^{\prime\cdots \prime} k^{\prime\cdots\prime} }  \big) \big| }           \bigg) \bigg\}  \\ \times \ket{\psi_{2\mathrm{XOR} \wedge \cdots \wedge 2\mathrm{XOR}}}              \bigg| \bigg|   < 18 \sqrt{N \epsilon^{\wedge}_{2\mathrm{XOR}}}   \text{. } 
\end{align*} }

\bigskip

\noindent \textbf{Lemma} \textit{2.5.3} (\textit{an arbitrary number of strong parallel repetition operations of the 3XOR game, akin to the error bound provided in} \textbf{Lemma} \textit{8}, {\color{blue}[37]}). Fix $\epsilon^{\wedge}_{3\mathrm{XOR}}$ sufficiently small. Denote,

{\small
 \[ \left\{\!\begin{array}{ll@{}>{{}}l}  
  A_k  \wedge A_{k^{\prime}} \wedge \cdots \wedge A_{k^{\prime\cdots \prime}}  , \\ \\       \ket{\psi_{3\mathrm{XOR} \wedge \cdots \wedge 3\mathrm{XOR}}}       , \\ \\              \underset{\sigma_1 \in S_3}{\sum} \big(  \wedge  C_{\sigma(ijk)} \big) , \\ \\   \underset{\sigma_2, \sigma_3, \sigma_4, \sigma_5, \sigma_6  \in S_3}{\sum} \big( \wedge  C_{\sigma(ijk)} \big)      , \\ \\ \underset{\sigma_1 \in S_3}{\sum} \big( \wedge C_{\sigma(ijk)} \big)  , \\ \\   \underset{\sigma_2, \sigma_3, \sigma_4, \sigma_5, \sigma_6  \in S_3}{\sum} \big( \wedge C_{\sigma(ijk)}\big)    , 
\end{array}\right. 
\]  } 

\bigskip

\noindent as the parallel repetition operation $\wedge$ between $A_k, A_{k^{\prime}}, \cdots, A_{k^{\prime\cdots \prime}}$, the optimal 3XOR strategy corresponding to an arbitrary number of parallel repetition operations, the summation of $C_{\sigma ( ijk)}$ for , respectively. One has,

{\small \begin{align*}
   \bigg| \bigg|               \bigg\{  \big(  A_k  \wedge A_{k^{\prime}} \wedge \cdots \wedge A_{k^{\prime\cdots \prime}} \big) \otimes \textbf{I} \otimes \textbf{I} \bigg\}  \ket{\psi_{3\mathrm{XOR} \wedge \cdots \wedge 3\mathrm{XOR}}}   \end{align*}

   \begin{align*} -  \bigg\{  \textbf{I} \otimes \textbf{I}  \otimes \bigg[     \frac{  \pm \underset{\sigma_1 \in S_3}{\sum} \big(  \wedge  C_{\sigma(ijk)} \big)  +  \underset{\sigma_2, \sigma_3, \sigma_4, \sigma_5, \sigma_6  \in S_3}{\sum} \big( \wedge  C_{\sigma(ijk)} \big)  }{  \bigg| \pm \underset{\sigma_1 \in S_3}{\sum} \big( \wedge C_{\sigma(ijk)} \big)  +  \underset{\sigma_2, \sigma_3, \sigma_4, \sigma_5, \sigma_6  \in S_3}{\sum} \big( \wedge C_{\sigma(ijk)}\big)   \bigg|  } \bigg]     \bigg\}  \ket{\psi_{3\mathrm{XOR} \wedge \cdots \wedge 3\mathrm{XOR}}}              \bigg| \bigg|  \\   < 18 \sqrt{N \epsilon^{\wedge}_{3\mathrm{XOR}}}   \text{, } 
\end{align*} }

\noindent where,

{\small \begin{align*}
 \wedge C_{\sigma(ijk)} \equiv  C_{\sigma(ijk)}  \wedge \cdots \wedge C_{\sigma(ijk)}  \text{. } \\ 
\end{align*} }

\noindent Expressing the optimal strategy of a game through a quantum state is related to the below tensor product expansions:

\bigskip

\noindent \textbf{Lemma} \textit{2.5.4} (\textit{odd n product expansion}, \textit{6.1}, \textbf{Lemma} \textit{6}, {\color{blue}[37]}). For odd $n$, one has an expansion for,

{\small \begin{align*}
 \underset{1 \leq i \leq n}{\prod} \widetilde{A}_i    \text{, } 
\end{align*} }

\noindent in terms of a signed block identity matrix,

{\small \[
\big( - 1 \big)^n \begin{bmatrix}
\textbf{I} & 0 \\ 0 & - \textbf{I}
\end{bmatrix}  \text{. } 
\] }

\noindent Hence, for,

{\small \begin{align*}
 \ket{\widetilde{\psi_{\mathrm{FFL}}}} = \frac{1}{\sqrt{2 \times  2^{\lfloor \frac{n}{2} \rfloor }}} \bigg[   \bigg\{       \underset{1 \leq j \leq 2^{\lfloor \frac{n}{2}\rfloor}}{\sum}   + \underset{2^{\lfloor \frac{n}{2} \rfloor} + 1 \leq j \leq 2 \times 2^{\lfloor \frac{n}{2}\rfloor}}{\sum} \bigg\}  \bigg( \ket{j} \otimes \ket{j}  \bigg)   \bigg]    \text{, } 
\end{align*} }

\noindent one has,

{\small \begin{align*}
 \bigg( \bigg\{   \underset{1 \leq i \leq n}{\prod} \widetilde{A}_i   \bigg\}  \otimes \textbf{I} \bigg) \ket{\widetilde{\psi_{\mathrm{FFL}}}} = \frac{1}{\sqrt{2 \times  2^{\lfloor \frac{n}{2} \rfloor }}}  \bigg[     \bigg\{       \underset{1 \leq j \leq 2^{\lfloor \frac{n}{2}\rfloor}}{\sum}   + \underset{2^{\lfloor \frac{n}{2} \rfloor} + 1 \leq j \leq 2 \times 2^{\lfloor \frac{n}{2}\rfloor}}{\sum} \bigg\}  \\ \times \bigg\{  \bigg(       \bigg\{   \underset{1 \leq i \leq n}{\prod} \widetilde{A}_i   \bigg\}  \ket{j}       \otimes \ket{j}      \bigg)  \otimes \bigg( \bigg\{   \underset{1 \leq i \leq n}{\prod} \widetilde{A}_i   \bigg\}  \ket{j}    \otimes \ket{j} \bigg)   \bigg\}        \bigg]            \text{. } 
\end{align*} }

\bigskip

\noindent Straightforwardly, the adaptation of the above result is given below.

\bigskip

\noindent \textbf{Lemma} \textit{2.5.5} (\textit{an adaptation of Lemma 6 of} {\color{blue}[37]} \textit{for parallel repetition of expanded games}). For odd n, one has an expansion for,

{\small \begin{align*}
 \underset{1 \leq i \leq n}{\prod} \big( \widetilde{A}_i  \wedge \widetilde{A_{i^{\prime}}} \big)    \text{, } 
\end{align*}
 } 

\noindent in terms of a signed block identity matrix,

{\small \[
\big( - 1 \big)^n \begin{bmatrix}
\textbf{I} \wedge \textbf{I} & 0 \\ 0 & - \textbf{I} \wedge \textbf{I}
\end{bmatrix}  \text{. } 
\] }

\noindent Hence, for,

{\small \begin{align*}
 \ket{\psi^{\prime}} \wedge \ket{\psi^{\prime}} = \frac{1}{\sqrt{2 \times  2^{\lfloor \frac{n}{2} \rfloor }}} \bigg[   \bigg\{       \underset{1 \leq j \leq 2^{\lfloor \frac{n}{2}\rfloor}}{\sum}   + \underset{2^{\lfloor \frac{n}{2} \rfloor} + 1 \leq j \leq 2 \times 2^{\lfloor \frac{n}{2}\rfloor}}{\sum} \bigg\} \bigg( \big( \ket{j} \wedge \ket{j} \big)  \otimes \big( \ket{j} \wedge \ket{j} \big)   \bigg) \text{ }  \bigg]    \text{, } 
\end{align*} }

\noindent one has,

{\small \begin{align*}
 \bigg( \bigg\{   \underset{1 \leq i \leq n}{\prod} \widetilde{A}_i   \bigg\}  \otimes \textbf{I} \bigg) \bigg( \ket{\psi^{\prime}} \wedge \ket{\psi^{\prime}} \bigg)   = \frac{1}{\sqrt{2 \times  2^{\lfloor \frac{n}{2} \rfloor }}}  \bigg[    \bigg\{      \underset{1 \leq j \leq 2^{\lfloor \frac{n}{2}\rfloor}}{\sum}   + \underset{2^{\lfloor \frac{n}{2} \rfloor} + 1 \leq j \leq 2 \times 2^{\lfloor \frac{n}{2}\rfloor}}{\sum} \bigg\} \bigg\{  \bigg(       \bigg\{   \underset{1 \leq i \leq n}{\prod} \widetilde{A}_i   \bigg\}   \\ \times   \big( \ket{j} \wedge \ket{j} \big)     \otimes \big( \ket{j} \wedge \ket{j} \big)       \bigg) \otimes \bigg( \bigg\{   \underset{1 \leq i \leq n}{\prod} \widetilde{A}_i   \bigg\}  \big( \ket{j} \wedge \ket{j} \big)   \otimes\big( \ket{j} \wedge \ket{j} \big)    \bigg)   \bigg\}       \bigg]            \text{. } 
\end{align*} }

 \noindent Fortunately, the fact that $\pi_{\mathrm{XOR}}$ and $\pi_{\mathrm{FFL}}$ assign different probabilities over,

{\small

\begin{align*}
  \big( I , J \big) = \underset{i \neq j}{\underset{i,j}{\bigcup }} \big\{ \big( i , j \big) \big\} = \bigg\{  \underset{i \neq j}{\underset{i,j}{\bigcup }} \big( i , j \big) \bigg\}  = \big\{ \big( 0 , 0 \big) , \big( 1 , 0 \big) , \big(   0 , 1 \big) , \big( 1 , 1 \big)     \big\}  ,  
\end{align*}

}

 \noindent implies that one can compare the constants appearing in the below estimates,

{\small

\begin{align*}
    \bigg| \bigg|      \bigg\{  \text{ }  \bigg( \underset{1 \leq  i \leq n }{\prod} A^{j_i}_i \bigg) \otimes B_{kl} \bigg\}   \ket{\psi_{\mathrm{XOR}}}      -  \frac{1}{\sqrt{2}} \bigg[ \pm \mathrm{sign}  \big( i , j_1 , \cdots , j_n \big) \bigg\{  \text{ }  \bigg(   \underset{ i \equiv j_4+1 ,\text{ } \mathrm{set} \text{ } j_4+1 \equiv j_4 \oplus 1 }{\underset{1 \leq i \leq n}{\prod}}       A^{j_i}_i    \bigg)  
        \otimes \textbf{I}    \bigg\}  \text{ } \bigg] \\ \times \ket{\psi_{\mathrm{XOR}}}   \bigg| \bigg|_{\mathrm{F}}       , \\ 
\end{align*}

}

 \noindent to those with,

 {\small

\begin{align*}
    \bigg| \bigg|      \bigg\{  \text{ }  \bigg( \underset{1 \leq  i \leq n }{\prod} A^{j_i}_i \bigg) \otimes B_{kl} \bigg\}   \ket{\psi_{\mathrm{FFL}}}      -   \frac{2}{3} \bigg[ \pm \mathrm{sign}  \big( i , j_1 , \cdots , j_n \big) \bigg\{  \text{ }  \bigg(   \underset{ i \equiv j_4+1 ,\text{ } \mathrm{set} \text{ } j_4+1 \equiv j_4 \oplus 1 }{\underset{1 \leq i \leq n}{\prod}}       A^{j_i}_i    \bigg)  
        \otimes \textbf{I}    \bigg\}  \text{ } \bigg] \\ \times \ket{\psi_{\mathrm{FFL}}}   \bigg| \bigg|_{\mathrm{F}}     .  \\ 
\end{align*}

}

 \noindent The significance of computing the product expansion from the sequence,

{\small

\begin{align*}
    \bigg\{    \underset{ i \equiv j_4+1 ,\text{ } \mathrm{set} \text{ } j_4+1 \equiv j_4 \oplus 1 }{\underset{1 \leq i \leq i^{\prime}}{\prod}}       A^{j_i}_i    \bigg\}_{i^{\prime} \leq n }     , \\ 
\end{align*}

}

 \noindent in comparison to the expansion from the sequence,

{\small

\begin{align*}
  \bigg\{  \underset{1 \leq  i \leq i^{\prime} }{\prod} A^{j_i}_i \bigg\}_{i^{\prime} \leq n}   , \\ 
\end{align*}

}

 \noindent serves the purpose of determining, up to a $\pm$ sign of the optimal value, the term,

{\small

\begin{align*}
 n^2 \sqrt{\epsilon }   , \\ 
\end{align*}

}

 \noindent up to a constant. The above inequalities for the XOR and FFL games are related to additional inequalities of the form,
 
{\small

\begin{align*}
  \bigg| \bigg|    \bigg[   \bigg\{  \bigg(           \underset{1 \leq i \leq n}{\prod}       A^{j_i}_i     \bigg) \bigotimes B_{kl} \bigotimes \bigg(  \underset{1 \leq k \leq n-2}{\bigotimes} \textbf{I}_k     \bigg) \bigg\}   -   \omega_{N \mathrm{XOR}}         \bigg\{                    \pm \bigg( \mathrm{sign} \big( i_1 , j_1 , \cdots , j_n \big) \bigg[ \bigg\{         \bigg(   \underset{1 \leq i \leq n}{\prod} A^{j_i}_i \bigg)  \\  + \bigg(   \underset{i +1 \equiv i \oplus 1}{\underset{1 \leq i \leq n}{\prod}}  A^{j_i}_i \bigg)       \bigg\}  \bigg] \bigotimes \bigg( \underset{1 \leq k \leq n-1}{\bigotimes} \textbf{I}_k \bigg) \bigg)               \bigg\}     \bigg]               \ket{\psi_{N\mathrm{XOR}}}        \bigg| \bigg|    \text{, }  \\ 
  \end{align*}

  \begin{align*} 
\bigg| \bigg|    \bigg[   \bigg\{             A_i \bigotimes \textbf{I} \bigotimes \bigg(     \underset{\mathcal{Q}_1 , \mathcal{Q}_2 , \mathcal{Q}_3}{\prod}   C^{l_{ijk}}_{ijk}      \bigg) \bigotimes \bigg( \underset{1 \leq k \leq n-3}{\bigotimes}  \textbf{I}_k \bigg)                     \bigg\}   - \omega_{N\mathrm{XOR}} \bigg\{      \pm \mathrm{sign} \big( i_1 , j_1 , k_1  , \cdots , i_{111}  , \cdots \\ , j_{nm ( n+m )} , \cdots  , k_{nm(n+m)}  \big) \bigg[   \bigg( \textbf{I} \bigotimes \textbf{I} \bigotimes \bigg\{  \bigg(  \underset{\mathcal{Q}_3 , \mathcal{Q}_3 +1 \equiv \mathcal{Q}_3 \oplus 1}{\underset{\mathcal{Q}_2 , \mathcal{Q}_2 +1 \equiv \mathcal{Q}_2 \oplus 1}{\underset{\mathcal{Q}_1 , \mathcal{Q}_1 + 1 \equiv \mathcal{Q}_1 \oplus 1}{\prod}}}    C^{l_{ijk}}_{ijk}  \bigg) + \bigg(   \underset{\mathcal{Q}_1 , \mathcal{Q}_2 , \mathcal{Q}_3}{\prod}   C^{l_{ijk}}_{ijk}       \bigg) \bigg\}      \\  \bigotimes \bigg( \underset{1 \leq k \leq n-3} {\bigotimes} \textbf{I}_k  \bigg)      \bigg)   \bigg] \bigg\}                  \bigg]               \ket{\psi_{N\mathrm{XOR}}}        \bigg| \bigg|   \text{, } \\ \vdots \end{align*}

  \begin{align*}   \bigg| \bigg| \bigg[ \bigg\{              A_i \bigotimes  \bigg( \underset{1 \leq k \leq N-2}{\bigotimes}  \textbf{I}_k \bigg)  \bigotimes   \bigg(     \underset{\mathcal{Q}_1 , \mathcal{Q}_2 , \mathcal{Q}_3}{\prod}   C^{(N-2),l_{ijk}}_{ijk}      \bigg)                 \bigg\}   - \omega_{N\mathrm{XOR}}    \bigg\{                           \pm \mathrm{sign} \big( i_1 , j_1  , k_1 , \cdots , i_{111} \\  , \cdots , j_{nm ( n+m )} , \cdots  , k_{nm(n+m)}  \big) \bigg[   \bigg( \bigg( \underset{1 \leq k \leq N-1}{\bigotimes} \textbf{I}_k \bigg)  \bigotimes \bigg\{  \bigg(     \underset{\mathcal{Q}_1 , \mathcal{Q}_2 , \mathcal{Q}_3}{\prod}   C^{(N-2),l_{ijk}}_{ijk}      \bigg)   \\  + \bigg(  \underset{\mathcal{Q}_3 , \mathcal{Q}_3 +1 \equiv \mathcal{Q}_3 \oplus 1}{\underset{\mathcal{Q}_2 , \mathcal{Q}_2 +1 \equiv \mathcal{Q}_2 \oplus 1}{\underset{\mathcal{Q}_1 , \mathcal{Q}_1 + 1 \equiv \mathcal{Q}_1 \oplus 1}{\prod}}}    C^{(N-2), l_{ijk}}_{ijk}  \bigg)          \bigg\}  \bigg\}    \bigg]      \bigg)             \bigg]               \ket{\psi_{N\mathrm{XOR}}}        \bigg| \bigg|              \text{, } \\ 
\end{align*}

}

 \noindent in which the optimality constraint on the bias takes the form,

{\small \begin{align*}
  \big( 1 - \epsilon_{N\mathrm{XOR}} \big) \beta_{N\mathrm{XOR}} \big( G \big)      \leq      \underset{\mathcal{Q}_1 , \cdots , \mathcal{Q}_N}{\sum} \bra{\psi_{N\mathrm{XOR}}}  \bigg[           \underset{N \text{ players}}{\bigotimes}   \text{Tensors of player observables}     \bigg]      \ket{\psi_{N\mathrm{XOR}}}               \\  \leq \beta_{N\mathrm{XOR}} \big( G \big)          \text{, } \\ 
\end{align*} } 

\noindent for $\epsilon_{N\mathrm{XOR}}$ taken sufficiently small (refer to \textit{Table} \textbf{1} below for terms appearing in error bounds for the multiplayer XOR game which mirror aspects of the error bound provided for the FFL game). Under the associativity property of the ordinary tensor product $\bigotimes$, in error bounds one considers operators of the form,

{\small

\begin{align*}
     \bigotimes \bigg( \underset{1 \leq k \leq n-2}{\bigotimes}    \textbf{I}_k  \bigg) =  \underset{1 \leq k \leq n-1}{\bigotimes}    \textbf{I}_k      , \\  \end{align*}
     
     \begin{align*} A_i \bigotimes \bigg( \underset{1 \leq k       \leq n-2}{\bigotimes}    \textbf{I}_k \bigg) =   A_i \underset{1 \leq k       \leq n-2}{\bigotimes}    \bigg( \bigotimes   \textbf{I}_k \bigg)   =  \underset{1 \leq k       \leq n-2}{\bigotimes}    \bigg\{  A_i \bigotimes      \bigg( \bigotimes   \textbf{I}_k \bigg)     \bigg\}                  , \\ \end{align*}

     \begin{align*}       \bigg(   \frac{1}{\sqrt{\# \sigma}} \underset{\textit{permutations} \text{ } \sigma}{\sum} A_{\sigma (i  )}      \bigg) \bigotimes \bigg( \underset{1 \leq k       \leq n-2}{\bigotimes}    \textbf{I}_k \bigg)        =   \bigg\{  \bigg(   \frac{1}{\sqrt{\# \sigma}} \underset{\textit{permutations} \text{ } \sigma}{\sum} A_{\sigma (i  )}       \bigg)  \underset{1 \leq k       \leq n-2}{\bigotimes}  \\   \bigg( \bigotimes     \textbf{I}_k \bigg)    \bigg\} \end{align*}

     \begin{align*} =    \underset{1 \leq k       \leq n-2}{\bigotimes}    \bigg\{  \bigg(   \frac{1}{\sqrt{\# \sigma}}  \underset{\textit{permutations} \text{ } \sigma}{\sum} A_{\sigma (i  )}       \bigg)  {\bigotimes}    \bigg( \bigotimes     \textbf{I}_k \bigg)    \bigg\}           \\ \\ =  \frac{1}{\sqrt{\# \sigma}}       \underset{1 \leq k       \leq n-2}{\bigotimes}   \bigg\{  \bigg(    \underset{\textit{permutations} \text{ } \sigma}{\sum} A_{\sigma (i  )}       \bigg)  {\bigotimes}    \bigg( \bigotimes     \textbf{I}_k \bigg)    \bigg\}                                                           , \\ \\ 
\end{align*}

    \begin{align*}                        \bigg(   \frac{1}{\sqrt{\# \sigma}}  \underset{\textit{permutations} \text{ } \sigma}{\sum} A_{\sigma (i  )}      \bigg) \bigotimes    \bigg(   \frac{1}{\sqrt{\# \sigma^{\prime}}}  \underset{j \neq i }{\underset{\textit{permutations} \text{ } \sigma^{\prime}}{\sum}} B_{\sigma^{\prime} (j  )}     \bigg)  \bigotimes \bigg( \underset{1 \leq k       \leq n-2}{\bigotimes}    \textbf{I}_k \bigg)     \end{align*}
    
    \begin{align*} =    \bigg(   \frac{1}{\sqrt{\# \sigma}}  \underset{\textit{permutations} \text{ } \sigma}{\sum} A_{\sigma (i  )}      \bigg) \bigotimes    \bigg(   \frac{1}{\sqrt{\# \sigma^{\prime}}} \underset{j \neq i }{\underset{\textit{permutations} \text{ } \sigma^{\prime}}{\sum}} B_{\sigma^{\prime} (j  )}     \bigg)   \underset{1 \leq k       \leq n-2}{\bigotimes}    \bigg( \bigotimes     \textbf{I}_k \bigg)      \\ \\ =    \underset{1 \leq k       \leq n-2}{\bigotimes}    \bigg\{  \bigg(   \frac{1}{\sqrt{\# \sigma}} \underset{\textit{permutations} \text{ } \sigma}{\sum} A_{\sigma (i  )}       \bigg) \bigotimes    \bigg(   \frac{1}{\sqrt{\# \sigma^{\prime}}} \underset{j \neq i }{\underset{\textit{permutations} \text{ } \sigma^{\prime}}{\sum}} B_{\sigma^{\prime} (j  )}     \bigg)    {\bigotimes}  \\   \bigg( \bigotimes     \textbf{I}_k \bigg)    \bigg\}                                                          \\ \\ =     \frac{1}{\sqrt{\# \sigma}} \frac{1}{\sqrt{\# \sigma^{\prime}}}       \underset{1 \leq k       \leq n-2}{\bigotimes}    \bigg\{ \bigg(   \underset{\textit{permutations} \text{ } \sigma}{\sum} A_{\sigma (i  )}     \bigg) \bigotimes    \bigg(    \underset{j \neq i }{\underset{\textit{permutations} \text{ } \sigma^{\prime}}{\sum}} B_{\sigma^{\prime} (j  )}      \bigg)    {\bigotimes} \\    \bigg( \bigotimes     \textbf{I}_k \bigg)    \bigg\}  , \\ 
\end{align*}

}

\noindent as well as operators of the form,

{\small \[
 \left\{\!\begin{array}{ll@{}>{{}}l} 
          \\   \bigg(   \underset{1 \leq i \leq n}{\prod} A^{j_i}_i \bigg)    + \bigg(   \underset{i +1 \equiv i \oplus 1}{\underset{1 \leq i \leq n}{\prod}}  A^{j_i}_i \bigg)           , \\  \\   \bigg(  \underset{\mathcal{Q}_3 , \mathcal{Q}_3 +1 \equiv \mathcal{Q}_3 \oplus 1}{\underset{\mathcal{Q}_2 , \mathcal{Q}_2 +1 \equiv \mathcal{Q}_2 \oplus 1}{\underset{\mathcal{Q}_1 , \mathcal{Q}_1 + 1 \equiv \mathcal{Q}_1 \oplus 1}{\prod}}}    C^{l_{ijk}}_{ijk}  \bigg) + \bigg(   \underset{\mathcal{Q}_1 , \mathcal{Q}_2 , \mathcal{Q}_3}{\prod}   C^{l_{ijk}}_{ijk}       \bigg)    , \\  \vdots   \\ \bigg(     \underset{\mathcal{Q}_1 , \mathcal{Q}_2 , \mathcal{Q}_3}{\prod}   C^{(N-2),l_{ijk}}_{ijk}      \bigg)     + \bigg(  \underset{\mathcal{Q}_3 , \mathcal{Q}_3 +1 \equiv \mathcal{Q}_3 \oplus 1}{\underset{\mathcal{Q}_2 , \mathcal{Q}_2 +1 \equiv \mathcal{Q}_2 \oplus 1}{\underset{\mathcal{Q}_1 , \mathcal{Q}_1 + 1 \equiv \mathcal{Q}_1 \oplus 1}{\prod}}}    C^{(N-2), l_{ijk}}_{ijk}  \bigg)    , \\   
\end{array}\right.  
\]  } 

\bigskip 

\noindent can be understood as the operator for the responses of the first, second and $N$th players, for which the participant replaces a subset of questions of

\bigskip

 \noindent As previously mentioned, not only is the upper bound of the above form is not only obtained from the anticommutation of the terms,

{\small
\begin{align*}
     \big\{  A_k \otimes \textbf{I} \big\}  \ket{\psi}          , \end{align*}

     \begin{align*} \bigg\{    \textbf{I} \otimes   \bigg( \frac{\pm B_{kl} + B_{lk}}{\sqrt{2}} \bigg)   \bigg\}  \ket{\psi}        , \\ 
\end{align*} 
}

  \noindent but also from the anticommutation through the action of the representation-theoretic intertwiner, which is of the form,

{\small
\begin{align*}
    \bigg\{    \bigg( \frac{A_i + A_j}{2} \bigg)  \otimes \textbf{I}       \bigg\} \ket{\psi}  ,   \end{align*}

     \begin{align*}      \big\{ \textbf{I} \otimes B_{ij } \big\}  \ket{\psi}   .  
\end{align*} 
}

\subsection{Paper organization}

\noindent Following the overview provided in \textit{1.2} and \textit{1.3} after the overview provided in \textit{1.1}, we introduce a reference in the next section for the $\mathrm{XOR}^{\mathrm{*}}$ games, and establish connections between how optimal strategies for $\mathrm{XOR}$ games can be related to those of $\mathrm{XOR}^{\mathrm{*}}$ games. With such background on the $\mathrm{XOR}^{\mathrm{*}}$ game, we proceed to analyze how optimal, and approximately optimal, strategies for a different class of non-local games can be realized, from which other applications of the theory developed in {\color{blue}[37]} can be applied to other non-local $\mathrm{XOR}$ games rather than only the $\mathrm{XOR}^{\mathrm{*}}$ games including the FFL game. Given the nonequal probabilities that are assigned to each question $\big(i,j\big)$ from $\pi_{\mathrm{XOR}}$ and $\pi_{\mathrm{FFL}}$, we demonstrate how, as previously observed, the fact that,

{\small

\begin{align*}
    \omega_{\mathrm{XOR}  \wedge \mathrm{XOR}} \big( G_{\mathrm{XOR} \wedge  \mathrm{XOR}} \big)  \equiv \omega \big(  \mathrm{XOR} \wedge \mathrm{XOR} \big) \equiv \underset{1\leq j \leq 2}{\prod}   \omega \big( \mathrm{XOR}\big)^{j} \equiv \big\{ \omega \big( \mathrm{XOR} \big) \big\}^2 \equiv 1 / 2  \\ \neq \omega \big( \mathrm{XOR}  \big)  \neq 1 / {\sqrt{2}}   ,  \\ \end{align*}

    \begin{align*} \omega_{\mathrm{FFL} \wedge  \mathrm{FFL}} \big( G_{\mathrm{FFL}} \big) \equiv \omega_{\mathrm{FFL}} \big( G_{\mathrm{FFL}} \wedge G_{\mathrm{FFL}} \big)   \equiv \omega \big(  \mathrm{FFL}  \wedge \mathrm{FFL} \big) \equiv  \omega \big( \mathrm{FFL}\big) \equiv 2 / 3  \\ \neq \big( 2 / 3 \big)^2       , \\ 
\end{align*}

}

\noindent is related to the constants,

{\small

\[   \left\{\!\begin{array}{ll@{}>{{}}l} 
  9   , \\ \\  44 / 3 , \\ \\  8200\sqrt{2} / 27  , 
\end{array}\right. 
\] 
}

\bigskip 

\noindent of the Frobenius norm $\big| \big| T_{\mathrm{FFL}} \big| \big|_F$ for the FFL linear operator. Each of the above constants is obtained with respect to the standard Bell states,

{\small

\begin{align*}
\bigg(  \textbf{I} \otimes \textbf{I} \bigg) \bigg\{   \frac{\ket{00} + \ket{11}}{\sqrt{2}}\bigg\}   = \frac{\ket{00} + \ket{11}}{\sqrt{2}}   \text{ } \text{ , }     \bigg(     \sigma_x \otimes  \textbf{I} \bigg) \bigg\{  \frac{\ket{00} + \ket{11}}{\sqrt{2}} \bigg\}   = \frac{\ket{10} + \ket{01}}{\sqrt{2}}   \text{, } \\   \\   \bigg(   \sigma_z    \otimes  \textbf{I} \bigg) \bigg\{   \frac{\ket{00} + \ket{11}}{\sqrt{2}} \bigg\}   = \frac{\ket{00} - \ket{11}}{\sqrt{2}}       \text{ } \text{ , }    \bigg( \sigma_x \sigma_z     \otimes    \textbf{I}  \bigg)  \bigg\{  \frac{\ket{00}+\ket{11}}{\sqrt{2}} \bigg\}   = \frac{\ket{10}- \ket{01}}{\sqrt{2}}  , \\  
\end{align*}

}

\noindent in addition to the computation of the states,

{\small

\begin{align*}
 \bigg\{  A_i \bigg(  \underset{1 \leq i \leq n}{\prod} A^{j_i}_i \bigg) \otimes \textbf{I} \bigg\} \ket{\psi_{\mathrm{FFL}}}     , \\ \end{align*}

 \begin{align*} \mathrm{sign} \big( i , j_1 , \cdots , j_n \big) \bigg\{  \bigg( \underset{1 \leq i \leq n}{\prod} A^{j_i}_i  \bigg) \otimes \textbf{I} \bigg\}  \ket{\psi_{\mathrm{FFL}}}    . \\ 
\end{align*}

}

\noindent In comparison to the computation of the states,

{\small

\begin{align*}
 \bigg\{  A_i \bigg(  \underset{1 \leq i \leq n}{\prod} A^{j_i}_i \bigg) \otimes \textbf{I} \bigg\} \ket{\psi_{\mathrm{XOR}}}     , \\ \end{align*}

 \begin{align*} \mathrm{sign} \big( i , j_1 , \cdots , j_n \big) \bigg\{  \bigg( \underset{1 \leq i \leq n}{\prod} A^{j_i}_i  \bigg) \otimes \textbf{I} \bigg\}  \ket{\psi_{\mathrm{XOR}}}    , \\ 
\end{align*}

}

\noindent for the XOR game the prefactor,

{\small

\begin{align*}
  8200\sqrt{2} / 27   , \\ 
\end{align*}

}

\noindent is related to the computation of,

{\small

\[   \left\{\!\begin{array}{ll@{}>{{}}l} 
\bigg| \bigg|  \text{ } \bigg\{  \bigg( \underset{1 \leq i \leq n}{\prod} A^{j_i}_i     \bigg)    \otimes  \textbf{I}   \bigg\}        \ket{\psi_{\mathrm{FFL}}}    -   \bigg[ \mathrm{sign} \big( i , j_1 , \cdots , j_n \big) \text{ }\bigg\{   \bigg(   \underset{i \equiv j_1 + 1, \text{ } \mathrm{set}\text{ }  j_1 + 1 \equiv j_1 \oplus 1}{\underset{1 \leq i \leq n}{\prod}}        A^{j_i}_i         \bigg)     \otimes \textbf{I} \bigg\}   \bigg] \ket{\psi_{\mathrm{FFL}}}              \bigg| \bigg|_{\mathrm{F}}         , \\ \\           \bigg| \bigg|      \bigg\{    \bigg( \underset{1 \leq  i \leq n }{\prod} A^{j_i}_i \bigg) \otimes B_{kl} \bigg\}   \ket{\psi_{\mathrm{FFL}}}      -   \frac{2}{3} \bigg[ \pm \mathrm{sign}  \big( i , j_1 , \cdots , j_n \big) \bigg\{  \bigg(     \underset{ i \equiv j_4+1 ,\text{ } \mathrm{set} \text{ } j_4+1 \equiv j_4 \oplus 1 }{\underset{1 \leq i \leq n}{\prod}}       A^{j_i}_i    \bigg)    
       \otimes \textbf{I}    \bigg\}  \bigg] \ket{\psi_{\mathrm{FFL}}}   \bigg| \bigg|_{\mathrm{F}}                     , 
\end{array}\right. 
\] 
}

\noindent with respect to the Frobenius norm, where,

{\small

\begin{align*}
        \underset{i \equiv j_1 + 1, \text{ } \mathrm{set}\text{ }  j_1 + 1 \equiv j_1 \oplus 1}{\underset{1 \leq i \leq n}{\prod}}        A^{j_i}_i         =   A^{j_1 \oplus 1}_1 \cdot A^{j_2}_2 \times \cdots \times A^{j_n}_n                     , \\  \end{align*}

        \begin{align*} \underset{ i \equiv j_4+1 ,\text{ } \mathrm{set} \text{ } j_4+1 \equiv j_4 \oplus 1 }{\underset{1 \leq i \leq n}{\prod}}       A^{j_i}_i    =     A^{j_1}_1 A^{j_2}_2 A^{j_3}_3 A^{j_4 \oplus 1}_{4} \cdot A^{j_5}_5 \times \cdots \times A^{j_n}_n             . 
\end{align*}

}

\noindent The above product expansions of the responses that Alice provides for responding to iid samples of the referee's probability distribution mirror the product expansions,

 {\small

\[
 \left\{\!\begin{array}{ll@{}>{{}}l}  \underset{\mathcal{Q}_1 , \mathcal{Q}_2 , \mathcal{Q}_3}{\prod}   C^{l_{ijk}}_{ijk}  =     C^{l_{\mathcal{Q}_1, \mathcal{Q}_2, \mathcal{Q}_3}}_{\mathcal{Q}_1 \mathcal{Q}_2 \mathcal{Q}_3}            , 
\\ \\ \underset{\mathcal{Q}_1 , \mathcal{Q}_2 , \mathcal{Q}_3}{\prod}   C^{(N-2),l_{ijk}}_{ijk}  =      C^{(N-2),l_{\mathcal{Q}_1 \mathcal{Q}_2 \mathcal{Q}_3}}_{\mathcal{Q}_1 \mathcal{Q}_2 \mathcal{Q}_3}   , \\ \\ \\      \underset{\mathcal{Q}_3 , \mathcal{Q}_3 +1 \equiv \mathcal{Q}_3 \oplus 1}{\underset{\mathcal{Q}_2 , \mathcal{Q}_2 +1 \equiv \mathcal{Q}_2 \oplus 1}{\underset{\mathcal{Q}_1 , \mathcal{Q}_1 + 1 \equiv \mathcal{Q}_1 \oplus 1}{\prod}}}    C^{(N-2), l_{ijk}}_{ijk}  =  \underset{i \neq j \neq k}{\underset{\mathcal{Q}_3 , \mathcal{Q}_3 +1 \equiv \mathcal{Q}_3 \oplus 1}{\underset{\mathcal{Q}_2 , \mathcal{Q}_2 +1 \equiv \mathcal{Q}_2 \oplus 1}{\underset{\mathcal{Q}_1 , \mathcal{Q}_1 + 1 \equiv \mathcal{Q}_1 \oplus 1}{\prod}}}}   C^{(N-2), l_{\oplus i, \oplus j, \oplus k}}_{i,j,k}    , \end{array}\right.
\]

} 

\bigskip 

\noindent appearing in the multiplayer setting.

\section{Analyzing optimal strategies for the dual XOR game from optimal strategies of the XOR game}

\noindent We describe previous results underlying duality notions for quantum and classical games.

\subsection{The single bit of shared bipartite entanglement in the two-player nonlocal XOR game}

\noindent From the non-local $\mathrm{XOR}$ game, a corresponding $\mathrm{XOR}^{\mathrm{*}}$ game can be characterized by scenarios in which the bipartite state representing the strategies that Alice and Bob use to maximize a linear functional satisfies the following properties:

{\small

\begin{itemize}
    \item[$\bullet$] \textit{Obtaining the multiplicative factor of the Frobenius norm associated with Bob's strategy from the constant of the Frobenius norm associated with Alice's strategy}. In the setting of the FFL game, one has that,

{\small 

\begin{align*}
  9  , 
\end{align*}

}

    \noindent is used to obtain the remaining desired multiplicative factor,

{\small 

\begin{align*}
  44 / 3   , 
\end{align*}

}

    \noindent to the Frobenius norm of Alice's strategy.

    \bigskip 

  \item[$\bullet$] \textit{Expressing the superposition of Alice's observables for the FFL game independently of the action of suitably defined $T$ operators}. The action of the operator,

  {\small

  \begin{align*}
 \bigg\{   \frac{A_i A_j + A_j A_i}{2}   \otimes \textbf{I}  \bigg\}_{ \{ 1 \leq i \neq j  \leq d_A \} }  , \\ 
  \end{align*}
  
  }

\noindent independently of the operator $T_{\mathrm{FFL}}$ on states $\ket{\psi_{\mathrm{FFL}}}$ (the state associated with an optimal FFL strategy), has an upper bound of the form,

{\small

\begin{align*}
    2 \big(   \frac{7}{3}    \big)^2 n \big( n - 1 \big) \epsilon_{\mathrm{FFL}}   , \\ 
\end{align*}

}

\noindent for $\epsilon_{\mathrm{FFL}}$ taken sufficiently small.

  \bigskip 

    \item[$\bullet$] \textit{Expressing the superposition of Alice's observables for the FFL game under the action of suitably defined $T$ operators}. The action of the operators,

{\small

\begin{align*}
  \big\{  A_i \otimes \textbf{I} \big\}  T_{\mathrm{FFL}}     -          T_{\mathrm{FFL}} \big\{  \widetilde{A}_i  \otimes \textbf{I} \big\}     , \\  \\   \big\{  \textbf{I} \otimes B_{jk} \big\}  T_{\mathrm{FFL}}   - T_{\mathrm{FFL}} \big\{  \textbf{I} \otimes \widetilde{B_{jk}}    \big\}         , \\ 
\end{align*}

}

    \noindent dependently of the operator $T_{\mathrm{FFL}}$ on states $\ket{\psi_{\mathrm{FFL}}}$ has upper bounds of the form,

{\small

\begin{align*}
  9 n^2 \sqrt{\epsilon }  \big| \big| T_{\mathrm{FFL}} \big| \big|_F   , \\ \\ \frac{44}{3}  n^2 \sqrt{\epsilon }   \big| \big| T_{\mathrm{FFL}} \big| \big|_F   . 
\end{align*}

}

    \bigskip

    \item[$\bullet$] \textit{An intermediate inequality to the upper bounds on the Frobenius norm in the previous item}.

\end{itemize}

}

\noindent Besides $\textbf{Theorem}$ \textit{4} in {\color{blue}[37]}, there are several other inequalities involving bipartite states, possibly for strategies of an XOR game that are not optimized, which are adapted for proving characteristics of optimal, and approximately optimal, $\mathrm{CHSH}\big( n \big)$ strategies. In order to further explore the first type of game, the $\mathrm{XOR}^{\mathrm{*}}$ game, for which inequalities of a similar type to those involving the bipartite states 

\subsection{Further background on previous results of XOR optimal quantum strategies in the two-player setting}

\noindent \textbf{Lemma} \textit{1} (\textit{Two equalities between commuting observables, a bipartite state, and two vectors}, \textbf{Lemma} \textit{2}, {\color{blue}[37]}). For commuting $\pm 1$ observables $A_1 , \cdots , A_n$, the bipartite state $\psi \equiv \frac{1}{\sqrt{d}} \overset{d}{\underset{i=1}{\sum}} \ket{ii} \in \textbf{C}^d \otimes \textbf{C}^d$ and $u,v \in \textbf{R}^n$,

\begin{align*}
  \big\{  u "\cdot"  \vec{A} \big\}  \big\{  v "\cdot" \vec{A} \big\}  + \big\{  v "\cdot "  \vec{A} \big\} \big\{  u "\cdot" \vec{A} \big\}  = 2 \big\{  u^{\mathrm{T}} v \big\}  \textbf{I}  \text{, } 
\end{align*}

\noindent and, involving the bipartite state,

\begin{align*}
   \bra{\psi} \big\{   u "\cdot"  \vec{A}   \big\}  \otimes \big\{  v "\cdot"  \vec{A} \big\}^{\mathrm{T}} \ket{\psi}    \text{. } 
\end{align*}

\noindent \textbf{Theorem} \textit{2} (\textit{Frobenius norm upper bounds}, \textbf{Theorem} \textit{6}, {\color{blue}[37]}). Let $A_i$, $B_{jk}$, $\ket{\psi}$ be an $\epsilon$-approximate strategy on $\textbf{C}^{d_A} \otimes \textbf{C}^{d_B}$ and $\widetilde{A}_i$, $\widetilde{B}_{jk}$, $\ket{\widetilde{\psi}}$ be an optimal canonical strategy on $\textbf{C}^{2^{\lceil n / 2 \rceil}} \otimes \textbf{C}^{2^{\lceil n / 2 \rceil}}$. There exists a nonlinear operator $T$,

\begin{align*}
  T :  \big\{ \textbf{C}^{2^{\lceil \frac{n}{2} \rceil}}  \big\}  \otimes \big\{ \textbf{C}^{2^{\lceil \frac{n}{2} \rceil}}  \big\} \longrightarrow  \big\{ \textbf{C}^{d_A} \big\}  \otimes \big\{ \textbf{C}^{d_B} \big\}  \text{, } 
\end{align*}

\noindent for which:

\begin{itemize}
    \item[$\bullet$] {\textit{Frobenius norm for Alice's strategy}}. For all $i$,

\begin{align*}
  \big|\big|  \big\{  A_i \otimes \textbf{I} \big\}  T     -          T \big\{  \widetilde{A}_i  \otimes \textbf{I} \big\}     \big|\big|_{\mathrm{F}} \leq 12 n^2 \sqrt{\epsilon} \big| \big|     T   \big|\big|_{\mathrm{F}}   \text{. } 
\end{align*}

    \item[$\bullet$] {\textit{Frobenius norm for Bob's strategy}}. For all $j \neq k$,

\begin{align*}
    \big|\big|    \big\{  \textbf{I} \otimes B_{jk} \big\}  T   - T \big\{  \textbf{I} \otimes \widetilde{B_{jk}}    \big\}  \big|\big|_{\mathrm{F}} \leq  17 n^2 \sqrt{\epsilon}  \big| \big|     T   \big|\big|_{\mathrm{F}}  \text{. } 
\end{align*}

\end{itemize}

\noindent \textbf{Theorem} \textit{3} (\textit{matrix representations of semidefinite programs}, \textbf{Theorem} \textit{7}, {\color{blue}[37]}). For every $A_i$, $B_j$ and bipartite state $\psi$, there exists an $\big( n + m \big) \times \big( n + m \big)$ matrix $Z$ satisfying,

\begin{align*}
 G_{\mathrm{sym}} \cdot Z = \overset{n}{\underset{i=1}{\sum}} \overset{m}{\underset{j=1}{\sum}} G_{ij} \bra{\psi} A_i \otimes B_j \ket{\psi} \text{, } 
\end{align*}

\noindent in which $Z$ is said to be \textit{feasible}. Similarly, for every $A_i$, $B_j$ and bipartite state $\psi$ over $\textbf{C}^{2^{\lceil \frac{( n + m )}{2} \rceil}} \otimes \textbf{C}^{2^{\lceil \frac{( n + m )}{2} \rceil}}$, the same equality above holds for primal feasible solutions $Z$.

\bigskip

\noindent \textbf{Lemma} \textit{2} (\textit{Equality for Alice and Bob's strategy}, \textbf{Lemma} \textit{5}, {\color{blue}[37]}). Let $\{ A_i \}_{i \in \{1 , \cdots , n \}}$, $\{ B_j \}_{j \in \{ 1 , \cdots , m\}}$ and the bipartite state $\psi$ denote a quantum strategy. For vectors $u_1 , \cdots , u_r \in \textbf{R}^n$ and $v_1 , \cdots , v_r \in \textbf{R}^m$ satisfying,

\begin{align*}
   \overset{r}{ \underset{i=1}{\sum}  }   v_i v^{\mathrm{T}}_i = \overset{m}{\underset{i=1}{\sum}}     y_{n + i} E_{ii} =  \mathrm{diag} \big( y_{n+1} , \cdots , y_{n+m} \big)    \text{, } 
\end{align*}

\noindent one has that,

\begin{align*}
   \overset{r}{ \underset{k=1}{\sum}    }  \big| \big|     \big\{   u_k \cdot \vec{A} \otimes \textbf{I} \big\}  \ket{\psi} - \big\{ \textbf{I} \otimes v_k \cdot \vec{B} \big\}  \ket{\psi}    \big| \big|^2 =   \overset{m+n}{\underset{i=1}{\sum}} y_i -  \overset{n}{\underset{i=1}{\sum}} \overset{m}{\underset{j=1}{\sum}} G_{ij} \bra{\psi} A_i \otimes B_j \ket{\psi}  \text{. } 
\end{align*}

\bigskip

\noindent \textbf{Lemma} \textit{3} ($\epsilon$-\textit{approximate upper bound for} $\mathrm{CHSH}(n)$ \textit{strategies}), \textbf{Lemma} \textit{7}, {\color{blue}[37]}). For an $\epsilon$-approximate $\mathrm{CHSH}(n)$ strategy, $A_i$, $B_{jk}$, $n>0$, and $\ket{\psi}$, 

\begin{align*}
     \underset{1 \leq i < j \leq n}{ \sum}  \bigg| \bigg|  \bigg\{      \frac{A_i A_j + A_j A_i}{2}   \otimes \textbf{I} \bigg\}  \ket{\psi}   \bigg|\bigg|^2  \leq \big( 1 + \sqrt{2} \big)^2 n \big( n - 1 \big) \epsilon \text{. } 
\end{align*}

\bigskip

\noindent Given $A_i$ and $A_j$ introduced previously, permuting the operators acting on the tensor products below,

{\small \begin{align*}
   \big\{   A_i \otimes \textbf{I} \big\} \ket{\psi} =  \bigg\{  \textbf{I} \otimes \frac{B_{ij} + B_{ji}}{\sqrt{2}} \bigg\} \ket{\psi} \text{, }    \\     \\   \big\{  A_j \otimes \textbf{I} \big\}  \ket{\psi} = \bigg\{ \textbf{I} \otimes \frac{B_{ij} - B_{ji}}{\sqrt{2}} \bigg\}  \ket{\psi}    \text{, } 
\end{align*} }

\noindent which, in the case of the $\mathrm{CHSH}\big(2\big)$, $n=2$, strategy, is comprised of the Bell states,

{\small \begin{align*}      
\bigg(  \textbf{I} \otimes \textbf{I} \bigg) \bigg\{   \frac{\ket{00} + \ket{11}}{\sqrt{2}}\bigg\}   = \frac{\ket{00} + \ket{11}}{\sqrt{2}}  \text{, }     \bigg(     \sigma_x \otimes  \textbf{I} \bigg) \bigg\{  \frac{\ket{00} + \ket{11}}{\sqrt{2}} \bigg\}   = \frac{\ket{10} + \ket{01}}{\sqrt{2}}   \text{, }  \\  \\   \bigg(   \sigma_z    \otimes  \textbf{I} \bigg) \bigg\{   \frac{\ket{00} + \ket{11}}{\sqrt{2}} \bigg\}   = \frac{\ket{00} - \ket{11}}{\sqrt{2}}     \text{, }    \bigg( \sigma_x \sigma_z     \otimes    \textbf{I}  \bigg)  \bigg\{  \frac{\ket{00}+\ket{11}}{\sqrt{2}} \bigg\}   = \frac{\ket{10}- \ket{01}}{\sqrt{2}}   \text{. }  \\ 
\end{align*} }

\noindent Straightforwardly in the multiplayer basis over optimal strategies $\ket{\psi_{N\mathrm{XOR}}}$ (given the choice of a strictly positive $N$) the above Bell states generalize to a collection of relations of the form,

{\small

\begin{align*}
    \bigg( \frac{\textbf{I} \otimes \textbf{I} \otimes \overset{N-3}{\cdots} \otimes \textbf{I}}{\sqrt{N}} \bigg) \bigg\{     \underset{1 \leq j \leq N}{\sum} \ket{\text{Player } j \text{ state}}      \bigg\}   \equiv  \frac{1}{\sqrt{N}}  \bigg\{     \underset{1 \leq j \leq N}{\sum} \ket{\text{Player } j \text{ state}}      \bigg\}      \text{, } \\ \end{align*}
    
    \begin{align*} \bigg( \frac{\sigma_z \otimes \textbf{I} \otimes \overset{N-3}{\cdots} \otimes \textbf{I}}{\sqrt{N}} \bigg)   \bigg\{     \underset{1 \leq j \leq N}{\sum} \ket{\text{Player } j \text{ state}}      \bigg\}    \equiv  \frac{1}{\sqrt{N}}    \bigg\{     \ket{\text{Player 1 state}} \\ - \ket{\text{Player 2 state} } + \ket{\text{Player 3 state}} +  \cdots      + \ket{\text{Player } N \text{ state}}      \bigg\}                     \text{, } \\ \end{align*}
    
    \begin{align*}     \bigg(     \frac{\textbf{I} \otimes \sigma_z \otimes \textbf{I} \otimes \overset{N-3}{\cdots} \otimes \textbf{I}}{\sqrt{N}}     \bigg)          \bigg\{     \underset{1 \leq j \leq N}{\sum} \ket{\text{Player } j \text{ state}}      \bigg\}   \equiv  \frac{1}{\sqrt{N}}    \bigg\{                          \underset{1 \leq j \leq 2}{\sum}   \ket{\text{Player } j \text{ state}} \\ - \ket{\text{Player } 3 \text{ state}}       + \underset{4 \leq j \leq N}{\sum} \ket{\text{Player } j \text{ state}}       \bigg\}                 \text{, }  \end{align*}
    
    \begin{align*}       \vdots \\  \bigg(      \frac{\textbf{I} \otimes \textbf{I} \otimes \overset{N-4}{\cdots } \otimes \textbf{I} \otimes \sigma_z}{\sqrt{N}}                    \bigg)     \bigg\{     \underset{1 \leq j \leq N}{\sum} \ket{\text{Player } j \text{ state}}      \bigg\}  \equiv  \frac{1}{\sqrt{N}}    \bigg\{                          - \ket{\text{Player } 1 \text{ state}} \\ + \underset{2 \leq j \leq N}{\sum} \ket{\text{Player } j \text{ state}}              \bigg\}        \text{, } \\ \end{align*}
    
    \begin{align*}      \bigg(    \frac{\sigma_x \otimes \textbf{I} \otimes  \overset{N-2}{\cdots} \otimes \textbf{I}}{\sqrt{N}}  \bigg)     \bigg\{    \underset{1 \leq j \leq N}{\sum} \ket{\text{Player } j \text{ state}}      \bigg\}  \equiv  \frac{1}{\sqrt{N}}  \bigg\{     \widetilde{\ket{\text{Player } 1 \text{ state}}} + \widetilde{\ket{\text{Player } 2 \text{ state}}}    \\      + \underset{3 \leq j \leq N}{\sum}   \ket{\text{Player } j \text{ state}}                                 \bigg\}       \text{, }   \end{align*}
    
    \begin{align*}     \bigg(     \frac{\textbf{I} \otimes \sigma_x \otimes \overset{N-2}{\cdots} \otimes \textbf{I}}{\sqrt{N}}              \bigg)     \bigg\{     \underset{1 \leq j \leq N}{\sum} \ket{\text{Player } j \text{ state}}      \bigg\}  \equiv  \frac{1}{\sqrt{N}}    \bigg\{     \ket{\text{Player }1 \text{ state}} \\ + \widetilde{\ket{\text{Player } 2 \text{ state}}}  +         \widetilde{\ket{\text{Player } 3 \text{ state}}}       + \underset{4 \leq j \leq N}{\sum}   \ket{\text{Player } j \text{ state}}       \bigg\}       \text{, } \\          \vdots  
\\    \bigg( \frac{\textbf{I} \otimes \textbf{I} \otimes \overset{N-4}{\cdots} \otimes \sigma_z \otimes \sigma_x }{\sqrt{N}} \bigg)    \bigg\{     \underset{1 \leq j \leq N}{\sum} \ket{\text{Player } j \text{ state}}      \bigg\}    \equiv \frac{1}{\sqrt{N}}  \bigg\{       \widetilde{\ket{\text{Player } 1 \text{ state}}} \\ +   \ket{\text{Player } 2 \text{ state}}                        +     \underset{3 \leq j \leq N-1}{\sum}    \ket{\text{Player } j \text{ state}} -  \widetilde{\ket{\text{Player } N \text{ state}}}   \bigg\}   .  
\end{align*}

}

\subsection{Connecting the XOR game to the dual XOR, XOR*, game}

\noindent In the second subsection, to characterize optimality of strategies for $\mathrm{XOR}^{*}$ games from optimal strategies of $\mathrm{XOR}$ games, we also introduce background from {\color{blue}[5]} below. The following notion of duality between an XOR and XOR* game was leveraged for connecting optimal strategies in the original game to its dual. Moreover, by placing assumptions on the inputs and outputs that Alice and Bob can use in the XOR and XOR* settings the classical and quantum values for the Odd-Cycle game are obtained. As another game of interest separate to the XOR* and FFL games considered in this work with the optimal value,

{\small

\begin{align*}
      1 - \big\{ 2 n \big\}^{-1}    , \\ 
\end{align*}

}

\noindent for odd $n$ with $n \geq 3$ using quantum strategies and the action spaces,

{\small \begin{align*}
      \mathcal{A} \equiv  \big\{   x \in \big\{ 0 , \cdots , n -1 \big\}    :   \textit{x is distributed to Alice from an iid sample $\sim  \pi_{\mathrm{Odd-Cycle}}$}    \big\}   \in \big\{ 0 , 1 \big\}      , \\ \\         \mathcal{B} \equiv \big\{  y \in \big\{ x , \big( x+1 \big) \mathrm{mod} \text{ } n \big\}    :   \textit{y is distributed to Bob from iid sample $\sim  \pi_{\mathrm{Odd-Cycle}}$}    \big\}  \in \big\{ 0 , 1 \big\}  . \\ 
\end{align*}   }

\noindent of Alice and Bob, respectively, with the payoff function,

{\small

\[
 \left\{\!\begin{array}{ll@{}l} \underline{\textit{Payoff $+1$}}.  \text{ Alice and Bob receive a +1 score (they win the game) from the referee if: }  \bigg\{ \big\{ a = b \big\} \\ \Longleftrightarrow \big\{  y = x \big\} \bigg\}  ,  \\ \\ \underline{\textit{Payoff $+0$}}. \text{ Alice and Bob receive a +0 score (they lose the game) from the referee if: } \bigg\{ \big\{  a \neq b \big\}  \\ \Longleftrightarrow  \big\{             y = \big( x+1 \big)  \mathrm{mod} \text{ } n       \big\} \bigg\}  ,        \end{array}\right.  \]

}

\bigskip

\noindent and the referree's distribution,

{\small

\begin{align*}
    \pi_{\mathrm{Odd-Cycle}} \equiv \underset{i \in V ( \textbf{T}^2 )}{\bigcup} \big\{ i : \textit{the referee distributes i to Alice, or to Bob, for coloring in the Odd} \\ \textit{-Cycle game} \big\} , \\ 
\end{align*}

}

\noindent over the two-dimensional torus $\textbf{T}^2$. In the settings of the XOR and Odd-Cycle games alike, under the assumption that Alice and Bob only share a single bit of shared bipartite entanglement their strategies which are optimal for the original game are also optimal for the dual game, {\color{blue}[5]}. Within the context of optimality provided in {\color{blue}[37]}, particularly with regards to the relationship between the XOR optimal value and the constants in the Frobenius norm upper bounds stated in $\textbf{Theorem}$ $\textit{2}$, it is of interest to quantify how optimal strategies in XOR and XOR* games differ from those encountered in FFL games.

To reiterate we draw the attention of the reader back to the expressions provided in previous remarks for $\pi_{\mathrm{XOR}}$ and $\pi_{\mathrm{FFL}}$. Specifically, to relate XOR* optimality to FFL optimality, in which ways would changing the underlying probabilities that Alice and Bob respond to the iid sample $\big( i , j \big) \sim \pi \times \pi $, whether it be from,

{\small

\begin{align*}
     \pi \equiv    \pi_{\mathrm{XOR}} \big(  0 , 0   \big) =  \pi_{\mathrm{XOR}} \big( 0 , 1   \big) =    \pi_{\mathrm{XOR}} \big( 1 , 0   \big) =       \pi_{\mathrm{XOR}} \big( 1 , 1  \big)    =      1 / 4              , \\ 
\end{align*}

}

\noindent corresponding to the XOR game or from,

{\small

\begin{align*}
  \pi \equiv   \pi_{\mathrm{FFL}} \big( 0 , 0 \big) =  \pi_{\mathrm{FFL}} \big( 0 , 1 \big) = \pi_{\mathrm{FFL}} \big( 1 , 0 \big)    =  1 / 3  , \end{align*}

    \begin{align*} 
     \pi \equiv    \pi_{\mathrm{FFL}} \big( 1 , 1 \big)    =   0       , \\ 
\end{align*}

}

\noindent corresponding to the FFL game be related to the Frobenius norm of $T$ operators, such as ones of the form,

{\small

\begin{align*}
 \bigotimes T^{\mathrm{XOR}} : \underset{N \text{ copies}}{\bigotimes} \big\{   \textbf{C}^{2 \lceil \frac{n}{2} \rceil } \big\}   \longrightarrow        \underset{1 \leq i \leq N}{\bigotimes} \big\{  \textbf{C}^{d_i} \big\}  \text{, }  \end{align*}

 \begin{align*} \bigwedge T^{\mathrm{XOR}} :  \big\{ \textbf{C}^{2 \lceil \frac{n}{2} \rceil } \big\} \bigwedge     \big\{  \textbf{C}^{2 \lceil \frac{n}{2} \rceil } \big\}  \bigwedge \cdots \bigwedge  \big\{ \textbf{C}^{2 \lceil \frac{n}{2} \rceil }  \big\} \longrightarrow        \big\{ \textbf{C}^{d_A} \big\}  \bigwedge \big\{ \textbf{C}^{d_B} \big\}  \bigwedge \big\{ \textbf{C}^{d^{(1)}_B} \big\} \bigwedge\cdots \\ \bigwedge \big\{ \textbf{C}^{d^{(n-2)}_B}  \big\} \text{, } \end{align*}

 \begin{align*}    T^{ \{ \mathrm{XOR}\wedge \cdots \wedge \mathrm{XOR} \} } :     \textbf{C}^{ \{ 2 \lceil \frac{n}{2} \rceil \wedge 2 \lceil \frac{n}{2} \rceil \wedge \cdots \wedge 2 \lceil \frac{n}{2} \rceil \} }  \longrightarrow        \textbf{C}^{ \{ d_A \wedge d_B \wedge d^{(1)}_B \wedge \cdots \wedge {d^{(n-2)}_B } \} }    \text{, } \end{align*}

 \begin{align*}      T^{\mathrm{FFL}\wedge  \mathrm{FFL}} :     \textbf{C}^{ \{ 2 \lceil \frac{n}{2} \rceil \wedge 2 \lceil \frac{n}{2} \rceil        \}   }  \longrightarrow        \textbf{C}^{ \{ d_A \wedge d_B    \}   }    ? 
\end{align*}

}

\noindent Intuitively, while one could expect that the action of $T$ operators similar to those above can be related to the Frobenius norm, it is of great interest to quantify up to constants estimates for inequalities of the form,

{\small

\begin{align*}
    \big|\big|  \big\{  A_i \otimes \textbf{I} \big\}  T^{\mathrm{FFL}}     -          T^{\mathrm{FFL}} \big\{  \widetilde{A}_i  \otimes \textbf{I} \big\}     \big|\big|_{\mathrm{F}} /  \big| \big|     T^{\mathrm{FFL}}   \big|\big|_{\mathrm{F}}    \lesssim   n^2 \sqrt{\epsilon}        , \\ \\           \big|\big|    \big\{  \textbf{I} \otimes B_{jk} \big\}  T^{\mathrm{FFL}}   - T^{\mathrm{FFL}} \big\{  \textbf{I} \otimes \widetilde{B_{jk}}    \big\}  \big|\big|_{\mathrm{F}} /  \big| \big|     T^{\mathrm{FFL}}  \big| \big|_F \lesssim   n^2 \sqrt{\epsilon}      ,  \\ 
\end{align*}

}

\noindent for all $i$ such that $1 \leq i \leq d_A$ and all $j\neq i$ such that $1 \leq j \leq d_B$, respectively. 

\bigskip

\noindent We describe the specific criteria of duality developed in {\color{blue}[5]} with the following. Denote the set of possible inputs that Alice inputs into the XOR game with $\big| \mathcal{S} \big|$, and the set of possible inputs that Bob inputs with $\big| \mathcal{T} \big|$. Also, denote the Bell state with $\ket{\psi} = \frac{1}{\sqrt{2}} \big( \ket{00} + \ket{11} \big) $. To compare $\mathrm{XOR}$ games to the dual $\mathrm{XOR}^{*}$ game, we make use of the following items:

\begin{itemize}
    \item[$\bullet$] \big({The e-bit $\mathrm{XOR}$ game, \textbf{Lemma} \textit{1}, {\color{blue}[5]}}\big) An $\mathrm{XOR}$ game for which $\mathrm{min} \big\{ \big| \mathcal{S} \big| ,    \big| \mathcal{T} \big|   \big\} \leq 4$        is an e-bit $\mathrm{XOR}$ game.          

\bigskip 
        \item[$\bullet$] \big({Classical and quantum bounds for $\mathrm{XOR}$  and $\mathrm{{XOR}^{*}}$ games, \textbf{Theorem} \textit{2}, {\color{blue}[5]}}\big). Denote $a$ and $b$ as the two possible measurements that Alice and Bob can observe from some $s \in \mathcal{S}$ and $t \in \mathcal{T}$. Furthermore, denote the single qubit measurements from each possible $s$ and $t$ with $A_{a | s}$ and $B_{b|t}$, and the probability, conditional upon each input, as, $P \big(    a , b \big| s , t        \big)$, which can be expressed with the trace of the inner product $\big(  A_{a|s}       \otimes    B_{b|t}     \big) \ket{\psi} \bra{\psi} $. The output of the XOR game, $m = a \oplus b$, is such that the classical and quantum bounds of the $\mathrm{XOR}$ and $\mathrm{XOR^{*}}$games are equal.

        The same result holds for the converse.

\end{itemize}

\noindent In the context of the optimality framework provided in {\color{blue}[37]}, one could be interested in the following research directions:

{\small

\begin{itemize}
    \item[$\bullet$] \textit{Notions of optimality from bipartite entanglement of a single bit}. Is it possible to relate how quantum games besides those of XOR game in the two-player setting are related to other games which have different optimal values?

    \bigskip 

    \item[$\bullet$] \textit{Duality gaps}. What would be the expected gap between classical and quantum strategies for games obtained from the above duality notion? 

    \bigskip 

        \item[$\bullet$] \textit{Partial ordering with respect to the ordering positive semidefinite cone}. Along the lines of the background provided for semidefinite programs and their duals, as seen through the explicit representation of the dual objective function,

        {\small

\[
\vec{y_{3\mathrm{XOR}}}  \equiv  \left\{\!\begin{array}{ll@{}>{{}}l}
    {\tiny  y_{3\mathrm{XOR},1} \equiv \cdots \equiv y_{3\mathrm{XOR},n} \equiv \omega_{3\mathrm{XOR}} \big\{  {3! n} \big\}^{-1} } \text{, } \\  \\  {\tiny y_{3\mathrm{XOR},n+1} \equiv \cdots \equiv    y_{3\mathrm{XOR},n^2}  \equiv \omega_{3\mathrm{XOR}} \big\{  {3! n ( n-1 ) }  \big\}^{-1} } \text{, } \\ \\  {\tiny y_{3\mathrm{XOR},n^2+1} \equiv \cdots \equiv    y_{3\mathrm{XOR},n^3}  \equiv \omega_{3\mathrm{XOR}}  \big\{ {3! n ( n-1 ) ( n-2 ) } \big\}^{-1}  }  \text{, } 
      \end{array}\right.
\]

}

\bigskip

        \noindent for the multiplayer XOR game, what would be the expected form of other dual objective functions obtained from the above duality notion?

        \bigskip 

        \item[$\bullet$] \textit{Representations of the optimal strategy in Hilbert space}. Lastly, would there exist any expressions for states corresponding to the optimal strategy? For example, for the Odd-Cycle game discussed above in the context of the e-bit the optimal strategy takes the form, {\color{blue}[5]},

{\small \begin{align*}
\ket{\psi_{\text{Odd-Cycle}}} \equiv       \frac{1}{\sqrt{2}}  \bigg\{  {\tiny   \ket{0}_A \ket{1}_B + \mathrm{exp} \big( \theta \big)  \ket{1}_A \ket{0}_B    }     \bigg\}     \text{, } \\ 
\end{align*} }

\noindent for the rotation angle $\theta$ modulo $2\pi$ (for $0 \leq \theta \leq 2 \pi$), and,

{\small    \begin{align*}
     \ket{0}_A \neq  \ket{1}_A ,  \ket{0}_A ,   \ket{1}_A  \in \textbf{C}^{d^{\prime}_A}  \text{, } \\ \\   \ket{0}_B \neq  \ket{1}_B ,  \ket{0}_B , \ket{1}_B  \in \textbf{C}^{d^{\prime}_B}  \text{. }
\end{align*}  }

\noindent for $d^{\prime}_A \neq d_A$, $d^{\prime}_B \neq d_B$.

\end{itemize}

}

\subsubsection{FFL games}

\noindent In terms of the optimal, and approximately optimal strategies of $\mathrm{XOR}$ games introduced in {\color{blue}[37]}, the results can be recast from $\mathrm{XOR}$ games into $\mathrm{XOR^{*}}$ games from the duality notion described in the previous subsection. From previous remarks, the fact that,

{\small

\begin{align*}
 \pi_{\mathrm{FFL}} \big( 0 , 0 \big) =  \pi_{\mathrm{FFL}} \big( 0 , 1 \big) = \pi_{\mathrm{FFL}} \big( 1 , 0 \big)    =  1 / 3  , \end{align*}

    \begin{align*} 
      \pi_{\mathrm{FFL}} \big( 1 , 1 \big)    =   0       ,  
\end{align*}

}

\noindent implies, given the choice of bias,

{\small

\begin{align*}
\beta \big( G^* \big) \equiv   \beta_{\mathrm{FFL}} \big( G \big) =    \overset{n}{\underset{i=1}{\sum} } \overset{m}{\underset{j=1}{\sum}}  G^{*}_{ij} \bra{\psi^{*}} A_i \otimes B_j \ket{\psi^{*}} =   \overset{n}{\underset{i=1}{\sum} } \overset{m}{\underset{j=1}{\sum}}  G_{ij,\mathrm{FFL}} \bra{\psi_{\mathrm{FFL}}}  A_i \otimes B_j \ket{\psi_{\mathrm{FFL}}}                           , \\ 
\end{align*}

}

\noindent that an inequality of the form,

{\small \begin{align*}
         \big( 1 - \epsilon \big) \beta \big( G^{*} \big) \leq \overset{n}{\underset{i=1}{\sum} } \overset{m}{\underset{j=1}{\sum}}  G^{*}_{ij} \bra{\psi^{*}} A_i \otimes B_j \ket{\psi^{*}} \leq \beta \big( G^{*} \big)                     \text{, } 
\end{align*} }

\noindent should hold, given a strictly positive parameter $\epsilon$, for,

{\small \begin{align*}
 A_{S,\mathrm{FFL}} \equiv \underset{s \sim \pi_{\mathrm{FFL}}}{\underset{s \in S}{\bigcup}}  A_s \equiv \big\{   s  \in S :  A_s \in \big\{ - 1 , + 1 \big\}    \big\}     \text{, } \\  \\   B_{T,\mathrm{FFL}} \equiv \underset{t \sim \pi_{\mathrm{FFL}}}{\underset{t \in T}{\bigcup}} B_t  \equiv \big\{ t \in T  :  B_t \in \big\{ -1 , + 1 \big\}  \big\} , \\ \\ \widetilde{A}_{S,\mathrm{FFL}} \equiv \underset{s \sim \pi_{\mathrm{FFL}}}{\underset{s \in S}{\bigcup}}  \widetilde{A}_s \equiv \big\{   s  \in S :  \widetilde{A}_s \in \big\{ - 1 , + 1 \big\}    \big\}     \text{, } \\    \\ \widetilde{B}_{T,\mathrm{FFL}} \equiv \underset{t \sim \pi_{\mathrm{FFL}}}{\underset{t \in T}{\bigcup}} \widetilde{B}_t  \equiv \big\{ t \in T  :  \widetilde{B}_t \in \big\{ -1 , + 1 \big\}  \big\}   ;  
\end{align*}
 }

\noindent formally, the above inequality from the dual bias used to define the FFL bias is obtained by restricting the maximum cardinality of the set of inputs that Alice and Bob use for the game to be $\leq 4$, then the dual $\mathrm{XOR^{*}}$ game would have the same optimal strategy equaling $\frac{1}{\sqrt{2}}$ iff there exists a state $\ket{\psi^{*}}$, obtained from $\ket{\psi}$, for which in the multiplayer setting,

{\small

\begin{align*}
     \big( 1 - \epsilon \big) \beta \big( G \big)  \leq  \underset{\text{Questions}}{\sum} \bra{\text{Optimal Strategy}} \bigg[    \underset{\# \text{ Players }}{\bigotimes}   \text{Tensors of player observables}   \bigg]  \\ \times    \ket{\text{Optimal Strategy}}  \leq      \beta \big( G \big)          ,  
\end{align*}

}

\noindent

\bigskip 

\noindent where the optimal, and nearly optimal strategies, of the dual $G^{*}$ to $G$ are determined by entries of the dual game matrix $G^{*}$, and the corresponding bias. In light of the duality framework with the single entangled bit described in {\color{blue}[5]}, whether a dual primal feasible solution $Z^*$, distributed according to the correspondence,

{\small

\begin{align*}
 Z_{\mathrm{FFL}}  \overset{\mathrm{Duality}}{\longleftrightarrow} Z^* \overset{\mathrm{Existence}}{\Longleftrightarrow}    \underset{1 \leq i \leq d_A}{\bigcup }    \bigg\{   \bigg\{ \underset{\forall i = 1 , \cdots , m \text{ } :\text{ }  F^*_i \cdot Z^* = c^*_i}{\underset{Z^*  \succcurlyeq 0}{\mathrm{sup}}} \big\{   G^* \cdot Z^*   \big\}  \bigg\} \textit{ is well-posed for $c^*_i$} \bigg\}  \\ \\ {\Longleftrightarrow}  \bigg\{     \textit{For all $1 \leq i \leq d_A$ there exists a $c^*_i$ such that a dual primal feasi-} \\ \textit{ble solution $Z^{*}$ to the semidefinite program exists}     \bigg\}           , 
\end{align*}

}

\noindent from the semidefinite program,

{\small \begin{align*}
 \underset{\forall i = 1 , \cdots , m \text{ } :\text{ }  F^*_i \cdot Z^* = c^*_i}{\underset{Z^*  \succcurlyeq 0}{\mathrm{sup}}} \big\{   G^* \cdot Z^*   \big\}    \text{, } 
\end{align*} }

\noindent distributed according to,

{\small \begin{align*}
\underset{1 \leq i \leq d_A}{\bigcup }  \bigg\{  \underset{\forall i = 1 , \cdots , m \text{ } :\text{ }  F^*_i \cdot Z^* = c^*_i}{\underset{Z^*  \succcurlyeq 0}{\mathrm{sup}}} \big\{   G^* \cdot Z^*   \big\} \bigg\}    \text{, } 
\end{align*} }

\noindent can be related to the constraints of the dual semidefinite program,

{\small \begin{align*}
     \underset{\overset{m}{\underset{i=1}{\sum}} y^*_i F^*_i \succcurlyeq G^*}{\mathrm{inf}}   \big\{ \vec{c^*} \cdot \vec{y^*}    \big\}   \text{, }      
\end{align*} }

\noindent given the choice of the \textit{dual feasible objective function} $\vec{y^*}$. From the duality gap associated with the original and dual semidefinite programs above,

{\small \begin{align*}
  \bigg[   \overset{m}{\underset{i=1}{\sum}}   y^*_i F^*_i - G^*   \bigg]  \cdot Z^* \geq 0  \text{, }
\end{align*} }

\noindent one could be interested in how the previously mentioned notion of well-posedness for the semidefinite program with primal feasible solution $Z_{\mathrm{FFL}}$,

{\small

\begin{align*}
   \underset{1 \leq i \leq d_A}{\bigcup }    \bigg\{   \bigg\{ \underset{\forall i = 1 , \cdots , m \text{ } :\text{ }  F^*_i \cdot Z^* = c^*_i}{\underset{Z^*  \succcurlyeq 0}{\mathrm{sup}}} \big\{   G^* \cdot Z^*   \big\}  \bigg\} \textit{ is well-posed for $c^*_i$} \bigg\}  ,
\end{align*}

}

\noindent is related to well-posedness of the dual objective function,

{\small

\begin{align*}
   \underset{1 \leq i \leq d_A}{\bigcup }    \bigg\{  \bigg\{  \underset{\overset{m}{\underset{i=1}{\sum}} y^*_i F^*_i \succcurlyeq G^*}{\mathrm{inf}}   \big\{ \vec{c^*} \cdot \vec{y^*}    \big\}  \bigg\} \textit{ is well-posed for each $F^*_i$}                     \bigg\}  ,
\end{align*}

}

\noindent for the FFL game.

\bigskip

\noindent Besides the inequality above which provides lower and upper bound depending on the bias of dual $G^{*}$, two more inequalities for the dual optimal strategy $\ket{\psi^{*}}$, the first of which states,

{\small \begin{align*}
        \underset{1 \leq i < j \leq n}{\sum} \bigg[  \text{ } \bigg| \bigg|    \bigg[      \bigg\{  \frac{A_i + A_j}{\sqrt{2}} \bigg\}  \otimes \textbf{I} \bigg] \ket{\psi^{*}}    -  \big\{     \textbf{I} \otimes B_{ij} \big\}  \ket{\psi^{*}}   \bigg|\bigg|^2 + \bigg| \bigg| \bigg[  \bigg\{  \frac{A_i - A_j}{\sqrt{2}} \bigg\}   \otimes \textbf{I} \bigg] \ket{\psi^{*}}    -   \big\{   \textbf{I} \otimes B_{ji} \big\} \ket{\psi^{*}}     \bigg|\bigg|^2 \text{ }  \bigg] \\  \leq 2n \big( n - 1 \big) \epsilon              \text{, } 
\end{align*} }

\noindent corresponding to the {Approximately Optimal Observables} property given in the first item of \textbf{Theorem} \textit{1}, and the second of which states,

{\small \begin{align*}
       \underset{1 \leq i < j \leq n}{\sum} \bigg[  \text{ } \bigg| \bigg| \big\{  A_i \otimes \textbf{I} \big\} \ket{\psi^{*}}    -  \bigg[  \textbf{I} \otimes \bigg\{  \frac{B_{ij} + B_{ji}}{\sqrt{2}} \bigg\}   \bigg] \ket{\psi^{*}}   \bigg| \bigg|^2  + \bigg| \bigg|   \big\{  A_j  \otimes \textbf{I} \big\}  \ket{\psi^{*}}  -   \bigg[   \textbf{I} \otimes \bigg\{  \frac{B_{ij} - B_{ji}}{\sqrt{2}} \bigg\}  \bigg] \ket{\psi^{*}{}}    \bigg| \bigg|^2  \text{ }       \bigg] \\  \leq 2n \big( n - 1 \big) \epsilon            \text{, } 
\end{align*} } 

\noindent corresponding to reversing the order in which the tensor product is applied, which is given in the second item of \textbf{Theorem} \textit{2}. Under the assumption that the optimality parameter $\epsilon$ is taken small enough one would expect,

{\small

\begin{align*}
   \underset{1 \leq i < j \leq n}{\sum} \bigg[  \text{ } \bigg| \bigg| \big\{  A_i \otimes \textbf{I} \big\} \ket{\psi^{*}}    -  \bigg[  \textbf{I} \otimes \bigg\{  \frac{B_{ij} + B_{ji}}{\sqrt{2}} \bigg\}   \bigg] \ket{\psi^{*}}   \bigg| \bigg|^2 \bigg] \lesssim  \underset{1 \leq i < j \leq n}{\sum} \bigg[ \big\{ A_j \otimes \textbf{I} \big\}   \ket{\psi^{*}}  -   \bigg[   \textbf{I} \\ \otimes \bigg\{  \frac{B_{ij} - B_{ji}}{\sqrt{2}} \bigg\}  \bigg] \ket{\psi^{*}{}}    \bigg| \bigg|^2  \text{ }       \bigg]    , \\ 
\end{align*}

}

\noindent The above properties of error bounds parallel those in higher-dimensional games, particularly through the use of quantities of the form,

{\small

\begin{align*}
 \frac{1}{\sqrt{\# \sigma}} \underset{\textit{permutations} \text{ } \sigma}{\sum} A_{\sigma (i  )}       , \\ \\        \frac{1}{\sqrt{\# \sigma^{\prime}}}  \underset{j \neq i }{\underset{\textit{permutations} \text{ } \sigma^{\prime}}{\sum}} B_{\sigma^{\prime} (j  )}         , \\  \\           \frac{1}{\sqrt{\# \sigma^{\prime\prime}}}  \underset{k \neq j \neq i }{\underset{\textit{permutations} \text{ } \sigma^{\prime\prime}}{\sum}} C_{\sigma^{\prime} (k  )}                   , 
\end{align*}

}

\noindent in which the permutations $\sigma, \sigma^{\prime}, \sigma^{\prime\prime}$ are respectively taken over the symmetric groups of $1$, $2$ and $3$ letters. The above summations over $\sigma, \sigma^{\prime}, \sigma^{\prime\prime}$ reflect the behavior of superpositions of optimal states of the form,

{\small

\begin{align*}
          \bigg\{   \bigg(    \frac{1}{\sqrt{\# \sigma}} \underset{\textit{permutations} \text{ } \sigma}{\sum} A_{\sigma (i  )}   \bigg)   \bigotimes   \big\{  \textbf{I} \otimes \overset{(N-1) \text{ } \textit{terms}}{\cdots} \otimes \textbf{I} \big\}  \bigg\} \ket{\psi_{\mathrm{NXOR}}}                             , \end{align*}

          \begin{align*} \bigg\{   \textbf{I} \bigotimes           \bigg(    \frac{1}{\sqrt{\# \sigma^{\prime}}}  \underset{j \neq i }{\underset{\textit{permutations} \text{ } \sigma^{\prime}}{\sum}} B_{\sigma^{\prime} (j  )}  \bigg)          \bigotimes  \big\{  \textbf{I} \otimes \overset{(N-2) \text{ } \textit{terms}}{\cdots} \otimes \textbf{I} \big\}  \bigg\} \ket{\psi_{\mathrm{NXOR}}}         ,     \end{align*}

          \begin{align*}   \bigg\{  \big\{  \textbf{I} \otimes \textbf{I}      \big\} \bigotimes         \bigg(           \frac{1}{\sqrt{\# \sigma^{\prime\prime}}}  \underset{k \neq j \neq i }{\underset{\textit{permutations} \text{ } \sigma^{\prime\prime}}{\sum}} C_{\sigma^{\prime} (k  )}        \bigg)          \bigotimes  \big\{  \textbf{I} \otimes \overset{(N-3) \text{ } \textit{terms}}{\cdots} \otimes \textbf{I} \big\}  \bigg\} \ket{\psi_{\mathrm{NXOR}}}    , 
\end{align*}

}

\noindent given the optimal NXOR state $\ket{\psi_{\mathrm{NXOR}}}$ which are further developed in the FFL setting for superpositions of the form,

{\small

\begin{align*}
    \bigg\{      \bigg(    \frac{1}{\sqrt{\# \sigma}} \underset{\textit{permutations} \text{ } \sigma}{\sum} A_{\sigma (i  )}   \bigg)   \bigotimes   \textbf{I}        \bigg\} \ket{\psi_{\mathrm{FFL}}}       , \\ \\        \bigg\{         \textbf{I} \bigotimes           \bigg(    \frac{1}{\sqrt{\# \sigma^{\prime}}}  \underset{j \neq i }{\underset{\textit{permutations} \text{ } \sigma^{\prime}}{\sum}} B_{\sigma^{\prime} (j  )}  \bigg)              \bigg\} \ket{\psi_{\mathrm{FFL}}}      ,         
\end{align*}

}

\noindent given the optimal FFL state $\ket{\psi_{\mathrm{FFL}}}$.

\bigskip

\noindent Besides the inequalities above which are associated with dual state $\ket{\psi^{*}}$ that can be obtained by the optimization problem,

{\small

\begin{align*}
 \omega^* =   \underset{j\neq i }{\underset{B_{j}, j \sim \pi_{\mathrm{FFL}}}{\underset{A_i, i \sim \pi_{\mathrm{FFL}}}{\underset{\ket{\psi^{*}}}{\mathrm{sup}}}}} \big\{  \bra{{\psi^{*}}}                           A_i \otimes B_j           \ket{\psi^{*}}   \big\}  ,  
\end{align*}

}

\noindent of the optimal strategy for the dual optimal value $\omega^*$, for the Fortnow-Feige-Lovasz (FFL) game, the classical and quantum values coincide. Hence the associated classical-quantum duality gap, from the fact that,

\begin{align*}
\mathrm{Classical \text{ } FFL \text{ } value} \equiv   \omega_c \big(  \mathrm{FFL} \big) = \mathrm{quantum \text{ } FFL \text{ } value} \equiv \omega_q \big( \mathrm{FFL}  \big) = 2 / 3  \text{, } 
\end{align*}

\noindent is zero. As a result from the supremum of classical and quantum values equaling $2/3$, by following the arguments provided in {\color{blue}[37]}, one obtains the inequality,

\begin{align*}
2 / 3   \big( 1 - \epsilon \big) \leq \overset{n}{\underset{i=1}{\sum} } \overset{m}{\underset{j=1}{\sum}} \big(  G_{\mathrm{FFL}} \big)_{ij} \bra{\psi^{*}_{\mathrm{FFL}}} A_i \otimes B_j \ket{\psi^{*}_{\mathrm{FFL}}} \leq 2 / 3                 \text{, } 
\end{align*}

\noindent for $\epsilon \equiv \epsilon_{\mathrm{FFL}}$, which is equivalent to the condition,

{\small \begin{align*}
  \underset{1 \leq i < j \leq n}{\sum} \bigg[  \text{ } \bigg| \bigg|        \bigg[   \bigg\{  \frac{A_i + A_j}{\sqrt{2}} \bigg\}  \otimes \textbf{I} \bigg] \ket{\psi^{*}_{\mathrm{FFL}}}    -    \big\{   \textbf{I} \otimes B_{ij} \big\}  \ket{\psi^{*}_{\mathrm{FFL}}}   \bigg|\bigg|^2 +  \bigg| \bigg| \bigg[  \bigg\{  \frac{A_i - A_j}{\sqrt{2}} \bigg\}   \otimes \textbf{I} \bigg] \ket{\psi^{*}_{\mathrm{FFL}}}   \\  -    \big\{  \textbf{I} \otimes B_{ji} \big\}   \ket{\psi^{*}_{\mathrm{FFL}}}     \bigg|\bigg|^2  \text{ } \bigg]   \leq 2n \big( n - 1 \big) \epsilon     \text{, } \end{align*}   } 
  
  \noindent and to the condition, 
  
  {\small \begin{align*}     \underset{1 \leq i < j \leq n}{\sum} \bigg[ \text{ }  \bigg| \bigg|  \big\{  A_i \otimes \textbf{I} \big\} \ket{\psi^{*}_{\mathrm{FFL}}}    -   \bigg[ \textbf{I} \otimes \bigg\{  \frac{B_{ij} + B_{ji}}{\sqrt{2}} \bigg\}   \bigg] \ket{\psi^{*}_{\mathrm{FFL}}}   \bigg| \bigg|^2  + \bigg| \bigg|  \big\{    A_j  \otimes \textbf{I} \big\}  \ket{\psi^{*}_{\mathrm{FFL}}} \\  -     \bigg[ \textbf{I} \otimes \bigg\{  \frac{B_{ij} - B_{ji}}{\sqrt{2}} \bigg\}  \bigg] \ket{\psi^{*}_{\mathrm{FFL}}}    \bigg| \bigg|^2  \text{ }      \bigg]   \leq 2n \big( n - 1 \big) \epsilon         \text{. } 
\end{align*} }

\noindent Underlying the existence of,

\begin{align*}
     \mathrm{quantum \text{ } FFL \text{ } value} \equiv \omega_q \big( \mathrm{FFL}  \big) = 2 / 3  , 
\end{align*}

\noindent collections of strategies for which Alice and Bob can outperform the optimal value using classical strategies,

{\small

\begin{align*}
 \mathrm{Classical \text{ } FFL \text{ } value} \equiv   \omega_c \big(  \mathrm{FFL} \big) . 
\end{align*}

}

\noindent We quantitatively further expand on properties of classical and quantum strategies for the FFL game through assumptions of the existence of an operator $T$ for which,

{\small

\begin{align*}
    \big|\big|  \big\{  A_i \otimes \textbf{I} \big\}  T^{\mathrm{FFL}}     -          T^{\mathrm{FFL}} \big\{  \widetilde{A}_i  \otimes \textbf{I} \big\}     \big|\big|_{\mathrm{F}} /  \big|\big|  \big\{  A_i \otimes \textbf{I} \big\}  T^{\mathrm{FFL}}     -          T^{\mathrm{FFL}} \big\{  \widetilde{A}_i  \otimes \textbf{I} \big\}     \big|\big|_1        ,  \end{align*}
    
    \begin{align*} \big|\big|    \big\{  \textbf{I} \otimes B_{jk} \big\}  T^{\mathrm{FFL}}   - T^{\mathrm{FFL}} \big\{  \textbf{I} \otimes \widetilde{B_{jk}}    \big\}  \big|\big|_{\mathrm{F}} /  \big|\big|    \big\{  \textbf{I} \otimes B_{jk} \big\}  T^{\mathrm{FFL}}   - T^{\mathrm{FFL}} \big\{  \textbf{I} \otimes \widetilde{B_{jk}}    \big\}  \big|\big|_1       , \\ 
\end{align*}

}

\noindent namely whether the Frobenius norm of reversing the order in which Alice or Bob's answers for $i \sim \pi_{\mathrm{FFL}}$ is related to the $l-1$ norm. As a proxy for the above ratio of the Frobenius norm to the $l-1$ norm it is demonstrated that the computation of the norm,

{\small \begin{align*}
  \big| \big|   T^{\mathrm{FFL}} \big| \big|_F  \equiv \bigg| \bigg| \frac{1}{\sqrt{2^{n}}} \bigg[   \underset{(j_1 , \cdots , j_n) \in \{ 0 , 1  \}^n}{\sum}   \bigg\{  
 \text{ }         \bigg(       \underset{1 \leq i \leq n}{ \prod } A^{j_i}_{i}            \bigg)      \otimes \textbf{I} \bigg\}              \ket{\psi_{\mathrm{FFL}}} \bra{\widetilde{\psi_{\mathrm{FFL}}} } \bigg\{  \text{ } \bigg( \underset{1 \leq i \leq n}{\prod}            \widetilde{A}^{j_i}_i\bigg) \otimes \textbf{I}  \bigg\}^{\dagger}   \text{ }             \bigg]   \bigg| \bigg|_F     \text{, } 
\end{align*} } 

\noindent is intimately related to the norms,

{\small

\[   \left\{\!\begin{array}{ll@{}>{{}}l}
    \big|\big|  \big\{  A_i \otimes \textbf{I} \big\}  T^{\mathrm{FFL}}  \big| \big|_F   , \\ \\         \big|\big|      T^{\mathrm{FFL}} \big\{  \widetilde{A}_i  \otimes \textbf{I} \big\}          \big| \big|_F  , \\ \\     \big|\big|  \big\{  \textbf{I} \otimes B_{jk} \big\}  T^{\mathrm{FFL}}    \big| \big|_F   , \\ \\     \big|\big|  T^{\mathrm{FFL}} \big\{  \textbf{I} \otimes \widetilde{B_{jk}}    \big\}    \big| \big|_F   , \end{array}\right. 
\] 
}

\bigskip 

\noindent of the responses provided by Alice, or by Bob, respectively. In comparison to previous manipulations of a $T$ operator defined for the XOR and CHSH settings, {\color{blue}[37]},

{\small \begin{align*}
   T^{\mathrm{XOR}}   \equiv \frac{1}{\sqrt{2^{n}}} \bigg[   \underset{(j_1 , \cdots , j_n) \in \{ 0 , 1  \}^n}{\sum}   \bigg\{  
 \text{ }         \bigg(       \underset{1 \leq i \leq n}{ \prod } A^{j_i}_{i}            \bigg)      \otimes \textbf{I} \bigg\}              \ket{\psi_{\mathrm{XOR}}} \bra{\widetilde{\psi_{\mathrm{XOR}}} } \bigg\{  \text{ } \bigg( \underset{1 \leq i \leq n}{\prod}            \widetilde{A}^{j_i}_i\bigg) \otimes \textbf{I}  \bigg\}^{\dagger}   \text{ }             \bigg]     \text{, } 
\end{align*} } 

\noindent the replacement of the outer product $ \ket{\psi_{\mathrm{XOR}}} \bra{\widetilde{\psi_{\mathrm{XOR}}} } $ with $ \ket{\psi_{\mathrm{FFL}}} \bra{\widetilde{\psi_{\mathrm{FFL}}} } $, in addition to the action of the operator $\widetilde{\cdot} \equiv \widetilde{\cdot} \big|_{\mathrm{FFL}}$,

{\small

\begin{align*} 
  \widetilde{\cdot} :  \big\{   \textbf{C}^d \otimes \textbf{I} \big\} \longrightarrow  \big\{ \textbf{I} \otimes \textbf{C}^d \big\}  :  \big| \big|  \big(  \cdot \otimes \textbf{I} \big) \ket{\psi_{\mathrm{FFL}}} \big| \big|_F   \mapsto        \big| \big|  \big( \textbf{I} \otimes \widetilde{\cdot} \big) \ket{\psi_{\mathrm{FFL}}} \big| \big|_F   , 
\end{align*}

}

\noindent for some $d>0$, as opposed to the operation $\widetilde{\cdot} \equiv \widetilde{\cdot} \big|_{\mathrm{XOR}}$,

{\small

\begin{align*} 
  \widetilde{\cdot} :  \big\{   \textbf{C}^d \otimes \textbf{I} \big\} \longrightarrow  \big\{ \textbf{I} \otimes \textbf{C}^d \big\} :  \big| \big|  \big(  \cdot \otimes \textbf{I} \big) \ket{\psi_{\mathrm{XOR}}} \big| \big|_F   \mapsto        \big| \big|  \big( \textbf{I} \otimes \widetilde{\cdot} \big) \ket{\psi_{\mathrm{XOR}}} \big| \big|_F   , 
\end{align*}

}

\bigskip 

\noindent The existence of the quantum value for the FFL game implies that the dual optimal solution to the linear programming program for determining the $2/3 $ value of the optimal strategy is dependent upon,

{\small

\[   \left\{\!\begin{array}{ll@{}>{{}}l}
  i \sim \pi_{\mathrm{FFL}} , \\ \\ i \neq j \sim \pi_{\mathrm{FFL}}  , \end{array}\right. 
\] 
}

\noindent provided to Alice and Bob, respectively, by the referree and anticommutation of $\pm 1$ observables $A_i$, $A_j$, $B_{ij}$ and $B_{ji}$.

\bigskip

\noindent For the maximally entangled state, the same two identities appearing in \textbf{Lemma} \textit{2}, {\color{blue}[37]}, hold. Also, given collections of linear operators, $A_i$ and $B_i$, over some vector space for which $T A_i \equiv B_i T$, $\forall i \in I$, $T$ is called an \textit{intertwinning} operator that is supported over the union of the images of each linear operator.  Performing anticommutation is possible over the tensor product of the support of two subspaces, from the fact that for each possible $i$, there exists $j \neq k$ for which, under the mapping,

\begin{align*}
  T^{\mathrm{FFL}} :  \big\{ \textbf{C}^{2 \lceil \frac{n}{2} \rceil } \big\}  \otimes    \big\{   \textbf{C}^{2 \lceil \frac{n}{2} \rceil }    \big\}  \longrightarrow      \big\{   \textbf{C}^{d_A} \big\} \otimes \big\{ \textbf{C}^{d_B}  \big\}    \text{, 
 }
\end{align*}

\noindent an optimal, dual, FFL strategy, $\ket{\widetilde{\psi_{\mathrm{FFL}}}}$, satisfies, from the two conditions provided in the statement of \textbf{Theorem} \textit{2}:

\begin{itemize}
    \item[$\bullet$] {\textit{Frobenius norm for Alice's strategy}}. For all $i$,

\begin{align*}
  \big|\big|  \big\{  A_i \otimes \textbf{I} \big\} T^{\mathrm{FFL}}     -          T^{\mathrm{FFL}} \big\{  \widetilde{A}_i  \otimes \textbf{I} \big\}     \big|\big|_{\mathrm{F}} < 9  n^2 \sqrt{\epsilon} \big| \big|     T^{\mathrm{FFL}}   \big|\big|_{\mathrm{F}}  \text{. } 
\end{align*}

    \item[$\bullet$] {\textit{Frobenius norm for Bob's strategy}}. For all $j \neq k$,

\begin{align*}
    \big|\big|    \big\{  \textbf{I} \otimes B_{jk} \big\}  T^{\mathrm{FFL}}   - T^{\mathrm{FFL}} \big\{  \textbf{I} \otimes \widetilde{B_{jk}}    \big\}  \big|\big|_{\mathrm{F}} <   \frac{44}{3} n^2 \sqrt{\epsilon}  \big| \big|     T^{\mathrm{FFL}}   \big|\big|_{\mathrm{F}}   \text{. } 
\end{align*}

\end{itemize}

\noindent In the $\mathrm{XOR}$ setting, {\color{blue}[37]}, with respect to orthonormal bases of $\textbf{C}^{d_A}$ and $\textbf{C}^{d_B}$, the Schmidt decomposition has some blocks with entries that are arbitrarily small iff there exists countably many blocks, with each entry of the block being equal to $\lambda_i$. In the setting of the XOR* and FFL games, while one would still expect to make use of,

{\small

\begin{align*}
     \overset{s 2^{\lfloor \frac{n}{2} \rfloor}}{\underset{i=1}{\sum} } \sqrt{\lambda^{\mathrm{XOR}^{*}}_i}  \big\{   \ket{u_i} \otimes \ket{v_i}  \big\}  , \\ \\    \overset{s 2^{\lfloor \frac{n}{2} \rfloor}}{\underset{i=1}{\sum} } \sqrt{\lambda^{\mathrm{FFL}}_i}  \big\{   \ket{u_i} \otimes \ket{v_i}  \big\}  , 
\end{align*}

}

\noindent which correspond to the Schmidt decompositions for the XOR* and FFL games, respectively, satisfy the identities,

{\small

\begin{align*}
  \bra{\psi^*_{\mathrm{XOR}}}              A_i \otimes B_j      \ket{\psi^*_{\mathrm{XOR}}} =           \underset{1 \leq l \leq s}{\sum} \big\{   2^{\lfloor n / 2 \rfloor}        \lambda^{\mathrm{XOR}^{*}}_{l \times  2^{\lfloor n / 2 \rfloor} }        \bra{\psi^*_{\mathrm{XOR}}}              A_i \otimes B_j      \ket{\psi^*_{\mathrm{XOR}}} \big\} =    1 / {\sqrt{2}}          , \\ \\      \bra{\psi_{\mathrm{FFL}}}              A_i \otimes B_j      \ket{\psi_{\mathrm{FFL}}}   =             \underset{1 \leq l \leq s}{\sum} \big\{   2^{\lfloor n / 2 \rfloor}        \lambda^{\mathrm{FFL}}_{l \times  2^{\lfloor n / 2 \rfloor} }       \bra{\psi_{\mathrm{FFL}}}              A_i \otimes B_j      \ket{\psi_{\mathrm{FFL}}}    \big\} =     2 / {3}           , 
\end{align*}

}

\noindent as opposed to, {\color{blue}[37]},

{\small

\begin{align*}
    \bra{\psi_{\mathrm{XOR}}}              A_i \otimes B_j      \ket{\psi_{\mathrm{XOR}}} =           \underset{1 \leq l \leq s}{\sum} \big\{   2^{\lfloor n / 2 \rfloor}        \lambda^{\mathrm{XOR}}_{l \times  2^{\lfloor n / 2 \rfloor} }        \bra{\psi_{\mathrm{XOR}}}              A_i \otimes B_j      \ket{\psi_{\mathrm{XOR}}} \big\} =      1 / {\sqrt{2}}   , 
\end{align*}

}

\noindent for the XOR setting. To quantify how the condition, 

{\small

\begin{align*}
1 / \sqrt{2} \propto \overset{s 2^{\lfloor \frac{n}{2} \rfloor}}{\underset{i=1}{\sum} } \sqrt{\lambda^{\mathrm{XOR}^{*}}_i}  \big\{   \ket{u_i} \otimes \ket{v_i}  \big\}  >     \overset{s 2^{\lfloor \frac{n}{2} \rfloor}}{\underset{i=1}{\sum} } \sqrt{\lambda^{\mathrm{FFL}}_i}  \big\{   \ket{u_i} \otimes \ket{v_i}  \big\} \propto 2 / 3   ,
\end{align*}
}

\noindent is related to the upper bound,

{\small

\begin{align*}
   \bigg(       \frac{8200 \sqrt{2} }{27}   \bigg) n^2 \sqrt{\epsilon_{\mathrm{FFL}}}  , 
\end{align*}

}

\noindent for a parameter $\epsilon_{\mathrm{FFL}}$ taken sufficiently small and also to the system of upper bounds,

{\small

\[   \left\{\!\begin{array}{ll@{}>{{}}l}
 9  n^2 \sqrt{\epsilon_{\mathrm{FFL}}} , \\ \\           \frac{44}{3}  n^2 \sqrt{\epsilon_{\mathrm{FFL}}}   , \end{array}\right. 
\] 
}

\noindent we recall the following properties of the Schdmit decomposition associated with the XOR game. As a result of the duality notion between $\mathrm{XOR}$ and $\mathrm{XOR^{*}}$ games, {\color{blue}[5]}, which is predominantly reliant upon associations between nonlocality and preparation contextuality, optimal, and approximately optimal, quantum strategies for Alice and Bob can be characterized through the following differences between the XOR Schdmit decomposition with the XOR* and FFL Schdmit decompositions.

\bigskip

\noindent Before describing several properties of the XOR* and FFL Schmidt decompositions, we describe how optimality, and approximate optimality, of quantum strategies can be studied with the following steps in the XOR setting, {\color{blue}[37]}:

\begin{itemize}
    \item [$\bullet$] (1), \textit{Schmidt block partition for even n}: For $n \equiv 2k$, over $\textbf{C}^{2^k} \otimes \textbf{C}^{2^k}$,

\begin{align*}
    \overset{ 2^{\lfloor k \rfloor}}{\underset{i=1}{\sum} } \frac{1}{\sqrt{2^i}} \big\{  \ket{i} \otimes \ket{i} \big\} \text{. } 
\end{align*}

\item[$\bullet$] (2), \textit{Schmidt block partition for odd $n$}: For $n \equiv 2k+1$, over $\textbf{C}^{2^{k+1}} \otimes \textbf{C}^{2^{k+1}}$,

\begin{align*}
    \overset{ 2^{\lfloor k + 1 \rfloor}}{\underset{i=1}{\sum} } \frac{1}{\sqrt{2^{i+1}}} \big\{  \ket{i} \otimes \ket{i} \big\}    \text{. } 
\end{align*}

\end{itemize}

\noindent Regardless of whether $n$ is even or odd, the two decompositions are a special case of,

\begin{align*}
    \overset{s 2^{\lfloor \frac{n}{2} \rfloor}}{\underset{i=1}{\sum} } \sqrt{\lambda_i} \big\{  \ket{u_i} \otimes \ket{v_i}  \big\}  \text{, } 
\end{align*}

\noindent for $s \equiv 1$, $\lambda_i \equiv  \frac{1}{2^i}$, and $u_i \equiv v_i \equiv i$, $\forall i$.

\bigskip

\noindent In comparison to the $\mathrm{XOR}$ and $\mathrm{XOR^{*}}$ games for which countably many elements along the diagnonal of the \textit{Schmidt decomposition} can be arbritrarily small, for other classes of games, including the XOR* and FFL games, to describe properties of the set of linear combinations,

{\small

\begin{align*}
  \underset{1 \leq i \leq d_A}{\underset{\textbf{C}^{d_A}}{\mathrm{span}}} \big\{     \ket{u^{\mathrm{XOR}^*}_i }  \big\}   , \\ \\   \underset{1 \leq j \leq d_B}{\underset{\textbf{C}^{d_B}}{\mathrm{span}}} \big\{   \ket{v^{\mathrm{XOR}^*}_j }  \big\}   , \\ \\  \underset{1 \leq j \leq d_B}{\underset{\textbf{C}^{d_B}}{\mathrm{span}}} \big\{   \ket{u^{\mathrm{FFL}}_i }  \big\}  , \\ \\ \underset{1 \leq j \leq d_B}{\underset{\textbf{C}^{d_B}}{\mathrm{span}}} \big\{   \ket{v^{\mathrm{FFL}}_j }  \big\}  , 
\end{align*}

}

\noindent subject to the orthonormality conditions,

{\small

\[   \left\{\!\begin{array}{ll@{}>{{}}l}
   \big| \big| \ket{u^{\mathrm{XOR}^*}_j }    \big| \big|_1 =   \big| \big|  \ket{v^{\mathrm{XOR}^*}_j } \big|   \big|_1  = 1      , \\ \\      \big| \big|       \ket{v^{\mathrm{FFL}}_i }      \big| \big|_1 =   \big| \big|  \ket{v^{\mathrm{FFL}}_j  }          \big| \big|_1  = 1         , \end{array}\right. 
\] 
}

\noindent for every $i$ and $j$ such that $1 \leq i \leq d_A$ and $1 \leq j \leq d_B$, respectively. Along the lines through which previous computations in {\color{blue}[37]} were performed with the state,

{\small \begin{align*}
 \ket{\psi^{\mathrm{XOR}}_L} \equiv  \ket{\psi_L} = \frac{1}{\sqrt{2^{\lfloor \frac{n}{2} \rfloor}}} \underset{(l-1) 2^{\rfloor \frac{n}{2} \lfloor } \leq i \leq l 2^{\rfloor \frac{n}{2} \lfloor }}{\sum} \big\{  \ket{u_i} \otimes \ket{u_i} \big\}    \text{, } 
\end{align*}}

\noindent for,

{\small \begin{align*}
 \ket{u_i } \equiv  \mathrm{span} \big\{ \mathrm{Im} \big(          \Psi_i                  \big)   \big\}   \text{, } 
\end{align*} } 

\noindent given,

{\small \begin{align*}
 \Psi_i : \textbf{C}^{d_i } \longrightarrow \textbf{C}^{d_i}   \text{, } 
\end{align*} }

\noindent for each $i$ such that $1 \leq i \leq d_A$ for the dual XOR setting, one introduces,

{\small \begin{align*}
  \ket{\psi^{\mathrm{XOR}^{*}}_L} = \frac{1}{\sqrt{2^{\lfloor \frac{n}{2} \rfloor}}} \underset{(l-1) 2^{\rfloor \frac{n}{2} \lfloor } \leq i \leq l 2^{\rfloor \frac{n}{2} \lfloor }}{\sum} \big\{  \ket{u^{\mathrm{XOR}^*}_i} \otimes \ket{u^{\mathrm{XOR}^*}_i} \big\}    \text{, } 
\end{align*}}

\noindent in addition for the FFL setting, one introduces,

{\small \begin{align*}
  \ket{\psi^{\mathrm{FFL}}_L}  = \frac{1}{\sqrt{2^{\lfloor \frac{n}{2} \rfloor}}} \underset{(l-1) 2^{\rfloor \frac{n}{2} \lfloor } \leq i \leq l 2^{\rfloor \frac{n}{2} \lfloor }}{\sum} \big\{ \ket{u^{\mathrm{FFL}}_i} \otimes \ket{u^{\mathrm{FFL}}_i} \big\}    \text{. } 
\end{align*}}

\noindent To describe how the above expressions for the states $\ket{\psi^{\mathrm{XOR}^{*}}_L}$ and $\ket{\psi^{\mathrm{FFL}}_L}$ are related to the finite-dimensional representations,

{\small

\[   \left\{\!\begin{array}{ll@{}>{{}}l}
      \begin{bmatrix}
 \sqrt{\lambda_{2^{\lfloor \frac{n}{2} \rfloor }}}              \textbf{I}   & 0 & \cdots & 0  \\ & \sqrt{\lambda_{2 \times 2^{\lfloor \frac{n}{2} \rfloor }}} \textbf{I} & \cdots  & 0   \\         & \ddots            \\ & & \sqrt{\lambda_{s 2^{\lfloor \frac{n}{2} \rfloor }}} \textbf{I} & 0 
\end{bmatrix}   , \\ \\     \begin{bmatrix}
 \sqrt{\lambda^{\mathrm{XOR}^*}_{2^{\lfloor \frac{n}{2} \rfloor }}}              \textbf{I}   & 0 & \cdots & 0  \\ & \sqrt{\lambda^{\mathrm{XOR}^*}_{2 \times 2^{\lfloor \frac{n}{2} \rfloor }}} \textbf{I} & \cdots  & 0   \\         & \ddots            \\ & & \sqrt{\lambda^{\mathrm{XOR}^*}_{s 2^{\lfloor \frac{n}{2} \rfloor }}} \textbf{I} & 0 
\end{bmatrix}   \equiv \begin{bmatrix}
 \sqrt{\lambda^*_{2^{\lfloor \frac{n}{2} \rfloor }}}              \textbf{I}   & 0 & \cdots & 0  \\ & \sqrt{\lambda^*_{2 \times 2^{\lfloor \frac{n}{2} \rfloor }}} \textbf{I} & \cdots  & 0   \\         & \ddots            \\ & & \sqrt{\lambda^*_{s 2^{\lfloor \frac{n}{2} \rfloor }}} \textbf{I} & 0 
\end{bmatrix}       , \\ \\     \begin{bmatrix}
 \sqrt{\lambda^{\mathrm{FFL}}_{2^{\lfloor \frac{n}{2} \rfloor }}}              \textbf{I}   & 0 & \cdots & 0  \\ & \sqrt{\lambda^{\mathrm{FFL}}_{2 \times 2^{\lfloor \frac{n}{2} \rfloor }}} \textbf{I} & \cdots  & 0   \\         & \ddots            \\ & & \sqrt{\lambda^{\mathrm{FFL}}_{s 2^{\lfloor \frac{n}{2} \rfloor }}} \textbf{I} & 0 
\end{bmatrix}         , \end{array}\right. 
\] 
}

\noindent corresponding to the XOR, XOR* and FFL games, respectively, for the Schmidt coefficients, observe, albeit the fact that the XOR setting the basis for each $\Psi_i$ can be represented as,

{\small \[
\begin{bmatrix}
 \sqrt{\lambda_{2^{\lfloor \frac{n}{2} \rfloor }}}              \textbf{I}   & 0 & \cdots & 0  \\ & \sqrt{\lambda_{2 \times 2^{\lfloor \frac{n}{2} \rfloor }}} \textbf{I} & \cdots  & 0   \\         & \ddots            \\ & & \sqrt{\lambda_{s 2^{\lfloor \frac{n}{2} \rfloor }}} \textbf{I} & 0 
\end{bmatrix} \text{, }
\] } 

\noindent for the XOR* setting the representation for the basis takes the form,

{\small \[
\begin{bmatrix}
 \sqrt{\lambda^*_{2^{\lfloor \frac{n}{2} \rfloor }}}              \textbf{I}   & 0 & \cdots & 0  \\ & \sqrt{\lambda^*_{2 \times 2^{\lfloor \frac{n}{2} \rfloor }}} \textbf{I} & \cdots  & 0   \\         & \ddots            \\ & & \sqrt{\lambda^*_{s 2^{\lfloor \frac{n}{2} \rfloor }}} \textbf{I} & 0 
\end{bmatrix} \text{, }
\] }

\noindent where,

{\small

\begin{align*}
   \Psi_i \overset{(\textit{Duality notion of } {\color{blue}[5]})}{\Longleftrightarrow} \Psi^*_i  . 
\end{align*}

}

\noindent Along similar lines, given a finite-dimensional representation for the Schmidt blocks of the FFL game,

{\small \[
\begin{bmatrix}
 \sqrt{\lambda^{\mathrm{FFL}}_{2^{\lfloor \frac{n}{2} \rfloor }}}              \textbf{I}   & 0 & \cdots & 0  \\ & \sqrt{\lambda^{\mathrm{FFL}}_{2 \times 2^{\lfloor \frac{n}{2} \rfloor }}} \textbf{I} & \cdots  & 0   \\         & \ddots            \\ & & \sqrt{\lambda^{\mathrm{FFL}}_{s 2^{\lfloor \frac{n}{2} \rfloor }}} \textbf{I} & 0 
\end{bmatrix} \text{, }
\] }

\noindent one performs computations with operators of the form,

{\small \begin{align*}
   T^{\mathrm{XOR^*}} \equiv \frac{1}{\sqrt{2^{n}}} \bigg[   \underset{(j_1 , \cdots , j_n) \in \{ 0 , 1  \}^n}{\sum}   \bigg\{  
      \bigg(       \underset{1 \leq i \leq n}{ \prod } A^{j_i}_{i}            \bigg)      \otimes \textbf{I} \bigg\}             \ket{\psi_{\mathrm{XOR^*}}} \bra{\widetilde{\psi_{\mathrm{XOR^*}}} } \bigg\{ \text{ } \bigg( \underset{1 \leq i \leq n}{\prod}            \widetilde{A}^{j_i}_i\bigg) \otimes \textbf{I}  \bigg\}^{\dagger}   \text{ }             \bigg]   \\ \\   T^{\mathrm{FFL}} \equiv \frac{1}{\sqrt{2^{n}}} \bigg[   \underset{(j_1 , \cdots , j_n) \in \{ 0 , 1  \}^n}{\sum}   \bigg\{  
     \bigg(       \underset{1 \leq i \leq n}{ \prod } A^{j_i}_{i}            \bigg)      \otimes \textbf{I} \bigg\}             \ket{\psi_{\mathrm{FFL}}} \bra{\widetilde{\psi_{\mathrm{FFL}}} } \bigg\{ \text{ } \bigg( \underset{1 \leq i \leq n}{\prod}            \widetilde{A}^{j_i}_i\bigg) \otimes \textbf{I}  \bigg\}^{\dagger}   \text{ }             \bigg]       \text{, } 
\end{align*} }

\noindent that will be further described in the next section.

\bigskip

\noindent Essential to the computations with the operator $T^{\mathrm{FFL}}$ shown above are

computations associated with the optimal strategy state for $\mathrm{XOR}$ games, for the numerical value of the supremum of optimal $\mathrm{FFL}$ strategies,

{\small \begin{align*}
        \bra{\psi_L}  \big\{  A^{(l)}_i \otimes B^{(l)}_{ij}               \big\}     \ket{\psi_L } \equiv \bra{\psi_L} \big\{     A^{(l)}_i          \otimes B^{(l)}_{ji} \big\}   \ket{\psi_L }  \equiv \bra{\psi_L} \big\{  A^{(l)}_j \otimes B^{(l)}_{ij} \big\}   \ket{\psi_L }    \equiv 2 / {3}        \text{, } 
\end{align*} }

\noindent and,

{\small \begin{align*}
    \bra{\psi_L} \big\{  A^{(l)}_j \otimes B^{(l)}_{ji} \big\}   \ket{\psi_L }  \equiv - 2 / {3}  \text{, } 
\end{align*} }

\noindent where each one of the expectation values is taken from states prepared for the optimal strategy. If we do not decompose each $A$ and $B$ according to the block representation, instead,

{\small \begin{align*}
   \bra{\psi}  \big\{   A_i \otimes B_{ij}               \big\}     \ket{\psi } \equiv \bra{\psi} \big\{    A_i          \otimes B_{ji} \big\}   \ket{\psi }  \equiv \bra{\psi} \big\{  A_j \otimes B_{ij} \big\}   \ket{\psi }    \equiv 2 / {3}       \text{, } 
\end{align*} }

\noindent and,

{\small \begin{align*}
   \bra{\psi} \big\{  A_j \otimes B_{ji} \big\}   \ket{\psi }  \equiv - 2 / {3}  \text{. } 
\end{align*} }

\noindent With $\ket{\psi_L}$, one also has another identity with the $2 / {3}$ supremum, in which,

{\small \begin{align*}
    \bra{\psi} \big( A_i \otimes B_{ij} \big) \ket{\psi} = \underset{1 \leq i \leq s}{\sum}       \big\{    2^{\lfloor  \frac{n}{2}\rfloor}  \lambda_{i 2^{\rfloor \frac{n}{2}\rfloor } }   \big\}     \bra{\psi_L} \big\{    A^{(l)}_i   \otimes B^{(l)}_{ij}           \big\}  \ket{\psi_L}              \equiv 2 / {3}      \text{, } 
\end{align*} }

\noindent for,

{\small \begin{align*}
 \ket{\psi} \equiv   \underset{1 \leq i \leq s}{ \sum}    \sqrt{2^{\lfloor \frac{n}{2} \rfloor} \lambda_{i 2^{\lfloor \frac{n}{2} \rfloor}}} \ket{\psi_L}     =  \underset{1 \leq i \leq s}{ \sum}    \sqrt{2^{\lfloor \frac{n}{2} \rfloor} \lambda_{i 2^{\lfloor \frac{n}{2} \rfloor}}}   \bigg\{         \frac{1}{\sqrt{2^{\lfloor \frac{n}{2} \rfloor}}} \underset{(i-1) 2^{\rfloor \frac{n}{2} \lfloor } \leq i^{\prime} \leq i 2^{\rfloor \frac{n}{2} \lfloor }}{\sum} \big\{  \ket{u_{i^{\prime}} } \otimes \ket{u_{i^{\prime}}} \big\}       \bigg\}      \text{. } 
\end{align*} }

\noindent To demonstrate how the optimal values for the $\mathrm{XOR^{*}}$ FFL games are related to the above $T$ operators we perform manipulations of the form described in the next section.

\subsubsection{FFL transformation}

\noindent Ultimately, the desired non-zero linear operator that must be constructed is normalized in the power set of $n$, has the form,

{\small \begin{align*}
    T^{\mathrm{FFL}} \equiv \frac{1}{\sqrt{2^{n}}} \bigg[   \underset{(j_1 , \cdots , j_n) \in \{ 0 , 1  \}^n}{\sum}   \bigg\{  
        \bigg(       \underset{1 \leq i \leq n}{ \prod } A^{j_i}_{i}            \bigg)      \otimes \textbf{I} \bigg\}             \ket{\psi_{\mathrm{FFL}}} \bra{\widetilde{\psi_{\mathrm{FFL}}} } \bigg\{  \bigg( \underset{1 \leq i \leq n}{\prod}            \widetilde{A}^{j_i}_i\bigg) \otimes \textbf{I}  \bigg\}^{\dagger}              \bigg]       \text{. } 
\end{align*} }

\noindent In a similar way to how the anchored game can be considered for removing correlations in quantum states that are entangled, products of the form,

{\small

\begin{align*}
    T^{\mathrm{FFL}}  \cdot \big\{ \textbf{I} \otimes B_{kl} \big\}  \equiv  \frac{1}{\sqrt{2^{n}}} \bigg[   \underset{(j_1 , \cdots , j_n) \in \{ 0 , 1  \}^n}{\sum}   \bigg\{  
        \bigg(       \underset{1 \leq i \leq l-1 }{ \prod } A^{j_i}_{i}   \times   A^{j_{l\oplus 1}}_l     \cdot      \underset{l+1  \leq i \leq n}{ \prod } A^{j_i}_{i}          \bigg)      \otimes \textbf{I} \bigg\}             \ket{\psi_{\mathrm{FFL}}} \bra{\widetilde{\psi_{\mathrm{FFL}}} } \\ \times \bigg\{  \bigg( \underset{1 \leq i \leq k-1 }{\prod}            \widetilde{A}^{j_i}_i  \times   \widetilde{A}^{j_{k \oplus 1}}_k    \cdot \underset{k+1  \leq i \leq n}{\prod}            \widetilde{A}^{j_i}_i       \bigg) \otimes \textbf{I}  \bigg\}^{\dagger}          \bigg]   \\     , 
\end{align*}

}

\noindent for $k \neq l$ between the intertwiner and the above tensor product are used for establishing that $ \big( A_i \otimes \textbf{I} \big) T - T \big( \widetilde{A}_i  \otimes \textbf{I} \big)$ and $\big( \textbf{I} \otimes B_{kl} \big) T - T \big( \textbf{I} \otimes \widetilde{B}_{kl} \big)$ hold, as stated in \textbf{Lemma} \textit{3}. 

\bigskip 

\noindent The corresponding dual state to the optimal strategy, $\ket{\widetilde{\Psi_{\mathrm{FFL}}}}$, which has the spanning set,

{\small \begin{align*}
 \mathrm{SPAN} \equiv  \mathrm{span} \bigg\{   \bigg\{   \bigg(  \underset{1 \leq i \leq n}{\prod} A^{j_i}_i  \bigg)  \otimes \textbf{I} \bigg\}  \ket{\widetilde{\Psi_{\mathrm{FFL}}}}           : \big( j_1 , \cdots , j_n \big) \in \big\{ 0 , 1 \big\}^n       \bigg\}  \text{, } 
\end{align*} }

\noindent yields the desired upper bounds, from the Frobenius norms of Alice's and Bob's strategies. The norm with respect to $T$ is captured with the following.

\bigskip

\noindent \textbf{Lemma} \textit{1} (\textit{the computation of the Frobenius norm for the anticommutation rule of $T$ yields the desired $\sqrt{\epsilon}$ approximate upper bound}). One has that,

{\small \begin{align*}
  \bigg| \bigg|  \text{ } \bigg\{  \bigg( \underset{1 \leq i \leq n}{\prod} A^{j_i}_i     \bigg)    \otimes  \textbf{I}   \bigg\}        \ket{\psi_{\mathrm{FFL}}}    -   \bigg[ \mathrm{sign} \big( i , j_1 , \cdots , j_n \big) \text{ }\bigg\{  \bigg(   \underset{i \equiv j_1 + 1, \text{ } \mathrm{set}\text{ }  j_1 + 1 \equiv j_1 \oplus 1}{\underset{1 \leq i \leq n}{\prod}}        A^{j_i}_i         \bigg)   \\  \otimes \textbf{I} \bigg\}   \bigg] \ket{\psi_{\mathrm{FFL}}}              \bigg| \bigg|_{\mathrm{F}}          \text{, } 
\end{align*} }

\noindent and,

{\small \begin{align*}
      \bigg| \bigg|      \bigg\{    \bigg( \underset{1 \leq  i \leq n }{\prod} A^{j_i}_i \bigg) \otimes B_{kl} \bigg\}   \ket{\psi_{\mathrm{FFL}}}      -   \frac{2}{3} \bigg[ \pm \mathrm{sign}  \big( i , j_1 , \cdots , j_n \big) \bigg\{   \bigg(     \underset{ i \equiv j_4+1 ,\text{ } \mathrm{set} \text{ } j_4+1 \equiv j_4 \oplus 1 }{\underset{1 \leq i \leq n}{\prod}}       A^{j_i}_i    \bigg)    
      \\   \otimes \textbf{I}    \bigg\} \bigg] \ket{\psi_{\mathrm{FFL}}}   \bigg| \bigg|_{\mathrm{F}}         \text{, } 
\end{align*} }

\noindent have upper bounds, $9 n^2 \sqrt{\epsilon}$ and $\frac{44}{3} n^2 \sqrt{\epsilon}$, respectively.

\bigskip

\noindent To demonstrate that the above inequalities hold with respect to the Frobenius norm, in the following subsections we relate properties of the Frobenius norm to properties of the operator norm.

\subsubsection{FFL identites from the transformation}

\noindent Equipped with the nonlinear transformation $T$, to demonstrate that the Frobenius norms of the tensor product between $T$ and $A_i \otimes \textbf{I}$, or with $\widetilde{A}_i  \otimes \textbf{I}$, are upper bounded with an appropriately chosen multiplicative factor of Frobenius norm of $T$, optimal, and approximately optimal, FFL strategies we introduce the following properties.

\bigskip

\noindent \textbf{Lemma} \textit{2} (\textit{orthogonality of the canoncial FFL strategy}). The vectors contained in the set of linear combinatiuons SPAN are orthonormal.

\bigskip

\noindent \textit{Proof of Lemma 2}. It suffices to consider the computation of products of the form,

{\small

\begin{align*}
  A_i \cdot  A^{j_1}_1 \times \cdots \times A^{j_n}_n  , \\ \\   \widetilde{A}_i \cdot  A^{j_1}_1 \times \cdots \times A^{j_n}_n  , \\ \\   A_i \cdot \widetilde{A}^{j_1}_1 \times \cdots \times \widetilde{A}^{j_n}_n  , \\ \\  \widetilde{A}_i \cdot \widetilde{A}^{j_1}_1 \times \cdots \times \widetilde{A}^{j_n}_n  , 
\end{align*}

}

\noindent by making use of product expansions for the FFL game, \textbf{Lemma} \textit{2.5.4} and \textbf{Lemma} \textit{2.5.5},

{\small \begin{align*}
 \underset{1 \leq i \leq n}{\prod} \widetilde{A}_i    \text{, } 
\end{align*} }

\noindent in terms of a signed block identity matrix,

{\small \[
\big( - 1 \big)^n \begin{bmatrix}
\textbf{I} & 0 \\ 0 & - \textbf{I}
\end{bmatrix}  \text{, } 
\] }

 \noindent with,

{\small \begin{align*}
 \ket{\widetilde{\psi_{\mathrm{FFL}}}} = \frac{1}{\sqrt{2 \times  2^{\lfloor \frac{n}{2} \rfloor }}} \bigg[   \bigg\{       \underset{1 \leq j \leq 2^{\lfloor \frac{n}{2}\rfloor}}{\sum}   + \underset{2^{\lfloor \frac{n}{2} \rfloor} + 1 \leq j \leq 2 \times 2^{\lfloor \frac{n}{2}\rfloor}}{\sum} \bigg\}  \bigg( \ket{j} \otimes \ket{j}  \bigg)   \bigg]    \text{. } 
\end{align*} }

\noindent As stated in \textbf{Lemma} \textit{8} the above decomposition for $ \ket{\widetilde{\psi_{\mathrm{FFL}}}}$ satisfies,

{\small \begin{align*}
 \bigg\{  \bigg(  \underset{1 \leq i \leq n}{\prod} \widetilde{A}_i   \bigg) \otimes \textbf{I} \bigg\}  \ket{\widetilde{\psi_{\mathrm{FFL}}}} = \frac{1}{\sqrt{2 \times  2^{\lfloor \frac{n}{2} \rfloor }}}  \bigg[    \text{ } \bigg(      \underset{1 \leq j \leq 2^{\lfloor \frac{n}{2}\rfloor}}{\sum}   + \underset{2^{\lfloor \frac{n}{2} \rfloor} + 1 \leq j \leq 2 \times 2^{\lfloor \frac{n}{2}\rfloor}}{\sum} \bigg) \bigg\{  \bigg(       \bigg(  \underset{1 \leq i \leq n}{\prod} \widetilde{A}_i   \bigg) \ket{j}      \otimes \ket{j}      \bigg) \\ \otimes \bigg( \bigg(  \underset{1 \leq i \leq n}{\prod} \widetilde{A}_i   \bigg) \ket{j}    \otimes \ket{j} \bigg)   \bigg\}  \text{ }      \bigg]            \text{. } 
\end{align*} }

\noindent To make use of the product expansion for the FFL games for odd $n$ first through the above identity, from,

{\small

\begin{align*}
    T \cdot \big\{ \textbf{I} \otimes B_{kl} \big\}  \equiv  \frac{1}{\sqrt{2^{n}}} \bigg[   \underset{(j_1 , \cdots , j_n) \in \{ 0 , 1  \}^n}{\sum}   \bigg\{  
        \bigg(       \underset{1 \leq i \leq l-1 }{ \prod } A^{j_i}_{i}   \times   A^{j_{l\oplus 1}}_l     \cdot      \underset{l+1  \leq i \leq n}{ \prod } A^{j_i}_{i}          \bigg)      \otimes \textbf{I} \bigg\}             \ket{\psi_{\mathrm{FFL}}} \bra{\widetilde{\psi_{\mathrm{FFL}}} } \\ \times \bigg\{  \bigg( \underset{1 \leq i \leq k-1 }{\prod}            \widetilde{A}^{j_i}_i  \times   \widetilde{A}^{j_{k \oplus 1}}_k    \cdot \underset{k+1  \leq i \leq n}{\prod}            \widetilde{A}^{j_i}_i       \bigg) \otimes \textbf{I}  \bigg\}^{\dagger}        \bigg]        ,   \\   
\end{align*}

}

\noindent introduce $\big( k_1, \cdots, k_n\big) \neq \big( l_1, \cdots, l_n \big) \in \big\{ 0 , 1 \big\}^n$. From a choice of each $k$ and $l$ observe that a computation of the form,

{\small

\begin{align*}
     \bra{\widetilde{\psi_{\mathrm{FFL}}}}                \big\{    \widetilde{A}^{k_1} \times \cdots \times \widetilde{A}^{k_n}_n                 \big\}^{\dagger} \big\{   \widetilde{A}^{l_1} \times \cdots \times \widetilde{A}^{l_n}_n      \big\}                        \ket{\widetilde{\psi_{\mathrm{FFL}}} }              =      \big\{  \big\{   \widetilde{A}^{k_1} \times \cdots \times \widetilde{A}^{k_n}_n               \big\}    \ket {\widetilde{\psi_{\mathrm{FFL}}}}                \big\}^{\dagger}      \big\{   \widetilde{A}^{l_1} \\ \times \cdots \times \widetilde{A}^{l_n}_n      \big\}                        \ket{\widetilde{\psi_{\mathrm{FFL}}} }   \end{align*}

     \begin{align*} =   \ket{\widetilde{\psi_{\mathrm{FFL}}}} "\cdot" \bigg[ \bigg\{   \big\{               \widetilde{A}^{j_1}_1 \times \cdots \times \widetilde{A}^{j_n}_n       \big\}      \otimes \textbf{I}  \bigg\} \ket{\widetilde{\psi_{\mathrm{FFL}}}}  \bigg]  \end{align*}

     \begin{align*}    = 0                                        , 
\end{align*}

}

\noindent would imply that SPAN is orthonormal iff, 

{\small

\begin{align*}
      \ket{\widetilde{\psi_{\mathrm{FFL}}}} = \frac{1}{\sqrt{2 \times  2^{\lfloor \frac{n}{2} \rfloor }}} \bigg[   \bigg\{       \underset{1 \leq j \leq 2^{\lfloor \frac{n}{2}\rfloor}}{\sum}   + \underset{2^{\lfloor \frac{n}{2} \rfloor} + 1 \leq j \leq 2 \times 2^{\lfloor \frac{n}{2}\rfloor}}{\sum} \bigg\}  \bigg( \ket{j} \otimes \ket{j}  \bigg)   \bigg]        , \end{align*}

\begin{align*}
    \Updownarrow 
\end{align*}

      \begin{align*}
      \bigg[ \bigg\{  \underset{1 \leq i \leq n}{\prod} \widetilde{A}_i     \bigg\} \otimes \textbf{I} \bigg] \ket{\widetilde{\psi_{\mathrm{FFL}}}}    =   \bigg[ \bigg\{  \underset{1 \leq i \leq n}{\prod} \widetilde{A}_i     \bigg\} \otimes \textbf{I} \bigg]            \times   \bigg[  \frac{1}{\sqrt{2 \times  2^{\lfloor \frac{n}{2} \rfloor }}}  \bigg\{       \underset{1 \leq j \leq 2^{\lfloor \frac{n}{2}\rfloor}}{\sum}   + \underset{2^{\lfloor \frac{n}{2} \rfloor} + 1 \leq j \leq 2 \times 2^{\lfloor \frac{n}{2}\rfloor}}{\sum} \bigg\}  \bigg( \ket{j} \otimes \ket{j}  \bigg)   \bigg]    \\ \\   =     \frac{1}{\sqrt{2 \times  2^{\lfloor \frac{n}{2} \rfloor }}}         \bigg\{       \underset{1 \leq j \leq 2^{\lfloor \frac{n}{2}\rfloor}}{\sum}   + \underset{2^{\lfloor \frac{n}{2} \rfloor} + 1 \leq j \leq 2 \times 2^{\lfloor \frac{n}{2}\rfloor}}{\sum} \bigg\}          \bigg(  \bigg[ \bigg\{  \underset{1 \leq i \leq n}{\prod} \widetilde{A}_i     \bigg\} \otimes \textbf{I} \bigg]   \ket{j} \otimes  \bigg[ \bigg\{  \underset{1 \leq i \leq n}{\prod} \widetilde{A}_i     \bigg\} \otimes \textbf{I} \bigg]   \ket{j}            \bigg)            \end{align*}

      \begin{align*} =  0          ,
      \end{align*}

}

\noindent for every $\big( j_1, \cdots , j_n \big)$ and $n$ odd. To demonstrate that the identical conclusion holds for $n$ even observe that the state,

{\small

\begin{align*}
 \bra{\widetilde{\psi_{\mathrm{FFL}}}}         \bigg\{ \big\{   \widetilde{A}^{k_1}_1 \times \cdots \times \widetilde{A}^{k_n}_n     \big\} \otimes \textbf{I}          \bigg\}^{\dagger}  \bigg\{   \big\{      \widetilde{A}^{l_1}_1 \times \cdots \times \widetilde{A}^{l_n}_n     \big\}    \otimes \textbf{I}    \bigg\}                              \ket{\widetilde{\psi_{\mathrm{FFL}}} }          ,  
\end{align*}

}

\noindent manipulated in the previous case for odd $n$, instead for even $n$, can be expressed from the relation,

{\small

\begin{align*}
   \widetilde{A}_i \bigg\{ \underset{1 \leq i \leq n}{\prod} \widetilde{A}^{j_i}_i \bigg\} \widetilde{A}_i = -  \bigg\{ \underset{1 \leq i \leq n}{\prod} \widetilde{A}^{j_i}_i \bigg\}          \Longleftrightarrow \textit{n even}          . 
\end{align*}

}

\noindent implies that the orthogonality condition is satisfied for even $n$ as,

  {\small  \begin{align*}
      \bigg[ \bigg\{  \underset{1 \leq i \leq n}{\prod} \widetilde{A}_i     \bigg\} \otimes \textbf{I} \bigg] \ket{\widetilde{\psi_{\mathrm{FFL}}}}    =   \bigg[ \bigg\{  \underset{1 \leq i \leq n}{\prod} \widetilde{A}_i     \bigg\} \otimes \textbf{I} \bigg]            \times   \bigg[  \frac{1}{\sqrt{2 \times  2^{\lfloor \frac{n}{2} \rfloor }}}  \bigg\{       \underset{1 \leq j \leq 2^{\lfloor \frac{n}{2}\rfloor}}{\sum}   + \underset{2^{\lfloor \frac{n}{2} \rfloor} + 1 \leq j \leq 2 \times 2^{\lfloor \frac{n}{2}\rfloor}}{\sum} \bigg\}  \bigg( \ket{j} \otimes \ket{j}  \bigg)   \bigg]    \\ \\   =     \frac{1}{\sqrt{2 \times  2^{\lfloor \frac{n}{2} \rfloor }}}         \bigg\{       \underset{1 \leq j \leq 2^{\lfloor \frac{n}{2}\rfloor}}{\sum}   + \underset{2^{\lfloor \frac{n}{2} \rfloor} + 1 \leq j \leq 2 \times 2^{\lfloor \frac{n}{2}\rfloor}}{\sum} \bigg\}          \bigg(  \bigg[ \bigg\{  \underset{1 \leq i \leq n}{\prod} \widetilde{A}_i     \bigg\} \otimes \textbf{I} \bigg]   \ket{j} \otimes  \bigg[ \bigg\{  \underset{1 \leq i \leq n}{\prod} \widetilde{A}_i     \bigg\} \otimes \textbf{I} \bigg]   \ket{j}            \bigg)            \end{align*}

      \begin{align*} =  0          ,
      \end{align*}
      }

      \noindent from which we conclude the argument. \boxed{}

\bigskip 

\noindent \textbf{Lemma} \textit{3} (\textit{two identities from the FFL transformation}). Denote the FFL transformation from the previous subsection with $T^{\mathrm{FFL}} \equiv T$. The first identity, from,

{\small \begin{align*}
 \big( A_i \otimes \textbf{I} \big) T - T \big( \widetilde{A}_i  \otimes \textbf{I} \big)    \text{, } 
\end{align*} }

\noindent asserts that the expression above is equal to,

{\small \begin{align*}
 \big(  A_i T  \otimes T \big) - \big( T \widetilde{A}_i  \otimes T \big) = \big( A_i  T - T \widetilde{A}_i  \big) \otimes T  \text{, } 
\end{align*} }

\noindent while the second identity, from,

{\small \begin{align*}
  \big( \textbf{I} \otimes B_{kl} \big) T - T \big( \textbf{I} \otimes \widetilde{B}_{kl} \big)  \text{, } 
\end{align*} }

\noindent asserts that the expression above is equal to,

{\small \begin{align*}
  \big( T \otimes B_{kl} T \big) - \big( T \otimes T \widetilde{B}_{kl} \big) = T \otimes \big( B_{kl} T - T \widetilde{B}_{kl} \big) \text{. } 
\end{align*} }

\noindent \textit{Proof of Lemma 3}. To argue that the two above identities holds, one makes use of expressions such as,

{\small

\begin{align*}
   A_i \cdot A^{j_1}_1 \times \cdots \times A^{j_n}_n    , 
\end{align*}

}

\noindent for inserting the term $A_i$ as the first term of the product,

{\small

\begin{align*}
 A^{j_1}_1 \times \cdots \times A^{j_n}_n   , 
\end{align*}

}

\noindent It suffices to demonstrate that the above identities take the form,

{\small

\begin{align*}
   \frac{1}{\sqrt{2^n}}        \underset{(j_1, \cdots , j_n ) \in \{ 0 , 1 \}^n}{\sum}         \bigg\{            \big\{ \big\{     A_i  \cdot A^{j_1}_1 \times \cdots \times   A^{j_n}_n \big\}    \otimes \textbf{I} \big\} \ket{\psi_{\mathrm{FFL}}}    - \mathrm{sign} \big( i , j_1 , \cdots , j_n \big)   \\ \times  \big\{  \big\{ A^{j_1}_1 \times \cdots \times A^{j_{i-1}}_{i-1} \cdot  A^{j_i \oplus 1}_{i } A^{j_{i+1}}_{i+1} \times \cdots \times A^{j_n}_n \big\}  \otimes \textbf{I} \big\} \ket{\psi_{\mathrm{FFL}}}                                       \bigg\}   \bigg\{ \bra{\widetilde{\psi_{\mathrm{FFL}}}}       \\ \times    \bigg\{ \big\{    \big\{ \widetilde{A}^{j_1}_1 \times \cdots \times \widetilde{A}^{j_n}_n  \big\} \otimes \textbf{I}           \big\}               \bigg\}^{\dagger}       \bigg\}                   , \\ \\ \tag{$*$} \end{align*}
   
   \begin{align*} \frac{1}{\sqrt{2^n}}            \underset{(j_1, \cdots , j_n ) \in \{ 0 , 1 \}^n}{\sum}            \bigg\{     \big\{ \big\{     A^{j_1}_1 \times \cdots \times A^{j_n}_n                          \big\} \otimes B_{kl} \big\} \ket{\psi_{\mathrm{FFL}}} -        \frac{1}{\sqrt{2} } \cdot  \bigg\{    \pm \mathrm{sign} \big( i , j_1, \cdots , j_n \big)    \\ \\     \times       \big\{ \big\{              A^{j_1}_1 \times \cdots \times A^{j_{k-1}}_{k-1} A^{j_k \oplus 1}_{k } \cdot A^{j_{k+1}}_{k+1} \times \cdots \times A^{j_n}_n             \big\}     \otimes  \textbf{I}  \big\} \ket{\psi_{\mathrm{FFL}}}            \\  \\ +  \big\{ \big\{              A^{j_1}_1 \times \cdots \times A^{j_{l-1}}_{l-1} A^{j_l \oplus 1}_{l} \cdot A^{j_{l+1}}_{l+1} \times \cdots \times A^{j_n}_n             \big\}     \otimes  \textbf{I}  \big\} \ket{\psi_{\mathrm{FFL}}}         \bigg\}                            \bigg\}           \\ \\ \times  \bigg\{     \bra{\widetilde{\psi_{\mathrm{FFL}}}}       \bigg\{    \big\{             \widetilde{A}^{j_1}_1 \times \cdots \times A^{j_n}_n  \big\} \otimes \textbf{I}                                     \bigg\}^{\dagger}            \bigg\}                                             ,  \\ \\ \\ \tag{$**$} \\  
\end{align*}

}

\noindent for $\big(  A_i T  \otimes T \big) - \big( T \widetilde{A}_i \otimes T \big) $ and $ \big( T \otimes B_{kl} T \big) - \big( T \otimes T \widetilde{B}_{kl} \big)$ respectively, by direct computation observe,

{\small

\begin{align*}
  T \big( \widetilde{A}_i \otimes \textbf{I} \big) =  \frac{1}{\sqrt{2^n}}     \underset{(j_1, \cdots , j_n ) \in \{ 0 , 1 \}^n}{\sum}     \bigg\{     \big\{   \big\{   A^{j_1}_1 \times \cdots \times A^{j_n}_n \big\} \otimes \textbf{I}      \big\}        \ket{\psi_{\mathrm{FFL}}}   \bigg\} \bigg\{  \bra{\widetilde{\psi_{\mathrm{FFL}}}}     \bigg\{         \big\{  \big\{ \widetilde{A}^{j_1}_1 \times \cdots \times \widetilde{A}^{j_n}_n \big\} \\ \otimes \textbf{I} \big\} \bigg\}^{\dagger}              \bigg\}   \big\{  \widetilde{A}_i \otimes \textbf{I} \big\}                          , 
\end{align*}

}

\noindent equals,

{\small

\begin{align*}
   \frac{1}{\sqrt{2^n}}     \underset{(j_1, \cdots , j_n ) \in \{ 0 , 1 \}^n}{\sum}     \bigg\{     \big\{   \big\{   A^{j_1}_1 \times \cdots \times A^{j_n}_n \big\} \otimes \textbf{I}      \big\}        \ket{\psi_{\mathrm{FFL}}}   \bigg\} \bigg\{  \big\{  \widetilde{A}_i \otimes \textbf{I} \big\}  \cdot      \bigg\{         \big\{  \big\{ \widetilde{A}^{j_1}_1 \times \cdots \times \widetilde{A}^{j_n}_n \big\} \otimes \textbf{I} \big\} \\ \times  \ket{\widetilde{\psi_{\mathrm{FFL}}}}\bigg\}^{\dagger}              \bigg\}   \\ \\ =  \frac{1}{\sqrt{2^n}}     \underset{(j_1, \cdots , j_n ) \in \{ 0 , 1 \}^n}{\sum}     \bigg\{     \big\{   \big\{   A^{j_1}_1 \times \cdots \times A^{j_n}_n \big\} \otimes \textbf{I}      \big\}        \ket{\psi_{\mathrm{FFL}}}   \bigg\} \bigg\{  \mathrm{sign} \big( i , j_1 , \cdots , j_n \big) \cdot  \bigg\{         \big\{  \big\{ \widetilde{A}^{j_1}_1 \times \cdots \\ \times A^{j_{i-1}}_{i-1} A^{j_{i \oplus 1}}_i \cdot A^{j_{i+1}}_{i+1} \times  \cdots \times \widetilde{A}^{j_n}_n \big\} \otimes \textbf{I} \big\}  \ket{\widetilde{\psi_{\mathrm{FFL}}}}\bigg\}^{\dagger}              \bigg\}              , 
\end{align*}

}

\noindent implying,

{\small

\begin{align*}
    \big(  A_i T  \otimes T \big) - \big( T \widetilde{A}_i  \otimes T \big)   =     \frac{1}{\sqrt{2^n}} \bigg[        \underset{(j_1 , \cdots , j_n ) \in \{ 0 , 1 \}^n}{\sum} \bigg\{ \bigg(   \bigg[    \big\{  A_i \cdot A^{j_1}_1 \times \cdots \times A^{j_n}_n  \big\} \otimes \textbf{I}         \bigg] \end{align*}
    
    \begin{align*} - \bigg[   \mathrm{sign} \big( i , j_1 , \cdots , j_n \big) \cdot       \big\{  \big\{ \widetilde{A}^{j_1}_1 \times \cdots  \times A^{j_{i-1}}_{i-1} A^{j_{i \oplus 1}}_i \cdot A^{j_{i+1}}_{i+1} \times  \cdots \times \widetilde{A}^{j_n}_n \big\} \otimes \textbf{I} \big\}                   \bigg]  \bigg) \ket{\psi_{\mathrm{FFL}}} \bigg\}    \\ \times  \bigg\{        \bra{\widetilde{\psi_{\mathrm{FFL}}}}  \bigg\{    \big\{      \widetilde{A}^{j_1}_1 \times \cdots \times \widetilde{A}^{j_n}_n        \big\} \otimes \textbf{I}       \bigg\}^{\dagger}     \bigg\}                      \bigg]             , 
\end{align*}

}
 
\noindent and hence, that (*) holds. To argue that the remaining identity holds, observe,

{\small

\begin{align*}
  \big( T \otimes T \widetilde{B}_{kl} \big) =   \frac{1}{\sqrt{2^n}} \underset{(j_1 , \cdots , j_n ) \in \{ 0 , 1 \}^n}{\sum}   \bigg\{  \big\{ \big\{   A^{j_1}_1 \times \cdots \times A^{j_n}_n   \big\} \otimes \textbf{I} \big\} \ket{\psi_{\mathrm{FFL}}}                 \bigg\}     \bigg\{           \bra{\widetilde{\psi_{\mathrm{FFL}}}} \big\{    \big\{ \widetilde{A}^{j_1}_1 \times \cdots \\ \times \widetilde{A}^{j_n}_n \big\}            \otimes \textbf{I}  \big\} \big\{ \textbf{I} \otimes \widetilde{B}_{kl}    \big\} \big\}           \bigg\}             , 
\end{align*}

}

\noindent equals,

{\small

\begin{align*}
   \frac{1}{\sqrt{2^n}} \underset{(j_1 , \cdots , j_n ) \in \{ 0 , 1 \}^n}{\sum}   \bigg\{  \big\{ \big\{   A^{j_1}_1 \times \cdots \times A^{j_n}_n   \big\} \otimes \textbf{I} \big\} \ket{\psi_{\mathrm{FFL}}}                 \bigg\}     \bigg\{           \bra{\widetilde{\psi_{\mathrm{FFL}}}} \bigg\{    \big\{ \widetilde{A}^{j_1}_1 \times \cdots  \times \widetilde{A}^{j_n}_n \big\}  \\           \otimes \textbf{I}  \big\} \bigg\{ \frac{\pm \widetilde{A}_k + \widetilde{A}_l}{\sqrt{2}} \otimes  \textbf{I}           \bigg\}   \bigg\}                   , 
\end{align*}

}

\noindent implying,

{\small

\begin{align*}
      \big( T \otimes B_{kl} T \big) - \big( T \otimes T \widetilde{B}_{kl} \big) =      \frac{1}{\sqrt{2^n}} \cdot \frac{1}{\sqrt{2}}  \bigg[        \underset{(j_1 , \cdots , j_n ) \in \{ 0 , 1 \}^n}{\sum} \bigg\{ \bigg(   \bigg[        \big\{   A^{j_1}_1 \times \cdots \times A^{j_n}_n \big\} \otimes  \textbf{I} \bigg] \\ \times \ket{\psi_{\mathrm{FFL}}}    \\ \times        \bigg[  \pm \mathrm{sign} \big( i , j_1 , \cdots , j_n \big) \cdot      \bra{\widetilde{\psi_{\mathrm{FFL}}}}  \big\{ \big\{   \widetilde{A}^{j_1}_1 \times \cdots \times \widetilde{A}^{j_{k-1}}_{k-1} A^{j_{k \oplus 1}}_{k} \cdot A^{j_{k+1}}_{k+1} \times \cdots \times A^{j_n}_n           \big\} \\ \otimes \textbf{I} \big\}     \\ +   \mathrm{sign} \big( i , j_1 , \cdots , j_n \big) \cdot      \bra{\widetilde{\psi_{\mathrm{FFL}}}} \big\{  \big\{   \widetilde{A}^{j_1}_1 \times \cdots \times \widetilde{A}^{j_{l-1}}_{l-1} A^{j_{l \oplus 1}}_{l} \cdot A^{j_{l+1}}_{l+1} \times \cdots \times A^{j_n}_n            \big\} \\ \otimes \textbf{I} \big\}    \bigg]                              , \\  \tag{$***$} 
\end{align*}

}

\noindent can be rearranged as the summation provided in (**) after consolidating the terms,

{\small

\begin{align*}
  A^{j_1}_1 \times \cdots \times A^{j_n}_n   ,
\end{align*}

}

\noindent with,

{\small

\begin{align*}
 \pm \mathrm{sign} \big( i , j_1 , \cdots , j_n \big) \cdot     \big\{   \widetilde{A}^{j_1}_1 \times \cdots \times \widetilde{A}^{j_{k-1}}_{k-1} A^{j_{k \oplus 1}}_{k} \cdot A^{j_{k+1}}_{k+1} \times \cdots \times A^{j_n}_n           \big\} +  \mathrm{sign} \big( i , j_1 , \cdots , j_n \big) \cdot    \big\{  \big\{   \widetilde{A}^{j_1}_1 \\ \times \cdots \times \widetilde{A}^{j_{l-1}}_{l-1} A^{j_{l \oplus 1}}_{l} \cdot A^{j_{l+1}}_{l+1} \times \cdots \times A^{j_n}_n            \big\}  ,
\end{align*}

}

\noindent upon distributing the terms $\widetilde{A}_k$ and $\widetilde{A}_l$ from,

{\small

\begin{align*}
    \frac{\pm \widetilde{A}_k + \widetilde{A}_l}{\sqrt{2}} ,
\end{align*}

}

\noindent in (***) above, where,

{\small

\begin{align*}
   \big\{  \pm \widetilde{A}_k = + \widetilde{A}_k \big\} \Longleftrightarrow    \big\{ k < l \big\}       , \\  \big\{  \pm \widetilde{A}_k = - \widetilde{A}_k \big\} \Longleftrightarrow    \big\{ k >  l \big\}        .  
\end{align*}

}

\noindent Hence (**) also holds, from which we conclude the argument. \boxed{}

\subsection{T has unit Frobenius norm}

\noindent \textbf{Lemma} \textit{4} (\textit{orthonormality of the FFL canonical basis from the fact that the Frobenius norm of the $T$ equals $1$}). With respect to the Frobenius norm, the norm of $T$ is $1$.

\bigskip

\noindent \textit{Proof of Lemma 4}. Directly apply the argument from \textit{6.2} in {\color{blue}[37]}. \boxed{}

\subsection{Error bounds}

\noindent For the $\mathrm{FFL}$ game, error bounds consist of statements the operator,

{\small \begin{align*}
 \frac{A_i A_j + A_j A_i }{2}     \text{, } \\ 
\end{align*} }

\noindent that is tensor produced with the identity $\textbf{I}$.

\bigskip

\noindent \textbf{Lemma} \textit{5} ($\epsilon$\textit{-optimality}, \textbf{Lemma} \textit{7}, {\color{blue}[37]}). For an $\epsilon$-optimal strategy $A_i$, $B_{jk}$ and $\ket{\psi_{\mathrm{FFL}}}$,

{\small \begin{align*}
 \underset{1 \leq i < j \leq n}{\sum} \bigg| \bigg| \bigg\{   \bigg(         \frac{A_i A_j + A_j A_i}{2}        \bigg) \otimes \textbf{I} \bigg\}  \ket{\psi_{\mathrm{FFL}}} \bigg| \bigg|^2 <   2 \big(   \frac{7}{3}    \big)^2 n \big( n - 1 \big) \epsilon \text{. } 
\end{align*} }

\noindent \textit{Proof of Lemma 5}. From \textit{6.5} in {\color{blue}[37]}, the same properties of the two operators hold, namely that they each have operator norm that is most $1 + \frac{3}{2}$. The desired inequality for $\epsilon$-optimality above holds from the fact that,

{\small \begin{align*}
   \bigg| \bigg| \bigg\{   \bigg( \frac{A_i A_j + A_j A_i}{2} \bigg) \otimes \textbf{I} \bigg\}     \ket{\psi_{\mathrm{FFL}}} \bigg| \bigg|^2 \leq \big( 1 + \frac{3}{2} \big) \cdot \bigg| \bigg|       \bigg\{    \bigg( \frac{A_i + A_j}{2} \bigg)  \otimes \textbf{I}   - \textbf{I} \otimes B_{ij }     \bigg\}  \ket{\psi_{\mathrm{FFL}}} \bigg| \bigg|   \text{. } 
\end{align*} }

\noindent As an anticommutation relation from the inequality above, one also has that,

{\small \begin{align*}
  \bigg| \bigg|              \bigg\{   \bigg(   \frac{A_i A_j + A_j A_i}{2} \bigg)  \otimes \textbf{I}  \bigg\}  \ket{\psi_{\mathrm{FFL}}}  \bigg| \bigg|^2  \leq \big(  1 + \frac{3}{2} \big) \cdot \bigg|  \bigg|    \bigg\{   \bigg( \frac{A_i - A_j}{2} \bigg) \otimes \textbf{I} - \textbf{I} \otimes B_{ji}  \bigg\}  \ket{\psi_{\mathrm{FFL}}}    \bigg|                      \bigg|  \text{, } 
\end{align*} }

\noindent from which it also holds that,

{\small \begin{align*}
2 \cdot  \underset{1 \leq i < j \leq n }{\sum} \bigg| \bigg| \bigg\{    \bigg(   \frac{A_i A_j + A_j A_i}{2}                          \bigg) \otimes \textbf{I} \bigg\}  \ket{\psi_{\mathrm{FFL}}}              \bigg| \bigg|^2  \leq       \big( 1 + \frac{3}{2} \big)^2  \cdot   \underset{1 \leq i < j  \leq n}{\sum}  \bigg[      \bigg| \bigg|            \bigg(  \frac{A_i + A_j}{\sqrt{2}} \bigg)  \otimes \textbf{I} \ket{\psi_{\mathrm{FFL}}}    
 - \textbf{I}  \\ \otimes B_{ij} \ket{\psi_{\mathrm{FFL}}}   \bigg| \bigg|^2  +      \bigg| \bigg|               \bigg( \frac{A_i - A_j}{\sqrt{2}} \bigg) \otimes \textbf{I} \ket{\psi_{\mathrm{FFL}}} - \textbf{I} \otimes B_{ji} \ket{\psi_{\mathrm{FFL}}}            \bigg| \bigg|^2          \bigg]         \text{, } 
\end{align*} }

\noindent from which we conclude the argument, as the final term in the inequality above is equal to,

{\small \begin{align*}
     2  \big( 1 + \frac{3}{2}  \big)^2           n \big( n -1 \big) \epsilon    \text{, }
\end{align*} }

\noindent which can be upper bounded with, 

{\small \begin{align*}
       2  \big(  \frac{7}{2}  \big)^2           n \big( n -1 \big) \epsilon   \text{, }                    
\end{align*} }

\noindent from which we conclude the argument. \boxed{}

\bigskip

\noindent From the above result one is also interested in characterizing the winning probability of the FFL game is dependent upon including a superposition of the form,

{\small

\begin{align*}
       \frac{\pm B_{kl} + B_{lk}}{\big| \pm B_{kl} + B_{lk}  \big| }              , 
\end{align*}

}

\noindent in the tensor product on states $\ket{\psi_{\mathrm{FFL}}}$, as opposed to a superposition of the form,

{\small

\begin{align*}
    \frac{A_i A_j + A_j A_i}{2}   .  
\end{align*}

}

\bigskip

\noindent \textbf{Lemma} \textit{6} ($\sqrt{\epsilon}$- \textit{approximality}, \textbf{Lemma} \textit{8}, {\color{blue}[37]}). From the same quantities introduced in the previous result, one has,

{\small \begin{align*}
   \bigg| \bigg|               \big\{   A_k \otimes \textbf{I} \big\}  \ket{\psi_{\mathrm{FFL}}}    -  \bigg\{  \textbf{I} \otimes \bigg(     \frac{\pm B_{kl} + B_{lk}}{\big| \pm B_{kl} + B_{lk}  \big| }           \bigg) \bigg\}  \ket{\psi_{\mathrm{FFL}}}              \bigg| \bigg| < 17 \sqrt{n \epsilon}   \text{. } 
\end{align*} }

\noindent The notation $\pm B_{kl}$ indicates,

{\small \begin{align*}
  \pm B_{kl} \equiv B_{kl}  \Longleftrightarrow     l > k  \text{, }  \\ \pm B_{kl} \equiv - B_{kl} \Longleftrightarrow  l \leq  k      \text{. } 
\end{align*}     }

\noindent \textit{Proof of Lemma 6}. To demonstrate that the above inequality holds, observe that different components of the norm above can be approximated with,

{\small \begin{align*}
        \big\{  A_k \otimes \textbf{I} \big\}  \ket{\psi_{\mathrm{FFL}}}   \approx  \bigg\{  \textbf{I} \otimes \bigg( \frac{\pm B_{kl} + B_{lk}}{\sqrt{2}} \bigg)  \bigg\}          \ket{\psi_{\mathrm{FFL}}}  \text{, } 
\end{align*} }

\noindent and with,

{\small \begin{align*}
       \bigg\{  \textbf{I} \otimes        \bigg(          \frac{\pm B_{kl } + B_{lk}}{\sqrt{2} }    \bigg)     \bigg\}   \ket{\psi_{\mathrm{FFL}}} \approx    \bigg\{  \textbf{I} \otimes        \bigg(          \frac{\pm B_{kl } + B_{lk}}{\big| \pm B_{kl} + B_{lk} \big| }    \bigg)     \bigg\}   \ket{\psi_{\mathrm{FFL}}}    \text{. } 
\end{align*} }

\noindent The existence of the two approximations above is a result of the existence of operators in Hilbert space for which,

{\small \begin{align*}
         \frac{\big| A_k \otimes \textbf{I} \big|}{\bigg| \textbf{I} \otimes \bigg( \frac{\pm B_{kl} + B_{lk}}{\sqrt{2}} \bigg)    \bigg| }             \approx 1  \text{, } 
\end{align*} }

\noindent and for which,

{\small \begin{align*}
          \frac{\bigg| \textbf{I} \otimes \bigg( \frac{\pm B_{kl} + B_{lk}}{\sqrt{2}} \bigg)      \bigg| }{\bigg| \textbf{I} \otimes \bigg( \frac{\pm B_{kl} + B_{lk}}{\big| \pm B_{kl } + B_{lk} \big| } \bigg)     \bigg| }    \equiv \frac{\bigg|   \bigg[ \textbf{I} \otimes \bigg( \frac{\pm B_{kl} + B_{lk}}{\sqrt{2}}  \bigg)  \bigg]  \bigg|  \cdot  \frac{\big| \pm B_{kl} + B_{lk} \big|}{\big| \pm B_{kl} + B_{lk} \big|}   }{\bigg|   \textbf{I} \otimes \bigg( \frac{\pm B_{kl} + B_{lk}}{\big| \pm B_{kl } + B_{lk} \big| } \bigg)          \bigg| }  \approx          \frac{\big| \pm B_{kl} + B_{lk} \big| }{\sqrt{2}}      \approx 1    \text{. } 
\end{align*} }

\noindent Besides the approximation of operators above, a previous relation,

{\small \begin{align*}
  \underset{1 \leq i < j \leq n}{\sum} \bigg[ \bigg| \bigg|              \big\{  A_k \otimes \textbf{I} \big\}  \ket{\psi_{\mathrm{FFL}}} - \bigg\{  \textbf{I} \otimes     \bigg( \frac{B_{kj} + B_{jk}}{\sqrt{2}}   \bigg) \bigg\}  \ket{\psi_{\mathrm{FFL}}}    \bigg| \bigg|^2 \bigg] \leq  n \big( n - 1 \big) \epsilon  \text{, } 
\end{align*} }

\noindent and,

{\small \begin{align*} 
  \underset{1 \leq i < j \leq n}{\sum} \bigg[ \bigg| \bigg| \big\{  A_k \otimes \textbf{I} \big\}  \ket{\psi_{\mathrm{FFL}}}   -  \bigg\{    \textbf{I} \otimes   \bigg(    - \frac{B_{kj} - B_{jk}}{\sqrt{2}}       \bigg) \bigg\}  \ket{\psi_{\mathrm{FFL}}}    \bigg| \bigg|^2 \bigg] \leq  n \big( n - 1 \big) \epsilon \text{, } 
\end{align*} }

\noindent together imply,

{\small \begin{align*}
    \underset{1 \leq i < j \leq n}{\sum} \bigg[ \bigg| \bigg|              \big\{  A_k \otimes \textbf{I} \big\}  \ket{\psi_{\mathrm{FFL}}} - \bigg\{  \textbf{I} \otimes     \bigg( \frac{B_{kj} + B_{jk}}{\sqrt{2}}   \bigg) \bigg\}  \ket{\psi_{\mathrm{FFL}}}    \bigg| \bigg|^2  +      \bigg| \bigg| \big\{  A_k \otimes \textbf{I} \big\}  \ket{\psi_{\mathrm{FFL}}}   \\ -  \bigg\{    \textbf{I} \otimes   \bigg(    - \frac{B_{kj} - B_{jk}}{\sqrt{2}}       \bigg) \bigg\}  \ket{\psi_{\mathrm{FFL}}}    \bigg| \bigg|^2     \bigg]       \leq 2 n \big( n-1 \big) \epsilon   \text{. }
\end{align*} }

\noindent Next, observe,

{\small \begin{align*}
\bigg| \bigg|   \bigg\{  \textbf{I} \otimes \bigg( \frac{\pm B_{kj} + B_{jk}}{\sqrt{2}} \bigg)  \bigg\}  \ket{\psi_{\mathrm{FFL}}}   -    \bigg\{    \textbf{I} \otimes  \bigg( \frac{\pm B_{kl} + B_{lk}}{\big| \pm B_{kl} + B_{lk} \big|}     \bigg)    \bigg\}  \ket{\psi_{\mathrm{FFL}}}       \bigg| \bigg|   \leq \bigg| \bigg|  \bigg\{   \textbf{I} \otimes \bigg(            \frac{B_{kl} B_{lk} + B_{lk} B_{kl}}{2}    \bigg) \bigg\}   \\ \times \ket{\psi_{\mathrm{FFL}}}         \bigg| \bigg|   \text{, } 
\end{align*} }

\noindent because,

{\small \begin{align*}
 \bigg[ \textbf{I} \otimes \bigg\{  \frac{\pm B_{kj} + B_{jk}}{\sqrt{2}} \bigg\}   \bigg]  \ket{\psi_{\mathrm{FFL}}}   -    \bigg[   \textbf{I} \otimes  \bigg\{  \frac{\pm B_{kl} + B_{lk}}{\big| \pm B_{kl} + B_{lk} \big|}     \bigg\}     \bigg] \ket{\psi_{\mathrm{FFL}}}   \\    \\ \approx   \bigg[ \textbf{I} \otimes \bigg\{   \frac{\pm B_{kj} + B_{jk}  \mp B_{kl} - B_{lk} }{2}    \bigg\}    \bigg]      \ket{\psi_{\mathrm{FFL}}}   \text{. } \\ 
\end{align*} }

\noindent Moreover, the operations associated with,

{\small \begin{align*}
     \bigg| \bigg|  \bigg\{  \textbf{I} \otimes \bigg( \frac{B_{kl} B_{lk} + B_{lk} B_{kl}}{2} \bigg) \bigg\}  \ket{\psi_{\mathrm{FFL}}}        \bigg| \bigg|          \text{, } \\ 
\end{align*} } 

\noindent can be analyzed from a factorization,

{\small \begin{align*}
     \bigg| \bigg|  \bigg[     \bigg[   \big\{  A_k \otimes \textbf{I} \big\}  + \bigg\{  \textbf{I} \otimes   \bigg(   \frac{\pm B_{kl} + B_{lk}}{\sqrt{2}}    \bigg) \bigg\} \bigg]  \cdot   \bigg[   \big\{  A_k \otimes \textbf{I} \big\}  - \bigg\{  \textbf{I} \otimes \bigg( \frac{\pm B_{kl} + B_{lk}}{\sqrt{2}} \bigg)        \bigg\}    \bigg]   \bigg] \ket{\psi_{\mathrm{FFL}}}                    \bigg| \bigg|              \text{, } 
\end{align*} } 

\noindent which can be upper bounded with,

{\small \begin{align*}
\bigg| \bigg|    \bigg[ \big\{  A_k \otimes \textbf{I} \big\}  + \bigg\{  \textbf{I} \otimes   \bigg(   \frac{\pm B_{kl} + B_{lk}}{\sqrt{2}}  \bigg)   \bigg\}  \bigg]        \bigg| \bigg|  \cdot              \bigg| \bigg|   
      \bigg[              \big\{  A_k \otimes \textbf{I} \big\}  - \bigg\{  \textbf{I} \otimes \bigg( \frac{\pm B_{kl} + B_{lk}}{\sqrt{2}}      \bigg)  \bigg\}      \bigg]  \ket{\psi_{\mathrm{FFL}}}                    \bigg| \bigg|  \text{, } 
\end{align*} }

\noindent which can be further upper bounded with,

{\small \begin{align*}
       \big( 1 + \frac{3}{2} \big)  \frac{3}{2} \sqrt{n \epsilon }   \text{. } 
\end{align*} } 

\noindent Altogether,

{\small \begin{align*}
  \bigg| \bigg|       \bigg[         \big\{  A_k \otimes \textbf{I} \big\}  - \bigg\{  \textbf{I} \otimes   \bigg( \frac{\pm B_{kl} + B_{lk}}{\big| \pm B_{kl} + B_{lk} \big| } \bigg) \bigg\}                 \bigg]   \ket{\psi_{\mathrm{FFL}}}  \bigg| \bigg|  \leq 3   \big( \frac{9}{4} \big) \big( 1 +  \frac{3}{2}   \big)             \sqrt{n \epsilon }  =    \frac{135}{8}     \sqrt{n \epsilon }     < 17 \sqrt{n \epsilon}      \text{, } 
\end{align*} }

\noindent from which we conclude the argument. \boxed{}

\subsection{First Error bound for permuting indices of A operators}

\noindent In addition to the previous error bound, another error bound follows from the fact that the tensor product,

{\small \begin{align*}
 \bigg\{  A_i \bigg(  \underset{1 \leq i \leq n}{\prod} A^{j_i}_i \bigg) \otimes \textbf{I} \bigg\}  \ket{\psi_{\mathrm{FFL}}}   \text{, } 
\end{align*} } 

\noindent can be placed into corresponding the operation,

{\small \begin{align*}
    \mathrm{sign} \big( i , j_1 , \cdots , j_n \big) \bigg\{  \bigg( \underset{1 \leq i \leq n}{\prod} A^{j_i}_i  \bigg) \otimes \textbf{I} \bigg\}  \ket{\psi_{\mathrm{FFL}}}      \text{, } 
\end{align*} }

\noindent by introducing the sequence of manipulations:

\begin{itemize}
    \item[$\bullet$] \textit{Switching a tensor product of $A^{j_i}_i$ terms with AB-switches}.

    \item[$\bullet$] \textit{Switching the last observable from the B side to the A side}.

    \item[$\bullet$] \textit{Perform an odd, or even, number of anticommutation swaps to permute $A^{j_i}_i$ to the desired position}.
\end{itemize}

\noindent The statement below provides an upper bound on the error associated from switching the sign of $i , j_1 , \cdots , j_n$.

\bigskip

\noindent \textbf{Lemma} \textit{7} (\textit{error bound from permuting indices}). One has,

{\small \begin{align*}
 \bigg| \bigg|  \bigg\{    \bigg( \underset{1 \leq i \leq n}{\prod} A^{j_i}_i  \bigg)   -    \bigg( \underset{if i \equiv j_1+1, \text{ } \mathrm{set} \text{ } j_1 + 1 \equiv j_1 \oplus 1}{ \underset{1 \leq i \leq n}{\prod} }A^{j_i}_i        \bigg) \otimes \textbf{I} \bigg\}  \ket{\psi_{\mathrm{FFL}}}   \bigg| \bigg|  \leq \frac{100}{9} n^2 \sqrt{\epsilon}  \text{. } 
\end{align*} }

\bigskip

\noindent \textit{Proof of Lemma 7}. The desired upper bound follows from the observation that the quantitative error from performing the three operations listed above scales as,

{\small \begin{align*}
n^2 \sqrt{\epsilon} \bigg\{  \bigg(  \big( \frac{5}{3} \big)^2 + \frac{5}{3} \bigg)  + \frac{1}{\sqrt{n}} \bigg(  \big( \frac{5}{3} \big)^2 + \frac{5}{3} \bigg)              \bigg\}    \text{, } 
\end{align*} }

\noindent which has an upper bound,

{\small \begin{align*}
   2  n^2 \sqrt{\epsilon}   \bigg\{  \big( \frac{5}{3} \big)^2 + \frac{5}{3} \bigg\}  \leq     4  n^2 \sqrt{\epsilon}       \big( \frac{5}{3} \big)^2         \leq      \frac{100}{9} n^2 \sqrt{\epsilon }    \text{, } 
\end{align*} }

\noindent less than or equal to the quantitative error. We conclude the argument. \boxed{}

\bigskip

\noindent Under $\widetilde{\cdot}$ associated with the objects,

{\small \begin{align*}
\widetilde{A}_S \equiv \underset{s \in S}{\bigcup} \widetilde{A}_s \equiv \big\{   s  \in S :  \widetilde{A}_s \in \big\{ - 1 , + 1 \big\}    \big\}  \subsetneq \textbf{C}^{2^{\lceil n / 2 \rceil}} , \\ \\  \widetilde{B}_T \equiv \underset{t \in T}{\bigcup} \widetilde{B}_t \equiv \big\{   t  \in T :  \widetilde{B}_t \in \big\{ - 1 , + 1 \big\}    \big\}  \subsetneq \textbf{C}^{2^{\lceil n / 2 \rceil}}   ,    
\end{align*}

\begin{align*}
   \widetilde{A}_{i,\mathrm{FFL}}  , \\ \\   \widetilde{B}_{jk, \mathrm{FFL}}      , \\ \\         \ket{\widetilde{\psi_{\mathrm{FFL}}}}  , 
\end{align*}

}

\noindent for $j \neq k$ products of $A_i$ operations takes the following form.

\bigskip

\noindent \textbf{Lemma} \textit{8} (\textit{odd n}, \textit{6.1}, {\color{blue}[37]}). For odd $n$, one has an expansion for,

{\small \begin{align*}
 \underset{1 \leq i \leq n}{\prod} \widetilde{A}_i    \text{, } 
\end{align*} }

\noindent in terms of a signed block identity matrix,

{\small \[
\big( - 1 \big)^n \begin{bmatrix}
\textbf{I} & 0 \\ 0 & - \textbf{I}
\end{bmatrix}  \text{. } 
\] }

\noindent Hence, for,

{\small \begin{align*}
 \ket{\widetilde{\psi_{\mathrm{FFL}}}} = \frac{1}{\sqrt{2 \times  2^{\lfloor \frac{n}{2} \rfloor }}} \bigg[   \text{ } \bigg(      \underset{1 \leq j \leq 2^{\lfloor \frac{n}{2}\rfloor}}{\sum}   + \underset{2^{\lfloor \frac{n}{2} \rfloor} + 1 \leq j \leq 2 \times 2^{\lfloor \frac{n}{2}\rfloor}}{\sum} \bigg) \bigg( \ket{j} \otimes \ket{j}  \bigg) \text{ }  \bigg]    \text{, } 
\end{align*} }

\noindent one has,

{\small \begin{align*}
 \bigg\{  \bigg(  \underset{1 \leq i \leq n}{\prod} \widetilde{A}_i   \bigg) \otimes \textbf{I} \bigg\}  \ket{\widetilde{\psi_{\mathrm{FFL}}}} = \frac{1}{\sqrt{2 \times  2^{\lfloor \frac{n}{2} \rfloor }}}  \bigg[    \text{ } \bigg(      \underset{1 \leq j \leq 2^{\lfloor \frac{n}{2}\rfloor}}{\sum}   + \underset{2^{\lfloor \frac{n}{2} \rfloor} + 1 \leq j \leq 2 \times 2^{\lfloor \frac{n}{2}\rfloor}}{\sum} \bigg) \bigg\{  \bigg(       \bigg(  \underset{1 \leq i \leq n}{\prod} \widetilde{A}_i   \bigg) \ket{j}      \otimes \ket{j}      \bigg) \\ \otimes \bigg( \bigg(  \underset{1 \leq i \leq n}{\prod} \widetilde{A}_i   \bigg) \ket{j}    \otimes \ket{j} \bigg)   \bigg\}  \text{ }      \bigg]            \text{. } 
\end{align*} } 

\noindent \textit{Proof of Lemma 8}. The result follows from arguments involving the representation-theoretic intertwining operation from \textit{6.1} in {\color{blue}[37]}. The final equality follows from the observation that,

{\small \begin{align*}
 \frac{1}{\sqrt{2 \times  2^{\lfloor \frac{n}{2} \rfloor }}}    \bigg\{  \bigg(  \underset{1 \leq i \leq n}{\prod} \widetilde{A}_i   \bigg) \otimes \textbf{I} \bigg\}     \bigg[   \text{ } \bigg(      \underset{1 \leq j \leq 2^{\lfloor \frac{n}{2}\rfloor}}{\sum}   + \underset{2^{\lfloor \frac{n}{2} \rfloor} + 1 \leq j \leq 2 \times 2^{\lfloor \frac{n}{2}\rfloor}}{\sum} \bigg) \bigg( \ket{j} \otimes \ket{j}  \bigg) \text{ }  \bigg]             \text{, } 
\end{align*} }

\noindent equals, 

{\small \begin{align*}
  \frac{1}{\sqrt{2 \times  2^{\lfloor \frac{n}{2} \rfloor }}}  \bigg[    \text{ } \bigg(      \underset{1 \leq j \leq 2^{\lfloor \frac{n}{2}\rfloor}}{\sum}   + \underset{2^{\lfloor \frac{n}{2} \rfloor} + 1 \leq j \leq 2 \times 2^{\lfloor \frac{n}{2}\rfloor}}{\sum} \bigg) \bigg\{  \bigg(       \bigg(  \underset{1 \leq i \leq n}{\prod} \widetilde{A}_i   \bigg) \ket{j}      \otimes \ket{j}      \bigg) \otimes   \bigg( \bigg(  \underset{1 \leq i \leq n}{\prod} \widetilde{A}_i   \bigg) \ket{j}    \otimes \ket{j} \bigg)   \bigg\}  \text{ }      \bigg]   \text{, } 
\end{align*} }

\noindent from which we conclude the argument. \boxed{}

\subsection{Second Error bound for permuting indices of the A operator}

\noindent In addition to the error bounds,

{\small \begin{align*}
   \bigg| \bigg|               \big\{   A_k \otimes \textbf{I} \big\}  \ket{\psi_{\mathrm{FFL}}}    -  \bigg\{  \textbf{I} \otimes \bigg(     \frac{\pm B_{kl} + B_{lk}}{\big| \pm B_{kl} + B_{lk}  \big| }           \bigg) \bigg\}  \ket{\psi_{\mathrm{FFL}}}              \bigg| \bigg| < 17 \sqrt{n \epsilon}   \text{, } 
\end{align*} } 

\noindent in \textbf{Lemma} \textit{5},

{\small \begin{align*}
 \underset{1 \leq i < j \leq n}{\sum} \bigg| \bigg| \bigg\{   \bigg(         \frac{A_i A_j + A_j A_i}{2}        \bigg) \otimes \textbf{I} \bigg\}  \ket{\psi_{\mathrm{FFL}}} \bigg| \bigg|^2 <   2 \big(   \frac{7}{3}    \big)^2 n \big( n - 1 \big) \epsilon \text{, } 
\end{align*} }

\noindent in \textbf{Lemma} \textit{6}, and,

{\small \begin{align*}
 \bigg| \bigg|  \bigg\{    \bigg( \underset{1 \leq i \leq n}{\prod} A^{j_i}_i  \bigg)   -    \bigg( \underset{if i \equiv j_1+1, \text{ } \mathrm{set} \text{ } j_1 + 1 \equiv j_1 \oplus 1}{ \underset{1 \leq i \leq n}{\prod} }A^{j_i}_i        \bigg) \otimes \textbf{I} \bigg\}  \ket{\psi_{\mathrm{FFL}}}   \bigg| \bigg|  \leq \frac{100}{9} n^2 \sqrt{\epsilon}  \text{, } 
\end{align*} }

\noindent in \textbf{Lemma} \textit{7}, the following error bound, through an estimate on the norm,

{\small \begin{align*}
 \bigg| \bigg|  \bigg\{  \bigg( \underset{1 \leq i \leq n}{\prod}   A^{j_i}_i \bigg)   \otimes B_{kl} \bigg\}   \ket{\psi_{\mathrm{FFL}}}  - \frac{2}{3} \bigg[ \pm \bigg\{  \mathrm{sign} \big( i , j_1 , \cdots , j_n \big) \bigg[ \text{ } \bigg(     \underset{i = j_k + 1 , \text{ } \mathrm{set} \text{ } j_k + 1 \equiv j_k \oplus 1 }{\underset{1 \leq i \leq n}{\prod}}     A^{j_i}_i    \bigg) \\ + \bigg( \underset{i = j_l + 1 , \text{ } \mathrm{set} \text{ } j_l + 1 \equiv j_l \oplus 1 }{\underset{1 \leq i \leq n}{\prod}}     A^{j_i}_i   \bigg) \text{ } \bigg]   \otimes \textbf{I} \bigg\}   \ket{\psi_{\mathrm{FFL}}} \bigg]           \bigg|    \text{, }
\end{align*} }

\noindent provides an upper bound between $A^{j_i}_i$ operators from Alice's strategy, and $B_{kl}$ operators from Bob's strategy.

\bigskip

\noindent \textbf{Lemma} \textit{9} (\textit{second error bound}, \textit{6.6}, {\color{blue}[37]}). From previously defined quantities, one has,

{\small \begin{align*}
 \bigg| \bigg|  \bigg\{  \bigg( \underset{1 \leq i \leq n}{\prod}   A^{j_i}_i \bigg)   \otimes B_{kl} \bigg\}   \ket{\psi_{\mathrm{FFL}}}  - \frac{2}{3} \bigg[ \pm \bigg\{  \mathrm{sign} \big( i , j_1 , \cdots , j_n \big) \bigg[ \text{ } \bigg(     \underset{i = j_k + 1 , \text{ } \mathrm{set} \text{ } j_k + 1 \equiv j_k \oplus 1 }{\underset{1 \leq i \leq n}{\prod}}     A^{j_i}_i    \bigg) \\ + \bigg( \underset{i = j_l + 1 , \text{ } \mathrm{set} \text{ } j_l + 1 \equiv j_l \oplus 1 }{\underset{1 \leq i \leq n}{\prod}}     A^{j_i}_i   \bigg) \text{ } \bigg]   \otimes \textbf{I} \bigg\}   \ket{\psi_{\mathrm{FFL}}} \bigg]           \bigg| \bigg|   <    \bigg(       \frac{8200 \sqrt{2} }{27}   \bigg)                   n^2 \sqrt{\epsilon }                   \text{. }
\end{align*} }

\noindent \textit{Proof of Lemma 9}. Up to multiplication of the quantum and Classical success biases which are both equal to $\frac{2}{3}$, and $\pm \mathrm{sign} \big( i , j_1 , \cdots j_n \big)$, the difference of terms with respect to the norm above is equivalent to,

{\small \begin{align*}
 \bigg| \bigg|    \bigg\{    \bigg[ \bigg(  
 \underset{1 \leq i \leq n}{\prod} A^{j_i}_i \bigg)       - \frac{2}{3} \bigg( \pm \mathrm{sign} \big( i , j_1 , \cdots , j_n \big) \bigg)   \bigg(  
\underset{i = j_k + 1 , \text{ } \mathrm{set} \text{ } j_k + 1 \equiv j_k \oplus 1 }{\underset{1 \leq i \leq n}{\prod}} A^{j_i}_i  \bigg)     \bigg] \otimes \textbf{I} \bigg\}  \ket{\psi_{\mathrm{FFL}}} \bigg| \bigg|   \text{. } 
\end{align*} } 

\noindent Furthermore, by the triangle inequality, there exists an upper bound,

{\small \begin{align*}
     \bigg| \bigg|       \bigg[       \bigg( \underset{1 \leq i \leq n}{\prod}  A^{j_i}_i \bigg)         - \bigg\{  \bigg( \underset{1 \leq i \leq n}{\prod} A^{j_i}_i \bigg) \bigg(  \frac{\pm A_k + A_l }{\sqrt{2}} \bigg) \bigg\}  \bigg]    \ket{\psi_{\mathrm{FFL}}}          \bigg| \bigg|   +  \frac{2}{3} \bigg[ \bigg| \bigg|   \bigg[      \bigg\{  \bigg( \underset{1 \leq i \leq n}{\prod} A^{j_i}_i \bigg) A_k \\ - \mathrm{sign} \big( i , j_1 , \cdots , j_n \big)         \bigg(         \underset{i = j_k + 1 , \text{ } \mathrm{set} \text{ } j_k + 1 \equiv j_k \oplus 1 }{\underset{1 \leq i \leq n}{\prod}}  A^{j_i}_i           \bigg)      \bigg\}         \otimes \textbf{I} \bigg]             \ket{\psi_{\mathrm{FFL}}}          \bigg| \bigg|  \\ +  \bigg| \bigg|       \bigg[      \bigg\{  \bigg( \underset{1 \leq i \leq n}{\prod} A^{j_i}_i \bigg) A_k - \mathrm{sign} \big( i , j_1 , \cdots , j_n \big)         \bigg(         \underset{i = j_l + 1 , \text{ } \mathrm{set} \text{ } j_l + 1 \equiv j_k \oplus 1 }{\underset{1 \leq i \leq n}{\prod}}  A^{j_i}_i           \bigg)      \bigg\}         \\ \otimes \textbf{I} \bigg]                     \ket{\psi_{\mathrm{FFL}}}          \bigg| \bigg| 
 \bigg]   \text{, } 
\end{align*} }

\noindent which can be further analyzed by observing that the first term is equal to,

{\small \begin{align*}
\bigg\{  \bigg[  \textbf{I}   - \textbf{I} \bigg(\frac{\pm A_k + A_l}{\sqrt{2}}  \bigg)  \bigg]     \otimes \textbf{I} \bigg\}  \ket{\psi_{\mathrm{FFL}}}  <        \sqrt{2 n \big( n-1 \big) \epsilon }         \text{, }
\end{align*} }

\noindent from the fact that the product of $A$ operations appearing in terms multiplying $\ket{\psi_{\mathrm{FFL}}}$ in the first term satisfy,

\begin{align*}
 \underset{1 \leq i \leq n}{\prod} A_i  = \textbf{I} \text{. } 
\end{align*}

\noindent To bound the second term, observe that it is equivalent to,

{\small \begin{align*}
   \bigg\{   \bigg[     \bigg( \underset{1 \leq i \leq n}{\prod} A^{j_i}_i \bigg)  -  \frac{2}{3} \big( \pm \mathrm{sign} \big( i , j_1 , \cdots , j_n \big)     \big) \bigg(  \underset{i \equiv j_1 + 1 , \text{ } \mathrm{set} \text{ }  j_1 + 1 \equiv j_1 \oplus 1}{ \underset{1 \leq i \leq n}{\prod} }       A^{j_i}_i      \bigg)       \bigg]    \otimes \textbf{I}   \bigg\}   \ket{\psi_{\mathrm{FFL}}} \text{, } 
\end{align*} } 

\noindent which can further be rearranged as,

{\small \begin{align*}
  \bigg\{  \bigg[  \text{ }    \textbf{I}  - \frac{2}{3} \big( \pm \mathrm{sign} \big( i , j_1 , \cdots , j_n \big) \big) \textbf{I}     \bigg] \otimes \textbf{I}
\bigg\}  \ket{\psi_{\mathrm{FFL}}}  =       \bigg\{  \textbf{I} \otimes \textbf{I} - \frac{2}{3} \big( \pm \mathrm{sign} \big( i , j_1 , \cdots , j_n \big) \big) \bigg( \textbf{I} \otimes \textbf{I} \bigg)  \bigg\}  \ket{\psi_{\mathrm{FFL}}} \\ <    n^2 \sqrt{\epsilon}   \text{. } 
\end{align*} }

\noindent Putting each of the previous estimates together, with arguments previous to those obtained from the previous subsection, implies,

{\small \begin{align*}
  \bigg| \bigg|  \bigg\{  \bigg( \underset{1 \leq i \leq n}{\prod}   A^{j_i}_i \bigg)   \otimes B_{kl} \bigg\}   \ket{\psi_{\mathrm{FFL}}}  - \frac{2}{3} \bigg[ \pm \bigg( \mathrm{sign} \big( i , j_1 , \cdots , j_n \big) \bigg[ \text{ } \bigg(     \underset{i = j_k + 1 , \text{ } \mathrm{set} \text{ } j_k + 1 \equiv j_k \oplus 1 }{\underset{1 \leq i \leq n}{\prod}}     A^{j_i}_i    \bigg) \text{ } \\ +      \bigg( \underset{i = j_l + 1 , \text{ } \mathrm{set} \text{ } j_l + 1 \equiv j_l \oplus 1 }{\underset{1 \leq i \leq n}{\prod}}     A^{j_i}_i   \bigg) \text{ } \bigg]   \otimes \textbf{I} \bigg)  \ket{\psi_{\mathrm{FFL}}} \bigg]           \bigg| \bigg|  \end{align*}
  
  \begin{align*} \leq       \sqrt{2  n \big( n - 1 \big) \epsilon} +  \bigg(  \frac{44}{6}     + \frac{100}{9} \bigg) n^2 \sqrt{\epsilon }  \end{align*}

  \begin{align*} <   \sqrt{2  n \big( n - 1 \big) \epsilon} + \frac{2200}{27}  n^2 \sqrt{\epsilon }  \end{align*}
  
  \begin{align*} \overset{(\mathrm{SEB})}{<} 5 \cdot \bigg\{   \sqrt{2 n \big( n -1 \big) } + \frac{2200}{27} n^2     \bigg\}  \sqrt{\epsilon}   \end{align*}

\begin{align*}
    < 5 \cdot \bigg\{                  \sqrt{2 n } +          \frac{2200}{27} n^2         \bigg\} \sqrt{\epsilon}
\end{align*}

\begin{align*}
    <   5 \cdot \bigg\{           \sqrt{2} +  \frac{2200}{27}       \bigg\} n^2 \sqrt{\epsilon }                   
\end{align*}

  \begin{align*}    <  \bigg(       \frac{8200 \sqrt{2} }{27}   \bigg) n^2 \sqrt{\epsilon}   \text{, }
\end{align*} }

\noindent where, in $(\mathrm{SEB})$ above,

{\small

\begin{align*}
    \sqrt{2  n \big( n - 1 \big) \epsilon} + \frac{2200}{27}  n^2 \sqrt{\epsilon } <   2 \cdot \sqrt{2  n \big( n - 1 \big) \epsilon + \bigg( \frac{2200}{27} \bigg)^2 n^4 \epsilon       }       <   2 \cdot \sqrt{ \bigg\{ 2 \cdot \bigg( \frac{2200}{27} \bigg)^2         n^4 -  2 n               \bigg\}    \epsilon                   }  \end{align*}
    
    \begin{align*} <     3 \cdot \sqrt{         2 \cdot \bigg( \frac{2200}{27} \bigg)^2         n^4  \epsilon        }         - 2 \cdot \sqrt{ 2 n \epsilon       }        \end{align*}
    
    \begin{align*} <     3 \cdot \sqrt{         2 \cdot \bigg( \frac{2200}{27} \bigg)^2         n^4  \epsilon        }   - 2 \cdot \sqrt{ 2 n \epsilon       } +    5  \cdot \sqrt{ 2 n \epsilon       }      \end{align*}
    
    \begin{align*}  <     3 \cdot \sqrt{         2 \cdot \bigg( \frac{2200}{27} \bigg)^2         n^4  \epsilon        }   - 2 \cdot \sqrt{ 2 \big( n - 1 \big)  \epsilon       } +   5  \cdot \sqrt{ 2 n \epsilon       }             \end{align*}

\begin{align*}
    =  2 \cdot \bigg\{ 3 \cdot \sqrt{         2 \cdot \bigg( \frac{2200}{27} \bigg)^2         n^4  \epsilon        }      +   \sqrt{n \big( n - 1 \big) \epsilon }   \bigg\}    
\end{align*}

    \begin{align*}   <        5  \cdot \bigg\{   \sqrt{2 n \big( n - 1 \big) }           +   \sqrt{       \bigg( \frac{2200}{27} \bigg)^2         n^4         }                     \bigg\} \sqrt{\epsilon }                   , 
\end{align*}

}

\noindent from which we conclude the argument. \boxed{}

\subsection{Frobenius norm upper bounds for Alice and Bob's strategies}

\noindent \textit{Proof of Lemma 1}. To obtain the first desired upper bound, observe that the expression with respect to the Frobenius norm is equivalent to,

{\small \begin{align*}
    \text{ }      \bigg| \bigg|      \bigg[           \bigg\{      \bigg(  \underset{1 \leq i \leq n}{\prod}     A^{j_i}_i   \bigg) - \bigg(    \mathrm{sign} \big( i , j_1 , \cdots , j_n \big) \times    \underset{ i \equiv j_4 + 1 , \text{ } \mathrm{set} \text{ } j+1 + 1 \equiv j_1 \oplus 1}{\underset{1 \leq i \leq n}{\prod}}     A^{j_i}_i            \bigg)           \bigg\}    \otimes \textbf{I}     \bigg]  \ket{\psi_{\mathrm{FFL}}}              \bigg| \bigg|_{\mathrm{F}}   \text{, } 
\end{align*} }

\noindent while to obtain the second desired upper bound, observe that the second expression that is taken with respect to the Frobenius norm is equivalent to,

{\small \begin{align*}
  \bigg| \bigg|  \bigg[    \bigg\{  \bigg(               \underset{1 \leq i \leq n}{\prod}   A^{j_i}_i       \bigg) -  \frac{2}{3} \bigg(   \pm \mathrm{sign} \big( i , j_1 , \cdots , j_n \big)  \times      \underset{i \equiv j_4 + 1 , \text{ } \mathrm{set} \text{ } j_4 + 1 \equiv j_4 \oplus 1}{\underset{1 \leq i \leq n}{\prod}}   A^{j_i}_i          \bigg)  \bigg\}   \otimes \textbf{I}             \bigg]  \ket{\psi_{\mathrm{FFL}}} \bigg| \bigg|_{\mathrm{F}}   \text{. } 
\end{align*} } 

\noindent In the first case, obtain the desired upper bound by writing,

{\small \begin{align*}
     \bigg\{                  6    +   4 \big( \frac{2}{3}  \big)       \bigg\}    n^2 \sqrt{\epsilon}    =      \frac{26}{3}    n^2 \sqrt{\epsilon}      < 9   n^2 \sqrt{\epsilon}   \text{, } 
\end{align*} }

\noindent while in the second case, obtain the desired upper bound by writing,

{\small \begin{align*}
 \bigg\{     \frac{17}{2}        +     6 \big( \frac{2}{3} \big)        \bigg\}    n^2 \sqrt{\epsilon}         =   \frac{87}{6}           n^2 \sqrt{\epsilon} < \frac{44}{3}  n^2 \sqrt{\epsilon}  \text{, } 
\end{align*} }

\noindent from which we conclude the argument. \boxed{}

\section{Conclusion}

\noindent In this paper we obtained error bounds for the $\mathrm{XOR}^{*}$ and FFL games. Motivated by recent developments in Quantum information theory presented in {\color{blue}[37]} regarding optimality we formally describe how entanglement in Quantum strategies can be developed for the FFL game. In particular, while the probabilities that Alice and Bob answer questions from the referee's probability are $1 \big/ 3$ in the FFL game as opposed to $1 \big/ 4 $ in the XOR game, the error bounds characterized in this work are related to the intertwining operation $T^{\mathrm{FFL}}$ and hence also to the canonical strategy $\ket{\widetilde{\psi_{\mathrm{FFL}}}}$. By leveraging several observations between XOR and $\mathrm{XOR}^{*}$ games obtained in {\color{blue}[5]} we not only drew the attention of the reader to how the optimal value can be used to obtain desired prefactors to the upper bounds provided in \textbf{Lemma} \textit{9}, but also to the upper bounds for the error bounds stated in \textbf{Lemma} \textit{5}, \textbf{Lemma} \textit{6} and \textbf{Lemma} \textit{7}. It continues to remain of interest to introduce alternatives to the representation-theoretic intertwining operation to formulate error bounds in games with less regular structure.            

\section{Data Availability Statement}

The manuscript has no associated data.

\section{Conflicts of Interest Statement}

On behalf of all authors, the corresponding author states that there is no conflict of interest.





\section{References}

\noindent [1] Asher, P., Terno, D.R.: Quantum information and relativity theory. Rev. Mod.
Phys. 76(1), 93–123 (2004). https://doi.org/10.1103/RevModPhys.76.93.

\bigskip

\noindent [2] Benedetti, M., Coyle, B., Fiorentini, M., Lubasch, M., Rosenkranz, M.: Variational Inference with a Quantum Computer. Phys Rev Applied 16(044057) (2021). https://doi.org/10.1103/PhysRevApplied.16.044057.

\bigskip

\noindent [3] Bittel, L., Kliesch, M.: Training variational quantum algorithms is np-hard. Phys-
ical Review Letters 127(120502) (2021) https://doi.org/10.1103/PhysRevLett.
127.120502.

\bigskip

\noindent [4] Bylicka, B., Chru´sci´nski, D., Maniscalco, S.: Non-markovianity and reservoir
memory of quantum channels: a quantum information theory perspective. Sci.
Rep. 4(5720) (2014). https://doi.org/10.1038/srep05720.

\bigskip

\noindent [5] Catani, L., Faleiro, R., Emeriau, P.E., Mansfield, S., Pappa, A.: Connecting
xor and xor* games. Phys. Rev. A 109(012427) (2024). https://doi.org/10.1103/
PhysRevA.109.012427.

\bigskip

\noindent [6] Chen, H., Vives, M., Metcalf, M.: Parametric amplification of an optomechanical
quantum interconnect. Physical Review Research 4(043119) (2022). https://doi.
org/10.1103/PhysRevResearch.4.043119.

\bigskip

\noindent [7] Cong, I., Duan, L.: Quantum discriminant analysis for dimensionality reduction
and classification. New Journal of Physics 18(073011) (2016). https://doi.org/10.
1088/1367-2630/18/7/073011.

\bigskip

\noindent [8] Cooney, T., Junge, M., Palazuelos, C., al.: Rank-one quantum games. comput.
complex. 24, 133–196 (2015). https://doi.org/10.1007/s00037-014-0096-x. 

\bigskip

\noindent [9] DiVicenzo, D.P., Loss, D.: Quantum information is physical. Superlatt.
Microstrut. 23(3-4), 419–432 (1998). https://doi.org/10.1006/spmi.1997.0520.

\bigskip

\noindent [10] Du, J., Li, H., Xu, X.: Experimental realization of quantum games on a quantum
computer. Phys. Rev. Lett. 88(137902). https://doi.org/10.1103/PhysRevLett.88.
137902.

\bigskip

\noindent [11] Du, J., Li, H., Xu, X., Zhou, X., Han, R.: Phase-transition-like behaviour of
quantum games. J. Phys. A: Math. Gen. 36(6551) (2003). https://doi.org/10.
1088/0305-4470/36/23/318.

\bigskip

\noindent [12] Eisert, J., Wilkens, M., Lewenstein, M.: Quantum games and quantum strategies.
Phys. Rev. Lett. 83(3077) (1999). https://doi.org/10.1103/PhysRevLett.83.3077.

\bigskip

\noindent [13] Ewe, W.-B., Koh, D.E., Goh, S.T., Chu, H.-S., Png, C.E.: Variational quantum-
based simulation of waveguide modes. IEEE Transactions on Microwave Theory
and Techniques 70(5), 2517–2525 (2022). https://doi.org/10.1109/TMTT.2022.
3151510.

\bigskip

\noindent [14] Flitney, A.P., Abbott, D.: An introduction to quantum game theory. Fluct. Noise
Lett. 2(4) (2002). https://doi.org/10.1142/S0219477502000981.

\bigskip

\noindent [15] Flitney, A.P., Abbott, D.: Quantum games with decoherence. J. Phys. A: Math.
Gen. 38(449) (2005). https://doi.org/10.1088/0305-4470/38/2/011.

\bigskip

\noindent [16] Garg, D., Ikbal, S., Srivastava, S.K., Vishwakarma, H., Karanam, H., Subrama-
niam, L.V.: Quantum embedding of knowledge for reasoning. Advance in Neural
Information Processing Systems 32 (2019).
\bigskip

\noindent [17] Genoni, M.G., Tufarelli, T.: Non-orthogonal bases for quantum metrology. Jour-
nal of Physics A: Mathematical and Theoretical 52(43) (2019). https://doi.org/
10.1088/1751-8121/ab3fe0.

\bigskip

\noindent [18] Gidi, J.A., Candia, B., Munoz-Moller, A.D., Rojas, A., Pereira, L., Munoz,
M., Zambrano, L., Delgado, A.: Stochastic optimization algorithms for quantum applications. Phys.Rev.A 108(032409) (2023). https://doi.org/10.1103/
PhysRevA.108.032409.

\bigskip

\noindent [19] Givi, P., Daley, A.J., Mavriplis, D., Malik, M.: Quantum speedup for aeroscience
and engineering. AIAA 58(8) (2020)
[20] Hadiashar, S.B., Nayak, A., Sinha, P.: Optimal lower bounds for quantum learning
via information theory. IEEE Transactions on Information Theory 70(3), 1876–
1896 (2024). https://doi.org/10.1109/TIT.2023.3324527.

\bigskip

\noindent [21] Head-Marsden, K., Flick, J., Ciccarino, C.J., Narang, P.: Quantum information
and algorithms for correlated quantum matter. Chem. Rev. 121(5), 3061–3120
(2021). https://doi.org/10.1021/acs.chemrev.0c00620.

\bigskip

\noindent [22] Horodecki, P., al.: Five open problems in quantum information theory. PRX
Quantum 3(1), 010101 (2022). https://doi.org/10.1103/PRXQuantum.3.010101.

\bigskip

\noindent [23] Hur, T., Kim, L., Park, D.K.: Quantum convolutional neural network for classical
data classification. Quantum Machine Intelligence 4(3) (2022). https://doi.org/
10.1007/s42484-021-00061-x.

\bigskip

\noindent [24] Holmes, Z., Coble, N.J., Sornborger, A.T., Subasi, Y.: On nonlinear transfor-
mations in quantum computation. Phys. Rev. Research 5(013105) (2023). https:
//doi.org/10.1103/PhysRevResearch.5.013105.

\bigskip

\noindent [25] Iqbal, A., Cheon, T.: Constructing quantum games from nonfactorizable
joint probabilities. Phys. Rev. E 76(061122) (2007). https://doi.org/10.1103/
PhysRevE.76.061122.

\bigskip

\noindent [26] Jing, H., Wang, Y., Li, Y.: Data-driven quantum approximate optimization
algorithm for cyber-physical power systems. arXiv: 2204.00738 (2022). https:
//doi.org/10.48550/arXiv.2204.00738.

\bigskip

\noindent [27] Kolokoltsov, V.N.: Dynamic quantum games. Dyn. Games Appl. 12, 552–573
(2022). https://doi.org/10.1007/s13235-021-00389-w.

\bigskip

\noindent [28] Kubo, K., Nakagawa, Y.O., Endo, S., Nagayama, S.: Variational quantum simula-
tions of stochastic differential equations. Physical Review A 103(052425) (2021). https://doi.org/10.1103/PhysRevA.103.052425.

\bigskip

\noindent [29] Kribs, D.W.: A quantum computing primer for operator theorists. Linear Algebra
and its Applications 400, 147–167 (2005). https://doi.org/10.48550/arXiv.math/
0404553.

\bigskip

\noindent [30] Li, R.Y., Di Felice, R., Rohs, R., Lidar, D.A.: Quantum annealing versus classi-
cal machine learning applied to a simplied computational biology problem. npj
Quantum Information 4(14) (2008). https://doi.org/10.1038/s41534-018-0060-8. 

\bigskip

\noindent [31] Mahdian, M., Yeganeh, H.D.: Toward a quantum computing algorithm to quan-
tify classical and quantum correlation of system states. Quantum Information
Processing 20(393) (2021). https://doi.org/10.1007/s11128-021-03331-6.

\bigskip

\noindent [32] Maldonado, T.J., Flick, J., Krastanov, S., Galda, A.: Error rate reduction of
single-qubit gates via noise-aware decomposition into native gates. Scientific
Reports 12(6379) (2022). https://doi.org/10.1038/s41598-022-10339-0.

\bigskip

\noindent [33] Manby, F.R., Stella, M., Goodpaster, J.D., Miller, T.F.: A simple, exact
density-functional-theory embedding scheme. Journal of Chemical Theory and
Computation 8(8), 2564–2568 (2012). https://doi.org/10.1021/ct300544e.

\bigskip

\noindent [34] Mensa, S., Sahin, E., Tacchino, F., Barkoutsos, P.K., Tavernelli, I.: Quantum
machine learning framework for virtual screening in drug discovery: a prospective
quantum advantage. Mach. Learn.: Sci. Technol. 4(015023) (2023). https://doi.
org/10.1088/2632-2153/acb900.

\bigskip

\noindent [35] Nan Sheng, H.M., Govono, M., Galli, G.: Quantum embedding theory for
strongly-correlated states in materials. J. Chem. Theory Comput. 17(4), 2116–
2125 (2021). https://doi.org/10.1021/acs.jctc.0c01258.

\bigskip

\noindent [36] Nawaz, A., Toor, A.H.: Quantum games with correlated noise. J. Phys. A: Math.
Gen. 39(29), 9321 (2006). https://doi.org/10.1088/0305-4470/39/29/022.

\bigskip

\noindent [37] Ostrev, D.: The structure of nearly-optimal quantum strategies for the non-local
xor games. Quantum Information and Computation 16(13-14), 1191–1211 (2016). https://doi.org/10.26421/QIC16.13-14-6.

\bigskip

\noindent [38] Paine, A.E., Elfving, V.E., Kyriienko, O.: Quantum kernel methods for solving
differential equations. Physical Review A 107(032428) (2023). https://doi.org/10.
1103/PhysRevA.107.032428.

\bigskip

\noindent [39] Paudel, H.P., Syamlal, M., Crawford, S.E., Lee, Y.-L., Shugayev, R.A., Lu, P.,
Ohodnicki, P.R., Mollot, D., Duan, Y.: Quantum computing and simulations for
energy applications: Review and perspective. ACS Eng. Au 3, 151–196 (2022). https://doi.org/10.1021/acsengineeringau.1c00033.

\bigskip

\noindent [40] Piispanen, L., Pfaffhauser, M., Wootton, J., al.: Defining quantum games.
EPJ Quantum Technol. 12(7) (2025). https://doi.org/10.1140/epjqt/
s40507-025-00308-7.

\bigskip

\noindent [41] Przhiyalkovskiy, Y.V.: Quantum process in probability representation of quan-
tum mechanics. Journal of Physics A: Mathematical and Theoretical 55(085301)
(2022). https://doi.org/10.1088/1751-8121/ac4b15.

\bigskip

\noindent [42] Rigas, P.: Variational quantum algorithm for measurement extraction from the
navier-stokes, einstein, maxwell, b-type, lin-tsien, camassa-holm, dsw, h-s, kdv-b,non-homogeneous kdv, generalized kdv, kdv, translational kdv, skdv, b-l and airy
equations (v4). arXiv: 2209.07714 (2022). https://doi.org/10.48550/arXiv.2209.
07714.

\bigskip

\noindent [43] Slofstra, W.: Lower bounds on the entanglement needed to play xor non-local
games. Journal of Mathematical Physics 52(10), 102202 (2011). https://doi.org/
10.1063/1.3652924.

\bigskip

\noindent [44] Tsakiroglou, M., Liliopoulos, I., Varsamis, G.D., al.: Graph-encoded strategic
interactions in multiplayer quantum games. Quantum Information Processing
25(109) (2026). https://doi.org/10.1007/s11128-026-05127-y.

\bigskip

\noindent [45] Van Dam, W., Sasaki, Y.: Quantum algorithms for problems in number theory,
algebraic geometry, and group theory. Diversities in Quantum Computation and
Quantum Information, 79–105 (2012). https://doi.org/10.1142/9789814425988
0003.

\bigskip

\noindent [46] Vedral, V.: The role of relative entropy in quantum information theory. Rev. Mod.
Phys. 74(1), 197–234 (2002). https://doi.org/10.1103/RevModPhys.74.197.

\bigskip

\noindent [47] Wang, Y., Krstic, P.S.: Multistate transition dynamics by strong time-dependent
perturbation in nisq era. J. Phys. Commun. 7(075004) (2023). https://doi.org/10.
1088/2399-6528/ace67a.

\bigskip

\noindent [48] Werner, R.F.: Quantum information theory — an invitation. Springer Tracts in
Modern Physics 173 (2001). https://doi.org/10.1007/3-540-44678-8 2.

\bigskip

\noindent [49] Whitaker, A.: Einstein, bohr and the quantum dilemma: From quantum theory to
quantum information. Cambridge University Press ISBN-13 978-0-521-67102-
6 (2006). https://doi.org/www.cambridge.org/9780521671026.

\bigskip

\noindent [50] Wilde, M.M.: Recoverability in quantum information theory. Proc. A 471(2182)
(2015). https://doi.org/10.1098/rspa.2015.0338.

\bigskip

\noindent [51] Yukalov, V.I., Sornette, D.: Quantum decision theroy as quantum theory of measurement. Phys. Lett. A 372(46), 6867–6871 (2008). https://doi.org/10.1016/j.
physleta.2008.09.053.

\bigskip

\noindent [52] Zhang, S.: Quantum strategic game theory. ITCS ’12’, 39–59 (2012). https://doi.
org/10.1145/2090236.2090241.

\bigskip

\noindent [53] Zhao, L., Zhao, Z., Rebentrost, P., Fitzsimons, J.: Compiling basic linear algebra
subroutines for quantum computers. Quantum Machine Intelligence 3(21) (2021). https://doi.org/10.1007/s42484-021-00048-8.


\end{document}